\documentclass{aastex6}

\shorttitle{IC 1805}
\shortauthors{Sung et al.}

\begin{document}

\title{An Optical and Infrared Photometric Study of the Young Open Cluster 
       IC 1805 in the Giant H II Region W4\footnote{The optical imaging data
       in this article were gathered with two facilities: the AZT-22 1.5m
       telescope at Maidanak Astronomical Observatory in Uzbekistan, and 
       the Canada-France-Hawaii Telescope (CFHT) which is operated 
       by the National Research Council of Canada, the Institut National des 
       Sciences de l'Univers of the Centre National de la Recherche 
       Scientifique of France, and the University of Hawaii.}\footnote{This work has made use of data from the European Space Agency (ESA)
mission {\it Gaia} (\url{http://www.cosmos.esa.int/gaia}), processed by
the {\it Gaia} Data Processing and Analysis Consortium (DPAC,
\url{http://www.cosmos.esa.int/web/gaia/dpac/consortium}). Funding
for the DPAC has been provided by national institutions, in particular
the institutions participating in the {\it Gaia} Multilateral Agreement.}}

\author{Hwankyung Sung\altaffilmark{1,2}, Michael S. Bessell\altaffilmark{3}, 
        Moo-Young Chun\altaffilmark{4}, Jonghyuk Yi\altaffilmark{5}, 
        Y. Naz\'e\altaffilmark{6}, Beomdu Lim\altaffilmark{4}, 
        R. Karimov\altaffilmark{7}, G. Rauw\altaffilmark{6},
        Byeong-Gon Park\altaffilmark{3}, and Hyeonoh Hur\altaffilmark{8}}

\altaffiltext{1}{Department of Astronomy and Space Science, Sejong University,
    209 Neungdong-ro, Kwangjin-gu, Seoul 05006, Korea; sungh@sejong.ac.kr}
\altaffiltext{2}{Visiting Researcher, Korea Astronomy and Space Science 
    Institute}
\altaffiltext{3}{Research School of Astronomy \& Astrophysics, The Australian
    National University, Canberra, ACT 2611, Australia}
\altaffiltext{4}{Korea Astronomy and Space Science Institute,
    776 Daedeokdae-ro, Yuseong-gu, Daejeon 34055, Korea}
\altaffiltext{5}{SELab, Inc., 8 Nonhyeon-ro 150-gil, Gangnam-gu, Seoul 06049, Korea}
\altaffiltext{6}{Groupe d'Astrophysique des Hautes Energies, Institut 
    d'Astrophysique et de G\'eophysique, Universit\'e de Li\`ege, 
    All\'ee du 6 Ao\^ut, 19c, B\^at B5c, 4000 Li\`ege, Belgium}
\altaffiltext{7}{Ulugh Beg Astronomical Institute, Uzbek Academy of Sciences,
    33 Astronomical Street, Tashkent 700052, Uzbekistan}
\altaffiltext{8}{Daegu National Science Museum, 20, Techno-daero-6-gil,
    Yuga-myeon, Dalseong-gun, Daegu 43023, Korea}

\begin{abstract}
We present deep wide-field optical CCD photometry and mid-infrared {\it
Spitzer}/IRAC and MIPS 24$\mu m$ data for about 100,000 stars in the young open cluster
IC 1805. The members of IC 1805 were selected from their location
in the various color-color and color-magnitude diagrams, and the presence of H$\alpha$
emission, mid-infrared excess emission, and X-ray emission.
The reddening law toward IC 1805 is nearly normal ($R_V$  = 3.05 $\pm$ 0.06).
However, the distance modulus
of the cluster is estimated to be 11.9 $\pm$ 0.2 mag ($d = 2.4 \pm
0.2$ kpc) from the reddening-free color-magnitude diagrams,
which is larger than the distance to the nearby massive star-forming region
W3(OH) measured
from the radio VLBA astrometry. We also determined the age of IC 1805
($\tau_{MSTO}$ = 3.5 Myr). In addition, we critically compared the age and mass
scale from two pre-main-sequence evolution models. The initial mass
function with a Salpeter-type slope of $\Gamma$  = -1.3 $\pm$ 0.2
was obtained and the total mass of IC 1805 was estimated to be about
 2700 $\pm$ 200 M$_\odot$. Finally, we found our distance determination
to be statistically consistent with the Tycho-Gaia Astrometric Solution Data Release 1,
within the errors. The proper motion of the B-type stars shows
an elongated distribution along the Galactic plane, which could be
explained by some of the B-type stars being formed in small clouds dispersed
by previous episodes of star formation or supernova explosions.

\end{abstract}


\section{INTRODUCTION}

The young open cluster IC 1805 is one of the core clusters of the Cas OB6 association
and is surrounded by the giant HII region W4, which is at the center of three
massive HII regions W3/W4/W5 in the Perseus spiral arm of the Galaxy.  
The Perseus region is one of most active star forming regions (SFRs) in the Galaxy.
The physical properties of the giant HII region or
the relation between W4 and the massive stars in IC 1805 are relatively well studied.
The relatively small masses of the molecular clouds in the region together with the lower metallicity,
higher gas temperature, and lower gas surface density are considered to be unfavorable
conditions for star formation. 
Although there are 8 known O-type stars in IC 1805, the largest
number of O stars in the northern young open clusters,
the stellar content, especially of low-mass stars, the star formation history, and the shape of
the initial mass function of the whole area of  IC 1805 are not well known, and therefore
are all interesting issues to study.

The giant HII region W4 is also known as a Galactic chimney/superbubble 
first proposed and discovered from the high resolution H I observations of
the Perseus arm by \citet{ntd96,ntd97}.
The reality of a Galactic superbubble was confirmed by \citet{dts97} from a 
wide-field H$\alpha$ image and analysis of the ionization balance. They
also estimated the age of the superbubble to be between 6.4 -- 9.6 Myr.
Based on the estimated age of the superbubble, they argued that the massive 
wind and supernova explosions from
an earlier generation of stars before the formation of IC 1805, was responsible
for the formation of the 230 pc superbubble.
Later, \citet{rsh01} found a much larger H$\alpha$ loop extending about 1.3 kpc
above the Galactic plane. They suggested that the formation of such a gigantic 
superbubble may take 10 -- 20 Myr or more, which implies the existence of an 
even older generation of stars in the region. \citet{gv89} also suggested
the existence of such an old group of stars from the broadening of the MS band
of early B type stars. \citet{chs00} found that about 39\% of the cluster population,
identified in the K' images of 32 IRAS point sources distributed in the Cas OB6
association, is embedded in small clouds located as far as 100 pc from the
W3/W4/W5 region, and speculated that these small clouds are fragments of
a cloud complex dispersed by previous episodes of massive star formation.

From the similarity in age
between stars in IC 1795 and IC 1805, \citet{owkw05} suggested that
the formation of IC 1795, as well as IC 1805, was triggered by the massive
stars of the earliest 
generation, and that IC 1805 may be located on the edge of the shell.
There are 8 (or 9 depending on the spectral type of MWC 50 = VSA 113)\footnote{
O9Ve by \citet{i70}, O9.5Ve by \citet{sh99},  Be by \citet{mjd95},and
B2 by \citet{wsr11}} known O stars in IC 1805, which are considered to be the triggering 
source of massive star formation in the high density layer of the eastern
part of W3 \citep{chs00,fms13}. Besides the star formation of the whole
W3/W4 regions, several small scale star formation events triggered by the hot
massive stars in IC 1805 were also investigated for the bright
rim clouds (BRCs) 5 and 7 \citep{osp02,fms13,pcp14}. 

The spatial distribution of the young stellar objects (YSOs) within a SFR, gives 
important information on the embedded physical processes that 
influence star formation in the region \citep{klb12}.
Although there are 8 O-type stars in IC 1805, 
the surface density of stars in IC 1805 is very sparse, and therefore the region
is called an aggregate \citep{gv89} or OB association \citep{owkw05}, 
rather than an open cluster. The stellar IMF of
IC 1805 is also another important issue given its many O type stars, 
although situated at a large Galacto-centric distance with unfavorable conditions
for star formation, such as lower surface density of molecular gas and 
relatively higher gas temperature due to lower metallicity.
The IMF of IC 1805 was investigated by
\citet{sl95,mjd95,nbes95}. \citet{sl95} derived a slightly shallow IMF
($\Gamma = -1.1 \pm 0.2$), while \citet{mjd95,nbes95} obtained
nearly normal IMF slopes; however, their work was limited to
massive stars. In addition, photometric studies with modern CCDs tend to be
relatively shallow and limited to the central region only.

The multiplicity fraction of massive stars is another important research topic as 
the binarity or multiplicity of stars is a direct result of star formation
processes \citep{dk13}. The multiplicity fraction of O stars is being actively
investigated by \citet{rdb04,dbrm06,hgb06,rn16}. Currently 
three O stars (HD 15558, BD+61 497, \& BD+61 498) have been identified
as double-lined spectroscopic binary (SB2) systems. 
 \citet{dbrm06} obtained a large mass ratio between
the primary and the secondary of HD 15558, and suspect
that HD 15558 is a massive triple system. However, the current value
of the binary fraction of IC 1805 is lower than that of other young open
clusters \citep{sdmdk12}. 

In this study we provide deep wide-field optical CCD photometry
of the young open cluster IC 1805. In addition, for a complete and
homogeneous census of low-mass pre-main sequence (PMS) stars
with thick circumstellar disks, we obtained the mid-infrared (MIR)
magnitudes of objects from the archival {\it Spitzer}/IRAC
\citep{irac} and MIPS \citep{mips} 24$\mu m$ images.
The published X-ray source lists were also used for the selection of
cluster members. Based on the large volume of photometric data over the $41'
\times 45'$ area of IC 1805, we investigated the reddening law, the
reddening, distance, age, and the IMF of IC 1805.

This paper is organized as follows. The optical photometry
 and photometry of {\it Spitzer}/IRAC and MIPS 24$\mu m$ images
are described and compared in section 2. Optical and mid-infrared
(MIR) photometric data for about 100000 stars are presented.
The cross match with X-ray
emission objects from {\it Chandra} and {\it XMM-Newton} X-ray 
observations is also performed in the section. Some properties of
X-ray emission objects were analyzed in the section.
Membership selection is described in section 3 including
a detailed description of
the selection of PMS stars with H$\alpha$ emission,
the classification of YSOs in the MIR diagrams and
selection of MIR excess emission stars,  X-ray emission stars, and
massive and intermediate-mass members. 
Fundamental parameters, such as reddening, the reddening law,
distance, and radius of IC 1805 are obtained in section 4.
The Hertzsprung-Russell diagram (HRD) is constructed in section
5. Also in section 5, the IMF and age of IC 1805 is derived,
the mass and age of PMS stars from two popular PMS evolution
models are compared, and the total mass of IC 1805 is estimated.
In section 6, we present some discussion on the distance of the W3/W4 regions,
the parallax and proper motion data from the {\it Gaia} astrometric mission are 
analyzed, and some discussion is made on the star formation process of massive 
O- and B-type stars in IC 1805. The star formation history of the IC 1805/Cas OB6
association and its relation with the high mass X-ray binary LS I +61 303
is also discussed in the section. The summary and conclusions are given in section 7.

\section{Observations}

\subsection{Optical Photometry}

For a study of the IMF and the star formation
history of the young open cluster IC 1805, we obtained deep wide-field $VRI$
and H$\alpha$ images of IC 1805 using the CFH12K mosaic CCD camera of the
CFHT on 2002 January 6 and 7. 
We also observed several regions in IC 1805, for a study of the reddening and 
massive star content, using the SITe 2000$\times$800 
CCD (Maidanak 2k CCD) and standard $UBVRI$ filters of the AZT-22 1.5m 
telescope at  the Maidanak Astronomical Observatory in Uzbekistan.
Later, we obtained additional images of the central region of IC 1805 with
the Fairchild 486 CCD (SNUCam - \citealt{lsb09,ikc10}) and $UBVI$
and H$\alpha$ filters of the AZT-22 telescope.
The optical observations are summarized in Table \ref{tab_log}.

\begin{deluxetable*}{c|l|l|c|l|c}
\tablecolumns{6}
\tabletypesize{\scriptsize}
\tablecaption{Observation Log \label{tab_log} }
\tablewidth{0pt}
\tablehead{
\colhead{Telescope} & \colhead{detector} & \colhead{Date of Obs.} & 
\colhead{ Region } & \colhead{Exposure Time} & \colhead{Seeing\tablenotemark{a} ($''$)} }

\startdata
     &        & 2002. 1. 6 & Center & $I$: 75s$\times$3, $R$: 150s$\times$3,
$V$: 150s$\times$3, H$\alpha$: 1875s$\times$3 & 0.86 $\pm$ 0.06 \\
CFHT & CFH12K & 2002. 1. 7 & North & $I$: 6s, 75s$\times$3, $R$: 10s, 
     150s$\times$3, $V$: 10s, 150s$\times$3, H$\alpha$: 900s & 0.82 $\pm$ 0.07 \\
     &        & 2002. 1. 7 & South & $I$: 6s, 75s$\times$3, $R$: 10s, 150s$\times$3,
$V$: 10s, 150s$\times$3, H$\alpha$: 900s$\times$3 & 0.82 $\pm$ 0.09 \\ \hline
       &                      & 2003. 8. 18 & F1, F2, F3, F4 & 
$U$: 600s, 15s, $B$: 300s, 7s, $V$: 180s, 5s, $R$: 90s, 5s, $I$: 60s, 5s & 
0.98 $\pm$ 0.02 \\
AZT-22 & SITe & 2004. 12. 25 & F5, F6, F7, F8 & 
$U$: 600s, 15s, $B$: 300s, 7s, $V$: 180s, 5s, $R$: 90s, 5s, $I$: 60s, 5s & 
1.21 $\pm$ 0.14 \\
       & 2000$\times$800  & 2004. 12. 30 & F9, F10, F11   & 
$U$: 600s, 15s, $B$: 300s, 7s, $V$: 180s, 5s, $R$: 90s, 5s, $I$: 60s, 5s & 
1.18 $\pm$ 0.17 \\ 
       &                      &              & F12, F13, F14  & &  \\ \hline
AZT-22 & Fairchild 486 & 2007. 10. 7 & C1 & $U$: 600s, 15s, $B$: 300s, 7s, 
$V$: 180s, 5s, $I$: 60s, 5s, H$\alpha$: 600s, 30s & 1.5 \\
       & (SNUCam)      & 2009. 1. 19 & C2 & $U$: 600s, 15s, $B$: 300s, 7s, 
$V$: 180s, 5s, $I$: 60s, 5s, H$\alpha$: 600s, 30s & 0.9 \\
\enddata
\tablenotetext{a}{Average and standard deviation of the FWHM of stellar 
profiles in long exposure $V$ images}

\end{deluxetable*}

\subsubsection{CFH12K Observations}

{\it 2.1.1.1. Observation and Standardization}

\begin{figure*}
\epsscale{0.9}
\plotone{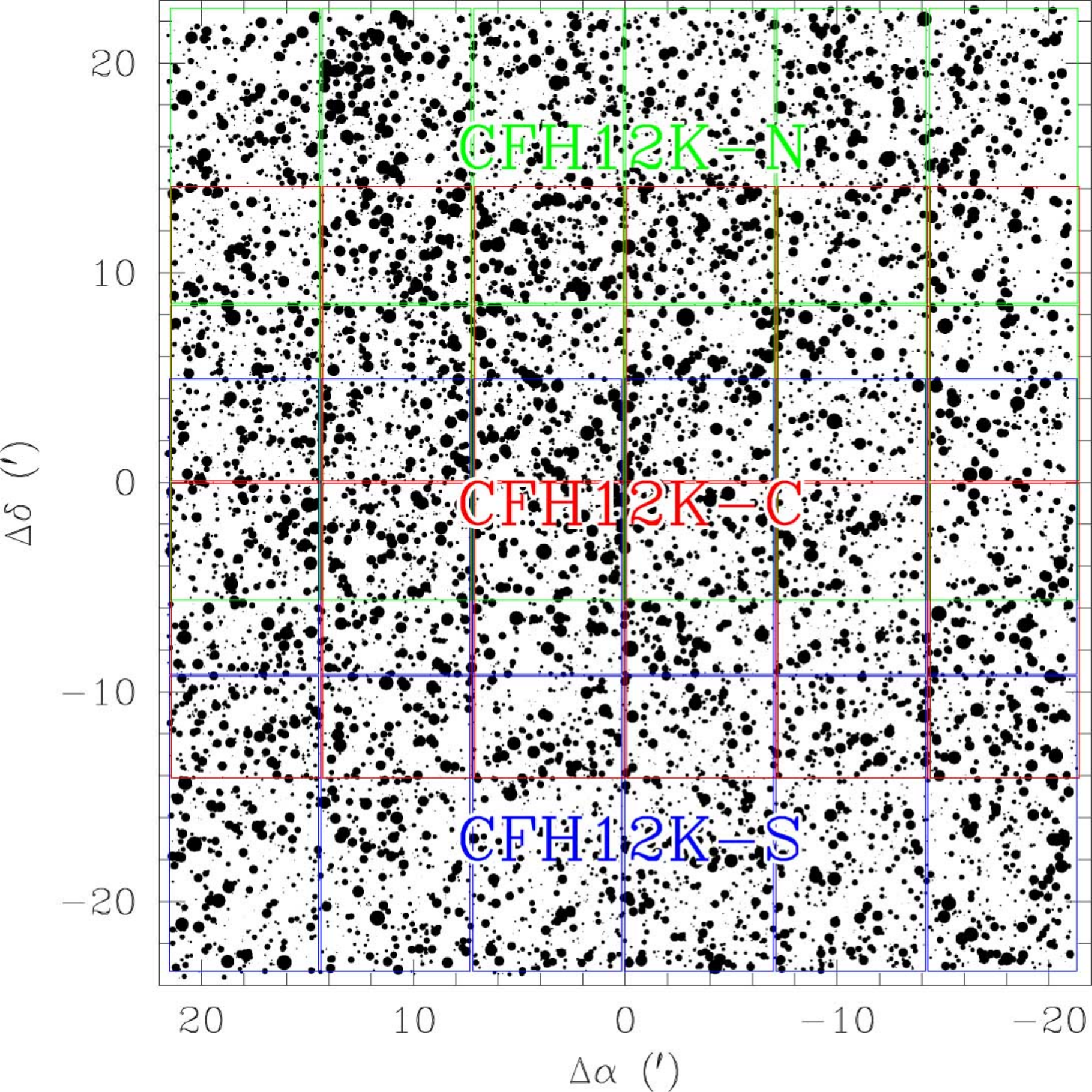}
\caption{Finder chart of IC 1805 for the stars brighter than $I$ = 18 from
CFH12K observations. The size of the dots is proportional to the brightness of the star. 
Squares represent a schematic view of the CFH12K mosaic CCD camera,
and three different pointings are drawn in different colors. The name
of each pointing is given near the center of each region.
The position of stars is relative to the brightest star HD 15558 [$\alpha$(
J2000) = 2$^h$ 32$^m$ 42.$^s$54, $\delta$(J2000) = +61$^\circ$ 27$'$ 21$\farcs$6].
\label{cfh_map} }
\end{figure*}

Deep wide-field $VRI$ and H$\alpha$ photometry was obtained for the young 
open cluster IC 1805 with the CFH12K, a $6 \times 2$ mosaic CCD camera
of the 3.6m CFHT. We obtained images of the central region  (we refer to
this region as ``CFH12K-C'') on 2002 January 6, and observed the North
(``CFH12K-N'') and South (``CFH12K-S'') regions
on 2002 January 7. The observed three regions largely overlap each other
and so the total surveyed area of IC 1805 is about $43' \times$ 45$'$, 
as shown in Figure \ref{cfh_map}. To fill in the gaps between CCD chips,
we used a three-point dithering pattern for a given pointing. The central
wavelength of the H$\alpha$ filter is 6584 $\AA$ with a bandwidth of 76 $\AA$.
The $VRI$ filters used at CFHT are Mould interference filters which have a more
rectangular responses than the standard colored glass filters \citep{msb90}.
Transformation to the standard Johnson-Cousins $VRI$ system requires
a multi-linear transformation in $R$, and is well documented in \citet{sbc08}.

The exposure times used on 2002 January 6 were 3 $\times$ 75s in $I$, 
3 $\times$ 150s in $V$ and $R$, and 3 $\times$ 1875s in H$\alpha$. 
On 2002 January 7, together with the long-exposure images,
we also obtained a short-exposure image for each 
$VRI$ filter to enable the photometry of bright stars. Unfortunately, due to the
limited observing time for IC 1805 (the main target was NGC 2264 -
see \citealt{sbc08}), only one image for CFH12K-N 
and three images for CFH12K-S 
were obtained in H$\alpha$. The exposure time used 
for H$\alpha$ on the second night was 900s, and therefore the photometric
depth in H$\alpha$ in the extreme north and extreme south ($|\Delta \delta|
\gtrsim 14'$) is shallower than that for the central region. 
The mean value of the seeing was about 0$\farcs$8
in the 150s-exposure $V$ images. The instrumental signatures were removed using
the IRAF/MSCRED package. Instrumental magnitudes were obtained using
the IRAF/DAOPHOT package via point-spread function (PSF) fitting. 


\floattable
\begin{figure*}
\epsscale{1.1}
\plotone{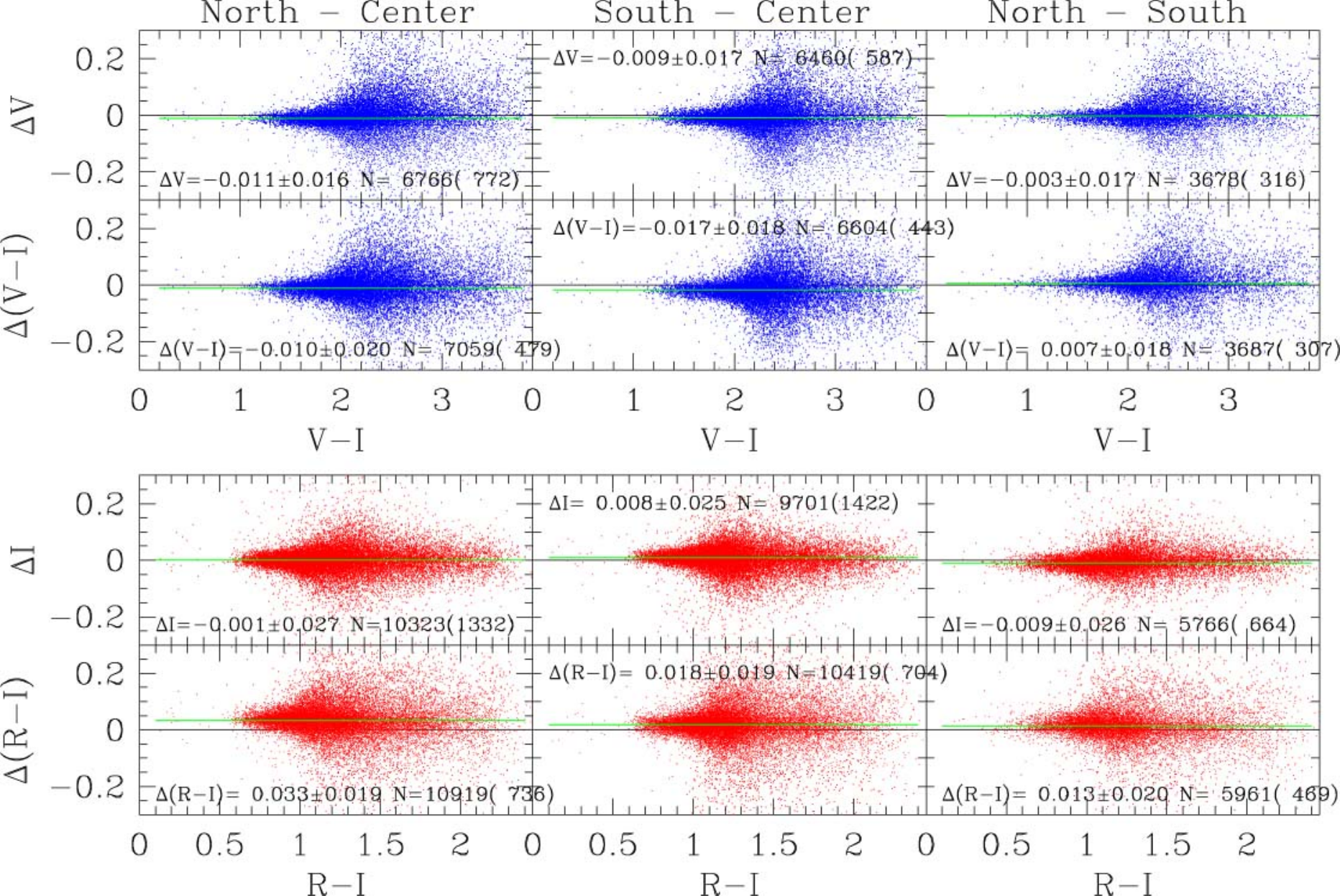}
\caption{Internal consistency of CFH12K data. The meaning of $\Delta$ is 
shown above the first panel of each column. The statistics were obtained
for the stars brighter than $V \leq 21$ mag for $\Delta V$ and $\Delta (V-I)$, 
and $I \leq 19.5$ mag for $\Delta I$ and $\Delta (R-I)$ from a successive 
exclusion of data with a large deviation ($> 2.5 \sigma$) from the mean.
The number in parenthesis represents the number of stars excluded from
the statistics.  \label{cfh_delta} }
\end{figure*}
\floattable
\begin{deluxetable*}{cccccccccccccc@{}c@{}c@{}c@{}c@{}c@{}c@{}cllrl}
\tablecolumns{26}
\tabletypesize{\scriptsize}
\rotate
\tablecaption{Photometric Data from CFHT Observations\tablenotemark{*} \label{tab_cfht}}
\tablewidth{0pt}
\tablehead{
\colhead{ID} & \colhead{$\alpha_{\rm J2000}$} & \colhead{$\delta_{\rm J2000}$} & \colhead{$V$} &
\colhead{$I$} & \colhead{$R-I$} & \colhead{$V-I$} & \colhead{$R$-H$\alpha$} &
\colhead{$\epsilon_{V}$} & \colhead{$\epsilon_{I}$} & \colhead{$\epsilon_{R-I}$} &
\colhead{$\epsilon_{V-I}$} & \colhead{$\epsilon_{R-{\rm H}\alpha}$} & \multicolumn{5}{c}{N$_{\rm obs}$} &
\colhead{D\tablenotemark{a}} & \colhead{M\tablenotemark{b}} & \colhead{Class\tablenotemark{c}} & \colhead{2MASS ID\tablenotemark{d}} &
\colhead{Spitzer ID\tablenotemark{e}} & \colhead{VSA\tablenotemark{f}}  & \colhead{Sp. Type} }

\startdata
C79360 &  2:34:56.37 & 61:17:17.1 & \nodata &  21.325 &   1.281 & \nodata & \nodata & \nodata &   0.010 &   0.022 & \nodata & \nodata &  0 &  6 &  6 &  0 &  0 &    &   & F  &                   & S084685  &  & \\
C79361 &  2:34:56.38 & 61:21:12.1 &  22.014 &  19.651 &   1.143 &   2.359 &  -3.179 &   0.007 &   0.013 &   0.003 &   0.018 &   0.013 &  6 &  7 &  7 &  6 &  5 &    &   &    &                   & S084690  &  & \\
C79362 &  2:34:56.39 & 61:44:44.0 &  22.420 &  20.191 &   1.065 &   2.226 &  -2.855 &   0.031 &   0.013 &   0.016 &   0.033 &   0.049 &  3 &  4 &  3 &  3 &  1 &    &   &    &                   &          &  & \\
C79363 &  2:34:56.39 & 61:15:43.5 &  17.723 &  16.180 &   0.719 &   1.535 &  -2.987 &   0.011 &   0.014 &   0.008 &   0.028 &   0.002 &  7 &  7 &  7 &  7 &  6 &    &   &    & 02345637+6115433  & S084683  &  & \\
C79364 &  2:34:56.39 & 61:44:01.2 &  18.650 &  17.064 &   0.779 &   1.585 &  -3.014 &   0.001 &   0.004 &   0.004 &   0.004 &   0.004 &  4 &  4 &  4 &  4 &  1 &    &   &    & 02345637+6144011  & S084695  &  & \\
C79365 &  2:34:56.39 & 61:29:05.4 & \nodata &  20.878 &   1.415 & \nodata & \nodata & \nodata &   0.007 &   0.030 & \nodata & \nodata &  0 & 11 &  9 &  0 &  0 &    &   &    &                   & S084696  &  & \\
C79366 &  2:34:56.39 & 61:38:09.3 &  21.250 &  19.261 &   1.004 &   2.004 &  -2.987 &   0.005 &   0.009 &   0.007 &   0.003 &   0.005 &  6 &  7 &  7 &  6 &  4 &    &   &    &                   & S084688  &  & \\
C79367 &  2:34:56.39 & 61:33:22.0 &  17.357 &  15.585 &   0.830 &   1.779 &  -3.005 &   0.007 &   0.001 &   0.008 &   0.003 &   0.009 &  7 &  7 &  7 &  7 &  4 &    &   & 4  & 02345640+6133220  & S084692  &  & \\
C79368 &  2:34:56.39 & 61:40:19.7 &  23.418 &  21.333 &   1.111 &   2.109 & \nodata &   0.091 &   0.039 &   0.051 &   0.107 & \nodata &  1 &  6 &  6 &  1 &  0 &    &   &    &                   &          &  & \\
C79369 &  2:34:56.40 & 61:27:28.6 & \nodata &  21.577 &   1.794 & \nodata & \nodata & \nodata &   0.004 &   0.061 & \nodata & \nodata &  0 &  9 &  2 &  0 &  0 &    &   &    &                   &          &  & \\
C79370 &  2:34:56.40 & 61:23:57.4 &  21.478 &  18.252 &   1.589 &   3.212 &  -2.123 &   0.069 &   0.015 &   0.027 &   0.081 &   0.029 & 10 & 11 & 11 & 10 &  7 &    & H & 3  & 02345637+6123576A & S084691A &  & \\
C79371 &  2:34:56.41 & 61:30:55.7 &  14.296 &  13.575 &   0.373 &   0.734 & \nodata &   0.016 &   0.022 &   0.024 &   0.000 & \nodata &  2 &  2 &  2 &  2 &  0 &    &   & 4  & 02345640+6130556  & S084694  & 281&B5\\
C79372 &  2:34:56.41 & 61:32:37.4 &  22.635 &  20.505 &   1.047 &   2.099 &  -3.020 &   0.028 &   0.006 &   0.034 &   0.027 &   0.044 &  6 &  7 &  6 &  6 &  4 &    &   &    &                   &          &  & \\
C79373 &  2:34:56.41 & 61:32:30.3 &  20.568 &  18.828 &   0.855 &   1.737 &  -3.001 &   0.011 &   0.006 &   0.034 &   0.005 &   0.011 &  7 &  7 &  7 &  7 &  4 &    &   &    &                   & S084697  &  & \\
C79374 &  2:34:56.42 & 61:19:51.5 &  23.375 &  20.943 &   1.201 &   2.408 & \nodata &   0.000 &   0.012 &   0.012 &   0.017 & \nodata &  5 &  7 &  6 &  5 &  0 &    &   &    &                   &          &  & \\
C79375 &  2:34:56.44 & 61:35:45.5 &  23.138 &  21.043 &   1.074 &   2.096 &  -2.829 &   0.013 &   0.005 &   0.012 &   0.017 &   0.070 &  6 &  7 &  6 &  6 &  3 &    &   &    &                   &          &  & \\
C79376 &  2:34:56.44 & 61:44:21.2 &  22.188 &  20.096 &   1.034 &   2.092 &  -2.954 &   0.018 &   0.013 &   0.015 &   0.022 &   0.035 &  3 &  4 &  3 &  3 &  1 &    &   &    &                   &          &  & \\
C79377 &  2:34:56.44 & 61:38:26.0 & \nodata &  21.410 &   1.642 & \nodata & \nodata & \nodata &   0.012 &   0.014 & \nodata & \nodata &  0 &  6 &  5 &  0 &  0 &    &   &    &                   &          &  & \\
C79378 &  2:34:56.45 & 61:19:20.5 & \nodata &  21.756 &   1.340 & \nodata & \nodata & \nodata &   0.002 &   0.104 & \nodata & \nodata &  0 &  6 &  3 &  0 &  0 &    &   &    &                   &          &  & \\
C79379 &  2:34:56.46 & 61:06:25.6 & \nodata &  21.181 &   1.372 & \nodata & \nodata & \nodata &   0.016 &   0.022 & \nodata & \nodata &  0 &  3 &  3 &  0 &  0 &    &   &    &                   & S084713  &  & \\
C79380 &  2:34:56.46 & 61:28:52.7 & \nodata &  22.194 & \nodata & \nodata & \nodata & \nodata &   0.075 & \nodata & \nodata & \nodata &  0 &  3 &  0 &  0 &  0 &    &   &    &                   &          &  & \\
\enddata

\tablenotetext{*}{Table \ref{tab_cfht} is presented in its entirety in
the electronic edition of the Astrophysical Journal Supplement Series. A portion is shown here
for guidance regarding its form and content. Units of right ascension are
hours, minutes, and seconds of time, and units of declination are degrees,
arcminutes, and arcseconds.}
\tablenotetext{a}{duplicity - D: stars whose PSF shows a double, but measures as a single star, G: galaxy. }
\tablenotetext{b}{membership - X: X-ray emission star, x: X-ray emission candidate, H: H$\alpha$ emission star, h: H$\alpha$ emission candidate, ``+'' =  X + H, ``-'' = X + h}
\tablenotetext{c}{YSO class - 1: Class I, F: flat spectrum, 2: Class II, 3: Class III, 4: Class IV, t: star with pre-transition disks, T: star with transition disks, P: stars with PAH emission, g: photometric galaxy candidates, ?: Two or more stars are identified as the optical counter parts of a Spitzer source}
\tablenotetext{d}{A, B, or b are added at the end of 2MASS ID if two or more stars are matched with a 2MASS source within a matching radius of 1$''$. A or B: The bright or faint component of a 2MASS source whose $I$ magnitude difference is less than 1 mag. b: The faint component of a 2MASS source whose $I$ magnitude difference is greater than 1 mag.}
\tablenotetext{e}{A, B, or C are added at the end of Spitzer ID if two or more stars are matched with a Spitzer source within a matching radius of 1$''$.}
\tablenotetext{f}{ID from Vasilevskis et al. (1965)}
\end{deluxetable*}

\floattable
\begin{deluxetable*}{cc|cccc@{}c@{}|cccc@{}c}
\tablecolumns{12}
\tabletypesize{\scriptsize}
\tablecaption{Internal Consistency of Optical Photometric Data\tablenotemark{*} \label{tab_int}}
\tablewidth{0pt}
\tablehead{
\colhead{Reference} & \colhead{Target} & \colhead{$\Delta V$} & 
\colhead{n(n$_{\rm ex}$)\tablenotemark{a}} & \colhead{$\Delta (V-I)$} &
\colhead{n(n$_{\rm ex}$)\tablenotemark{a}} & $V$ range & \colhead{$\Delta I$} &
\colhead{n(n$_{\rm ex}$)\tablenotemark{a}} & \colhead{$\Delta (R-I)$} &
\colhead{n(n$_{\rm ex}$)\tablenotemark{a}} & $I$ range }

\startdata
CFH12K-C & CFH12K-N & -0.011 $\pm$ 0.016 & 6766 (772) & 
-0.010 $\pm$ 0.020 & 7059 (479) & $\leq 21$ & -0.001 $\pm$ 0.027 & 
10323 (1332) & +0.033 $\pm$ 0.019 & 10919 (736) & $\leq 19.5$ \\
CFH12K-C & CFH12K-S & -0.009 $\pm$ 0.017 & 6460 (587) &
-0.017 $\pm$ 0.018 & 6604 (443) & $\leq 21$ & +0.008 $\pm$ 0.025 &
9701 (1422) & +0.018 $\pm$ 0.019 & 10419 (704) & $\leq 19.5$ \\
CFH12K-S & CFH12K-N & -0.003 $\pm$ 0.017 & 3678 (316) &
+0.007 $\pm$ 0.018 & 3687 (307) & $\leq 21$ & -0.009 $\pm$ 0.026 &
5766 (664) & +0.013 $\pm$ 0.020 & 5961 (469) & $\leq 19.5$ \\
CFH12K & Maidanak 2k & +0.000 $\pm$ 0.017 & 427 (37) & +0.000 $\pm$ 0.022 &
430 (34) & $\leq 17$ & +0.007 $\pm$ 0.022 & 991 (81) & -0.005 $\pm$ 0.036 &
1032 (27) & $\leq 16.5$ \\
CFH12K & SNUCam & +0.036 $\pm$ 0.020 & 388 (45) & +0.015 $\pm$ 0.026 &
416 (15) & $\leq 17$ & +0.031 $\pm$ 0.020 & 868 (128) & \nodata & \nodata &
$\leq 16.5$ \\
SNUCam & Maidanak2k & -0.033 $\pm$ 0.020 & 260 (24) & -0.012 $\pm$ 0.021 &
268 (16) & $\leq 17$ & -0.024 $\pm$ 0.024 & 516 (60) & \nodata & \nodata &
$\leq 16.5$ \\ \hline
            &             &  $\Delta (B-V)$ & n(n$_{\rm ex}$)\tablenotemark{a} &
$\Delta (U-B)$ & n(n$_{\rm ex}$)\tablenotemark{a} & $V$ range &$\Delta (U-B)$ &
n(n$_{\rm ex}$)\tablenotemark{a} & \multicolumn{2}{c}{range} & \\ \hline
SNUCam& Maidanak2k & +0.005 $\pm$ 0.022 & 165 (119) & -0.016 $\pm$ 0.053 & 165 (6) & \tablenotemark{b} & +0.013 $\pm$ 0.033 & 54 (2) & \multicolumn{2}{c}{$V \leq 17 ~\&~ (U-B) \leq 0.0$} & \\
\enddata
\tablenotetext{*}{$\Delta$ denotes the difference - reference minus target.}
\tablenotetext{a}{The number of stars excluded in the comparison is shown in parenthesis}
\tablenotetext{b}{$V \leq 17$ for $(B-V)$ and $V \leq 16$ for $(U-B)$}

\end{deluxetable*}

The instrumental
magnitudes were transformed to the standard magnitudes and colors using
the atmospheric extinction coefficients, transformation coefficients,
and photometric zero points summarized in Table 1 of \citet{sbc08}.
However, we did not apply the time variation of the photometric zero points on the first
half of 2002 January 6 because we found consistent magnitudes and colors
between those observed on 2002 January 6 and January 7 when we neglected 
the time variation coefficients. When we were checking the internal consistency
of photometric data between CFH12K-N and CFH12K-S, 
the photometric data obtained from chip 08 of CFH12K-N 
(N08) showed a large shift in zero points.
The  $V$ magnitude zeropoints were in relatively 
good agreement, but the $I$ and $R$ magnitudes were shifted by about
0.08 and 0.04 mag, respectively. 
We therefore corrected for the photometric zero points for N08 by these amounts.
The consistency of the photometric data from the three CFH12K
pointings after this correction are shown in Figure \ref{cfh_delta} and summarized in Table 
\ref{tab_int}. The difference between the data sets was calculated for relatively
bright stars with $V \leq 21$ for $\Delta V$ and $\Delta (V-I)$, $I \leq
19.5$ for $\Delta I$ and $\Delta (R-I)$, respectively. The statistics were
obtained from a successive exclusion scheme - the successive exclusion of data 
with a large difference ($> 2.5 \sigma$) from the mean. The number in
parenthesis indicates the number of stars excluded from the statistics.
The three independent data sets agreed very well with each other.\\

{\it 2.1.1.2 Astrometry and 2MASS Counterparts}

CCD coordinates were transformed to the equatorial coordinate system
using the Two Micron All Sky Survey (2MASS) point source catalog \citep{2mass},
and all data finally merged into a catalog (called CFH12K)
using the weighted averaging scheme described in \citet{sl95}.
The number of stars in the catalog CFH12K (Table \ref{tab_cfht}) is 91139.
We included in the table, duplicity from PSF photometry and membership information,
YSO class, 2MASS ID, Spitzer ID in Table \ref{tab_sst}, 
VSA ID \citep{vsa65}, and spectral types from various sources. 
Ten stars in Table \ref{tab_cfht} may be listed twice due to a large difference
in brightness ($\Delta I > 1$ mag) among three sets of data (e.g. C11519 \& 
C11524, C51759 \& C51760) or due to the problems in PSF deconvolution of a very close ($d  \lesssim 0\farcs15$) double
(e.g. C03096 \& C03097, C17752 \& C17755, C19384 \& C19387, C20486 \& C20491,
C22238 \& C22240, C23895 \& C23897, C39867 \& C39874, C53870 \& C53877).
Owing to the superior sky conditions
of the CFHT observations there are many cases of two or more stars being 
simultaneously identified as the optical counter-part of a 2MASS source within a matching radius of $1\farcs0$. 
If the difference in the $I$ magnitude of two or three stars being matched with a 2MASS
source was greater than 5 mag, we assigned the brightest star as the optical
counterpart of the 2MASS source. If the difference in $I$ was greater between 1 and 5 mags,
we assigned the brighter (brightest) star as the optical counterpart of 
the 2MASS source and added ``b'' after 2MASS ID for the fainter star(s).
If the difference in $I$ was less than 1 mag, we added ``A'' and ``B'' for
the brighter star and the fainter star, respectively. The same rule was
applied for the {\it Spitzer} ID.

\subsubsection{Maidanak AZT-22 1.5m Observations}

{\it 2.1.2.1. SITe $2000 \times 800$ CCD (Maidanak 2k) Observations}

For a comprehensive study of IC 1805, the photometry of the bright blue 
stars was very important and therefore we decided to observe several regions in IC 1805
with the AZT-22 1.5 m telescope at the Maidanak Astronomical Observatory in Uzbekistan.

\begin{figure*}
\gridline{\fig{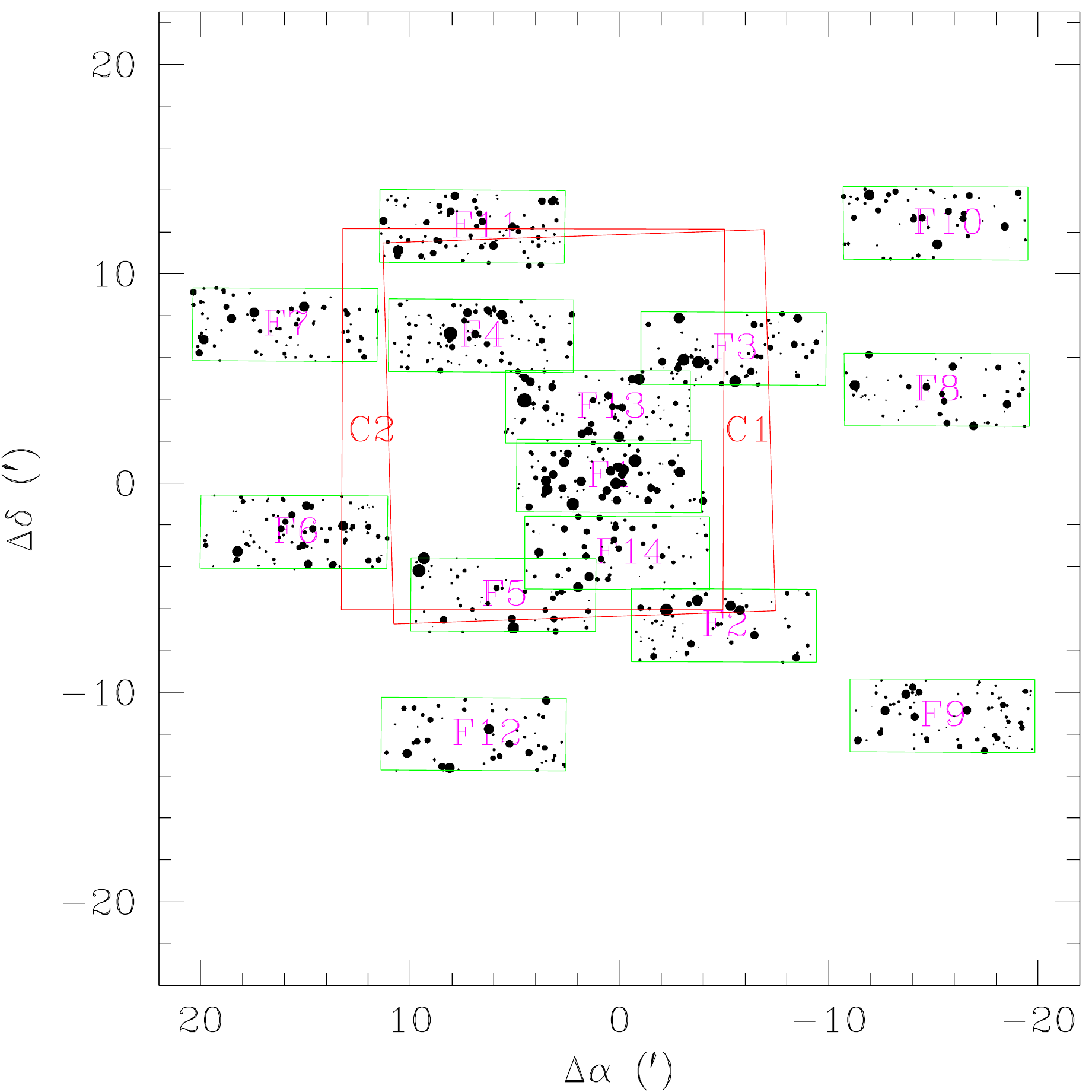}{0.60\textwidth}{(a)}
          \fig{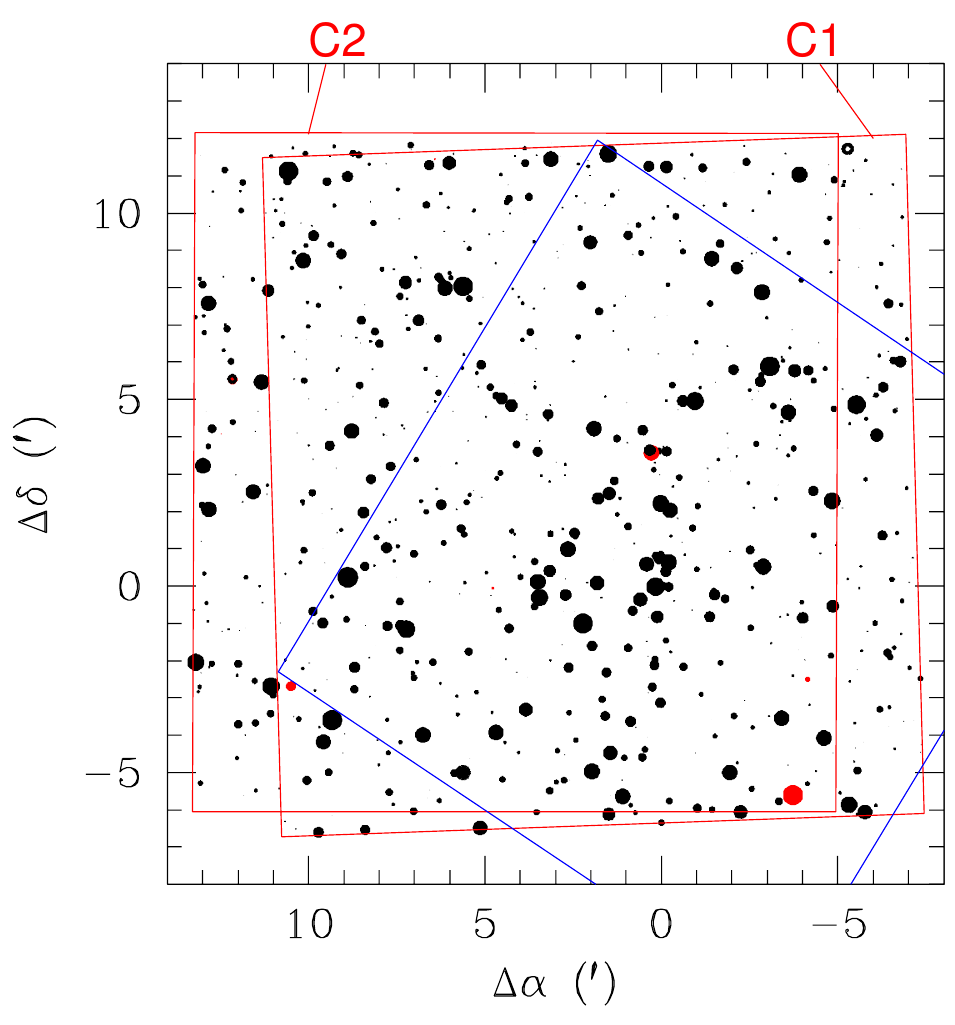}{0.35\textwidth}{(b)}}
\caption{Finder chart for the stars brighter than $I$ = 16 mag (or $V$ = 17.8
mag) from (a) Maidanak 2k observations and (b) SNUCam observations. 
The size of the dot is proportional to the brightness of the star. 
Red dots in (b) represent H$\alpha$ emission stars.
Fourteen green rectangles represent the field of views (FOVs) of 14 regions observed with 
the Maidanak 2k CCD, while two red squares denote the FOVs of two regions 
observed with the SNUCam CCD. The blue diamond in (b) represents 
the FOV of the {\it Chandra} X-ray observation. The origin ($\Delta
\alpha$ = 0.$'$0, $\Delta \delta$ = 0.$'$0) is the position of 
the brightest star in IC 1805 HD 15558. \label{mao_map} }
\end{figure*}

\begin{deluxetable*}{lc|ccccc}
\tablecolumns{6}
\tabletypesize{\scriptsize}
\tablecaption{Extinction coefficients, time-variation coefficients, and photometric zero-points at the Maidanak Astronomical Observatory\label{mao_atm} }
\tablewidth{0pt}
\tablehead{
\colhead{Date of Obs.} & \colhead{standard stars} & \colhead{Filter} & \colhead{$k_{1 \lambda}$} & 
\colhead{$k_{2 \lambda}$} & \colhead{$\alpha_{t,\lambda}$} & \colhead{$\zeta_\lambda$}\\
\colhead{CCD} & \colhead{or regions} & \colhead{} & \colhead{} & \colhead{} & \colhead{} & \colhead{} }
\startdata
            &   SA 111, SA 112     & $I$ & 0.081 $\pm$ 0.020 & \nodata & 0.0000 & 22.717 $\pm$ 0.011 \\
2003. 8. 18 &   SA 113, SA 114     & $R$ & 0.180 $\pm$ 0.034 & \nodata & \nodata & 23.095 $\pm$ 0.018 \\
SITe 2000$\times$800&   BD-11 162     & $V$ & 0.250 $\pm$ 0.020 & \nodata & -0.0060 & 23.167 $\pm$ 0.009 \\
            & (PG1633+099)     & $B$ & 0.303 $\pm$ 0.022 & 0.026  & 0.0000 & 22.969 $\pm$ 0.012 \\
            &                      & $U$ & 0.552 $\pm$ 0.018 & 0.023  & 0.0030 & 21.320 $\pm$ 0.005 \\ \hline
            &                      & $I$ & 0.114 $\pm$ 0.011 & \nodata & \nodata & 23.202 $\pm$ 0.006 \\
2004. 12. 24&                      & $R$ & 0.154 $\pm$ 0.012 & \nodata & \nodata & 23.373 $\pm$ 0.007 \\
SITe 2000$\times$800& SA 113, SA 114       & $V$ & 0.235 $\pm$ 0.018 & \nodata & \nodata & 23.606 $\pm$ 0.011 \\
            &              & $B$ & 0.359 $\pm$ 0.025 & 0.026  & \nodata & 23.538 $\pm$ 0.014 \\
            &                      & $U$ & 0.609 $\pm$ 0.036 & 0.023  & \nodata & 21.914 $\pm$ 0.023 \\ \hline
            &                      & $I$ & 0.110 $\pm$ 0.020 & \nodata & 0.0063 $\pm$ 0.0018 & 23.198 $\pm$ 0.012 \\
2004. 12. 30&  SA 92, SA 97        & $R$ & 0.174 $\pm$ 0.069 & \nodata & 0.0140 & 23.436 $\pm$ 0.019 \\
SITe 2000$\times$800& SA 98, SA 114        & $V$ & 0.229 $\pm$ 0.016 & \nodata & 0.0018 $\pm$ 0.0015 & 23.586 $\pm$ 0.010 \\
            &                   & $B$ & 0.369 $\pm$ 0.016 & 0.026  & \nodata & 23.525 $\pm$ 0.008 \\
            &                      & $U$ & 0.688 $\pm$ 0.043 & 0.026  & \nodata & 21.997 $\pm$ 0.023 \\ \hline \hline
            &  SA 92, SA 95        & $I$ & 0.024 $\pm$ 0.011 & \nodata & 0.0038 $\pm$ 0.0010 & 22.678 $\pm$ 0.014 \\
2007. 10. 7 &  SA 96, SA 98        & $V$ & 0.131 $\pm$ 0.006 & \nodata & 0.0023 $\pm$ 0.0007 & 23.192 $\pm$ 0.008 \\
Fairchild 486& SA 110, SA 113      & $B$ & 0.246 $\pm$ 0.009 & 0.035 $\pm$ 0.005 & 0.0039 $\pm$ 0.0011 & 23.021 $\pm$ 0.013 \\
            &  SA 114, BD-11 162   & $U$ & 0.405 $\pm$ 0.012 & 0.011 $\pm$ 0.007 & 0.0070 $\pm$ 0.0014 & 21.154 $\pm$ 0.016 \\
            &                      & H$\alpha$ & 0.043 $\pm$ 0.004 & \nodata & \nodata & 19.403 $\pm$ 0.036 \\ \hline
            & SA 93, SA 96         & $I$ & 0.042 $\pm$ 0.013 & \nodata & \nodata & 23.619 $\pm$ 0.011 \\
2009. 1. 19 & SA 97, SA 98$\times$3& $V$ & 0.139 $\pm$ 0.005 & \nodata & 0.0017 $\pm$ 0.0005 & 24.124 $\pm$ 0.009 \\
Fairchild 486& SA 99, SA 101       & $B$ & 0.249 $\pm$ 0.007 & 0.017 $\pm$ 0.004 & 0.0019 $\pm$ 0.0005 & 24.054 $\pm$ 0.007 \\
            &  SA 102, SA 104      & $U$ & 0.444 $\pm$ 0.015 & 0.027 $\pm$ 0.005 & \nodata & 22.388 $\pm$ 0.013 \\
            &                      & H$\alpha$ & 0.081 $\pm$ 0.005 & \nodata & \nodata & 20.412 $\pm$ 0.042 \\
\enddata
\end{deluxetable*}

$UBVRI$ CCD observations of IC 1805 were performed on 2003 August 18 (Region:
F1 -- F4), 2004 December 24 (F5 -- F8), and 2004 December 30 (F9 -- F14) at the
Maidanak Astronomical Observatory with the AZT-22 1.5 m telescope and
a thinned SITe 2000 $\times$ 800 CCD (15 $\mu m$ pixels; pixel scale = 0$\farcs$265
/pixel). The observed regions are shown in Figure \ref{mao_map}. Two sets of 
exposure times were used in the observations - long: 60 s in $I$, 90 s in $R$, 
180 s in $V$, 300 s in $B$, 600 s in $U$, and short: 3 s in $I$, $R$ and $V$, 
5 s in $B$, 15 s in $U$. The seeing was relatively good (1$\farcs$0 -- 1$\farcs$2).
As we did not observe many standard stars at various air masses, we used
as secondary standard stars, those stars common with the SNUCam FOV 
in the determination of atmospheric extinction coefficients and photometric
zero points. These coefficients are listed in Table \ref{mao_atm}. 
The transformation to the SAAO standard system was performed 
using the coefficients described in \citet{lsb09}. As can be seen in Table
\ref{tab_int}, the photometric zero-points of the CFH12K and SNUCam data
differ by about 3\% in $V$ and $I$. The photometric zero-points for $V$ and 
$I$ were adjusted to those of CFH12K, but those for ($B-V$) and ($U-B$) were
adjusted to those of the SNUCam data.

\floattable
\begin{deluxetable*}{cccrrrrrrcccccc@{}c@{}c@{}c@{}c@{}c@{}c@{}c@{}c@{}c@{}llrl}
\tablecolumns{26}
\tabletypesize{\scriptsize}
\rotate
\tablecaption{Photometric Data from the Maidanak AZT-22 1.5m Telescope and SITe 2000$\times$800 CCD\tablenotemark{*} \label{tab_m2k}}
\tablewidth{0pt}
\tablehead{
\colhead{ID} & \colhead{$\alpha_{\rm J2000}$} & \colhead{$\delta_{\rm J2000}$} & \colhead{$V$} &
\colhead{$I$} & \colhead{$R-I$} & \colhead{$V-I$} & \colhead{$B-V$} & \colhead{$U-B$} &
\colhead{$\epsilon_{V}$} & \colhead{$\epsilon_{I}$} & \colhead{$\epsilon_{R-I}$} & \colhead{$\epsilon_{V-I}$} &
\colhead{$\epsilon_{B-V}$} & \colhead{$\epsilon_{U-B}$} & \multicolumn{6}{c}{N$_{\rm obs}$} &
\colhead{D\tablenotemark{a}} & \colhead{M\tablenotemark{b}} & \colhead{Class\tablenotemark{c}} & \colhead{2MASS ID\tablenotemark{d}} &
\colhead{CFHT ID} & \colhead{VSA\tablenotemark{f}}  & \colhead{Sp. Type} }

\startdata
\enddata

\tablenotetext{*}{Table \ref{tab_m2k} is presented in its entirety in
the electronic edition of the Astrophysical Journal Supplement Series. A portion is shown here
for guidance regarding its form and content. Units of right ascension are
hours, minutes, and seconds of time, and units of declination are degrees,
arcminutes, and arcseconds.}
\end{deluxetable*}

\begin{figure*} 
\epsscale{0.9}
\gridline{\fig{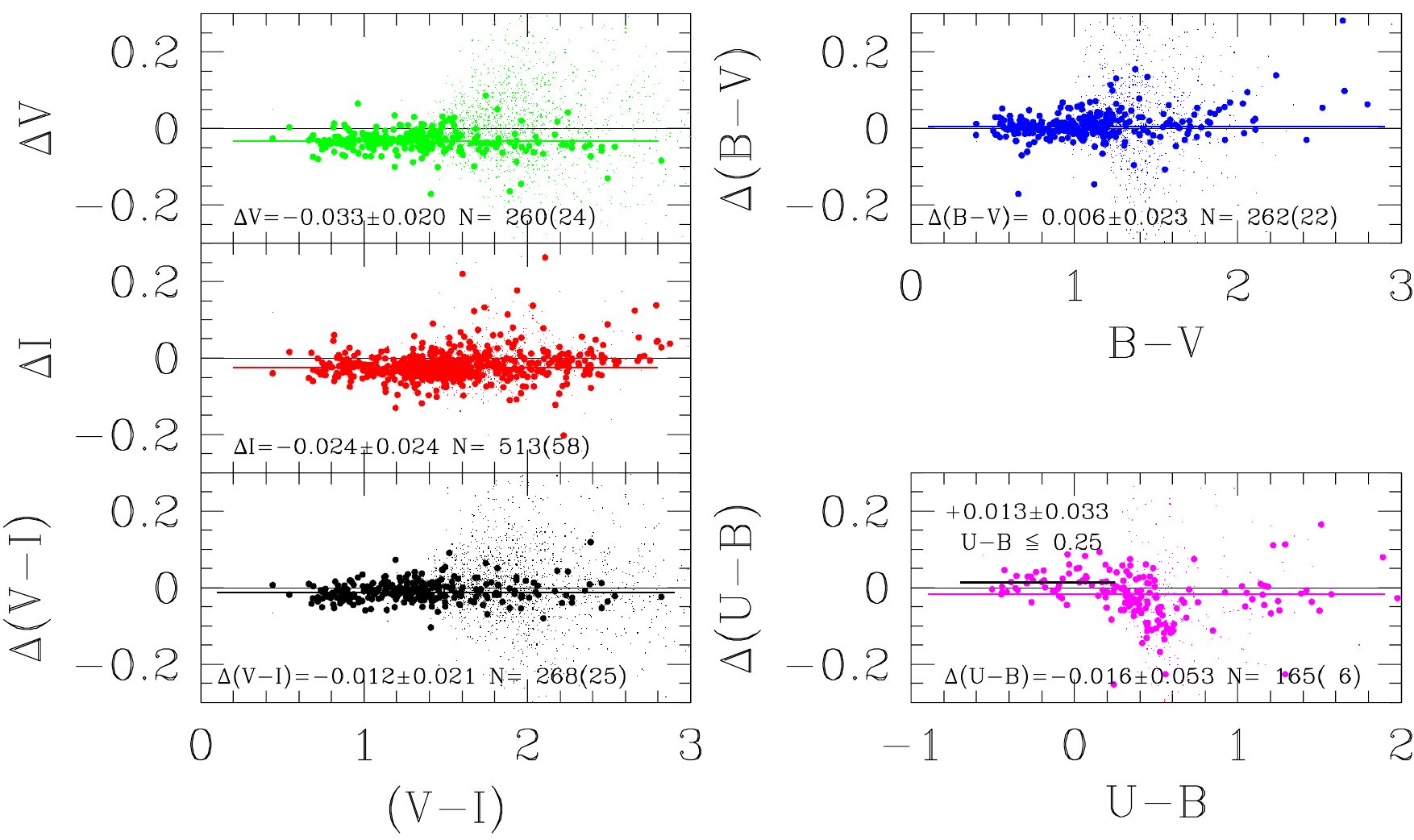}{0.62\textwidth}{(a)}
          \fig{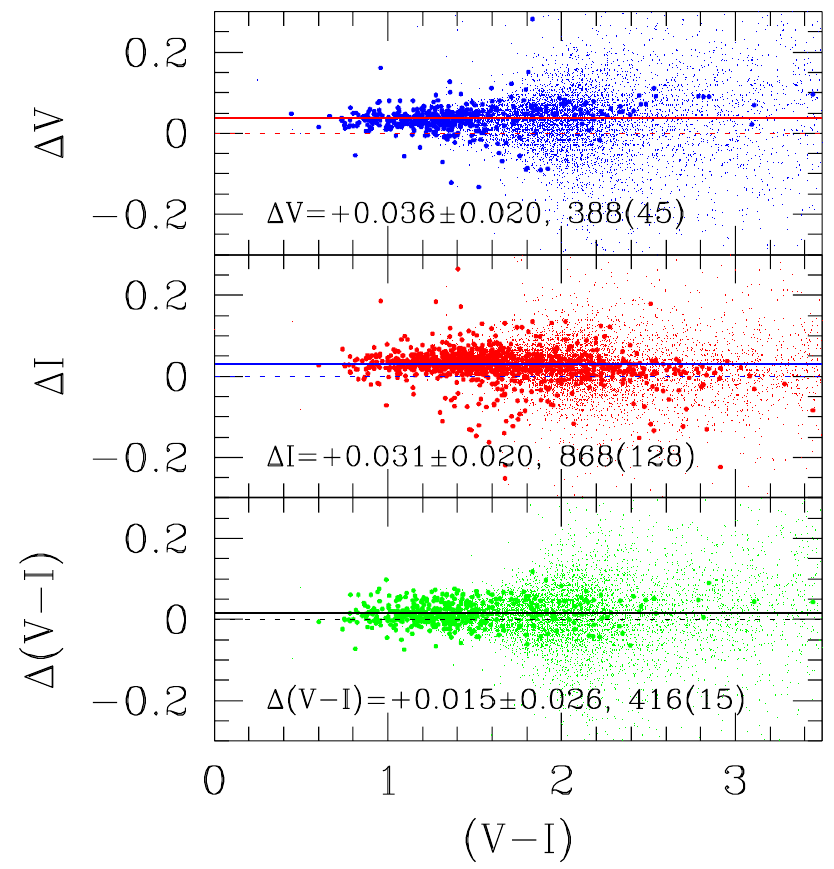}{0.35\textwidth}{(b)}}
\caption{Comparison of Photometry. $\Delta$ means (a) SNUCam data minus 
Maidanak2k data (left and middle panels) and (b) CFH12K data minus SNUCam data (right panel). 
Large dots represent data for relatively bright stars 
[$V \leq 17$ for $V$ and $(B-V)$, $V \leq 16$ for $(U-B)$, and $I \leq 16.5$ 
for $I$ and $(V-I)$].
\label{comp_m2km4k} }
\end{figure*}

A total of 5319 stars were measured and listed in Table \ref{tab_m2k} (we refer to this data set as ``Maidanak2k'').
Among them, 5121 stars were matched with a single object in Table
\ref{tab_cfht}, 145 stars were matched with two objects, and 2 stars were
matched with 3 stars within a matching radius of 0$\farcs$7, and 51 objects had
no counterpart in the CFH12K catalog. Forty nine objects were bright stars
($I < 15$ mag). The star M2k3582 is located just south of M2k3581 (= VSA 199, 
G7Ib), and was on the saturated portion of the CFH12K images. The other object
M2k4020 is an extended source, and was classified as a galaxy candidate ``G'' in
the Maidanak2k data. This source was rejected in the CFH12K images probably
because of its higher $\chi$ or $sharpness$ value. The difference in 
photometry between the SNUCam and Maidanak2k data is shown in 
Figure \ref{comp_m2km4k} (left and middle panels) and Table \ref{tab_int}. The difference in $V$ and 
$I$ was -0.033 and -0.024 mag, respectively. This means that the SNUCam data were
about 3\% brighter than the CFH12K data as the Maidanak2k data were adjusted 
to the CFH12K data.
On the other hand, the ($V-I$) and ($B-V$) colors for $V \leq$ 17 mag were well
consistent with each other. However, $\Delta (U-B)$ for $V \leq$ 16 mag showed
abnormal behaviour. Although the mean
value of the difference was very close to 0.0, there was a systematic difference
which was related to the Balmer jump of A -- F stars. As the SITe $2000
\times 800$ CCD and $U$ filter combination do not require a non-linear
correction \citep{lsb09}, the non-linear difference in $\Delta (U-B)$ is caused 
solely by the non-linear correction term in the $U$ transformation of the 
Fairchild 486 CCD and $U$ filter combination (see \citealt{lsb09,lsb15}).
More discussion on this issue will be dealt with in the following section.\\

{\it 2.1.2.2. Fairchild 486 CCD (SNUCam) Observations}

$UBVI$ and H$\alpha$ CCD photometry for the central region of IC 1805 was
obtained on 2007 October 7 (C1 region) and 2009 January 19 (C2 region) at the
Maidanak Astronomical Observatory with the AZT-22 (1.5 m) telescope (f/7.74) 
and a thinned Fairchild 486 CCD (15 $\mu m$ pixels) as a part of the Sejong
Open-cluster Survey (SOS - \citealt{sos}). The observed regions are shown in
Figure \ref{mao_map}.
The filters and exposure times used in the observations were the same as 
those used for the observations of IC 1848 \citep{lsk14a}. 
The seeing was good on 2009 January 19 (about $0\farcs9$ in $V$ 180 s
image), and moderate on 2007 October 7 ($1\farcs5$).

All the preprocessing needed to remove the instrumental signature was done 
using the IRAF/CCDRED package. Instrumental magnitudes were obtained using 
IRAF/DAOPHOT via PSF fitting for the target images and via simple 
aperture photometry for standard stars. All the instrumental magnitudes were 
transformed to the standard $UBVI$ system using SAAO photometry of
equatorial standard stars \citep{saao91} and blue and red standard stars in 
\citet{saao98}. Details of the transformations to the standard system can
be found in \citet{lsb09}. Individual data were compared with the CFH12K data,
and showed that the brightness of stars at $r > 5'$ from the center of the FOV
were fainter, especially in $I$, probably because 
of a large variation in the PSF at the edge of the CCD chip \citep{lsk08}. 
CCD coordinates were transformed 
to the equatorial coordinate system by identifying the optical counterpart 
of 2MASS point sources \citep{2mass}. Two sets of SNUCam data were 
merged into a photometric catalog (SNUCam data).


\floattable
\begin{deluxetable*}{cccrrrrrrccccccc@{}c@{}c@{}c@{}c@{}c@{}c@{}c@{}c@{}llrl}
\tablecolumns{26}
\tabletypesize{\scriptsize}
\rotate
\tablecaption{Photometric Data from the Maidanak AZT-22 1.5m Telescope and Fairchild 486 CCD (SNUCam) \tablenotemark{*} \label{tab_m4k}}
\tablewidth{0pt}
\tablehead{
\colhead{ID} & \colhead{$\alpha_{\rm J2000}$} & \colhead{$\delta_{\rm J2000}$} & \colhead{$V$} &
\colhead{$I$} & \colhead{$V-I$} & \colhead{$B-V$} & \colhead{$U-B$} & \colhead{H$\alpha$} &
\colhead{$\epsilon_{V}$} & \colhead{$\epsilon_{I}$} & \colhead{$\epsilon_{V-I}$} &
\colhead{$\epsilon_{B-V}$} & \colhead{$\epsilon_{U-B}$} & \colhead{$\epsilon_{{\rm H}\alpha}$} & \multicolumn{6}{c}{N$_{\rm obs}$} &
\colhead{D\tablenotemark{a}} & \colhead{M\tablenotemark{b}} & \colhead{Class\tablenotemark{c}} & \colhead{2MASS ID\tablenotemark{d}} &
\colhead{CFHT ID} & \colhead{VSA\tablenotemark{f}}  & \colhead{Sp. Type} }

\startdata
\enddata

\tablenotetext{*}{Table \ref{tab_m4k} is presented in its entirety in
the electronic edition of the Astrophysical Journal Supplement Series. A portion is shown here
for guidance regarding its form and content. Units of right ascension are
hours, minutes, and seconds of time, and units of declination are degrees,
arcminutes, and arcseconds.}
\end{deluxetable*}

We also identified the stars in Table \ref{tab_m4k} with those in Table
\ref{tab_cfht} using a matching radius of $0\farcs7$. Among 7011 stars in the
SNUCam data, 6804 stars had one counterpart in the CFH12K catalog.
Due to the superior sky condition at Mauna Kea, 162 objects in the SNUCam data
had two counterparts in the CFH12K catalog, and 2 objects had three counterparts. 
However 43 stars had no counterpart in the CFH12K catalog. Among them, 37
stars were bright ($I < 15$ mag) and not measured because of saturation.
Three objects were matched with two objects in the CFH12K catalog just outside
the matching radius, one object had a counterpart just outside the matching
radius, and the other two objects were missed because of their closeness
to bright saturated stars (M4k0305 on the spike of BD +60 497 and M4k2398 
near HD 15570).

We calculated the difference in photometry between the SNUCam data and CFH12K 
data or Maidanak2k data for the stars within a $5'$ radius from the center of the SNUCam FOV,
and presented them in Table \ref{tab_int} and Figure \ref{comp_m2km4k} (right panel).
SNUCam data were systematically brighter by about 3\% in $V$ and $I$
than CFH12K data. Such a small, but systematic difference in zero-points may be related to 
the standard stars used in the standard transformation (the Stetson version
of the Landolt standard system for CFH12K and the SAAO standard system for SNUCam 
data). A similar difference was found in \citet{sbc08}. The difference
in $(B-V)$ between the Maidanak 2k and SNUCam data was very small as 
the photometric zero-points of the Maidanak 2k data were adjusted to those of the
SNUCam data. On the other hand, as mentioned in section 2.2.1 the difference 
in $(U-B)$ was slightly curved, caused by the non-linear correction term
in the $U$ transformation of the SNUCam data. The size of the non-linear correction 
depends both on the size of the Balmer jump and the steep variation of the
quantum efficiency of the CCD chip between $\lambda\lambda 3000 - 4000\AA$,
and so is strongly affected by the amount of reddening. For early-type stars
it is very easy to calculate the amount of reddening, but is not as easy for
intermediate- or late-type field stars because their colors are affected both
by metallicity and gravity. As we have no information on the reddening of
field stars, we had to apply the mean reddening for these stars, which could cause
an over- or under-correction of the reddening effect.

\subsubsection{Comparison of Optical Photometry}

\begin{figure*}
\gridline{\fig{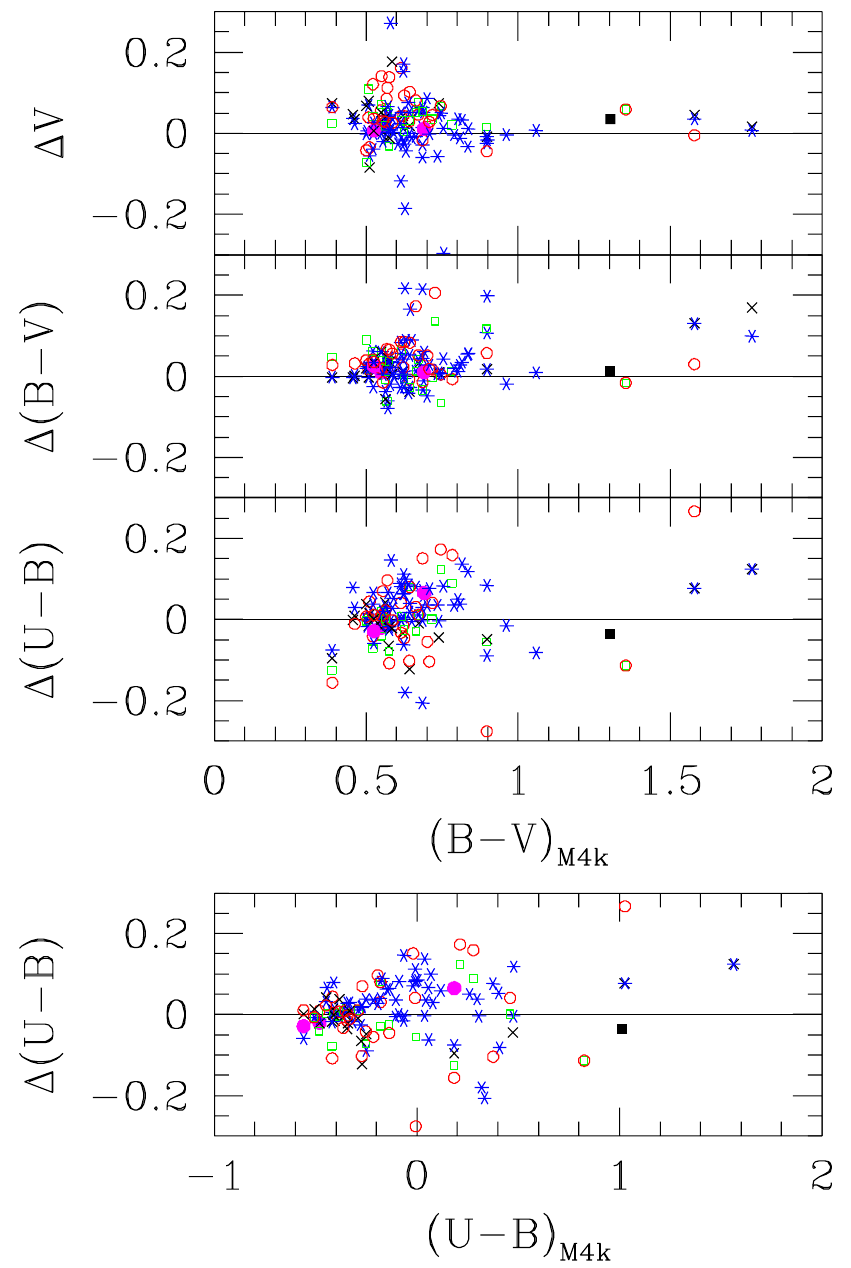}{0.23\textwidth}{(a)}
          \fig{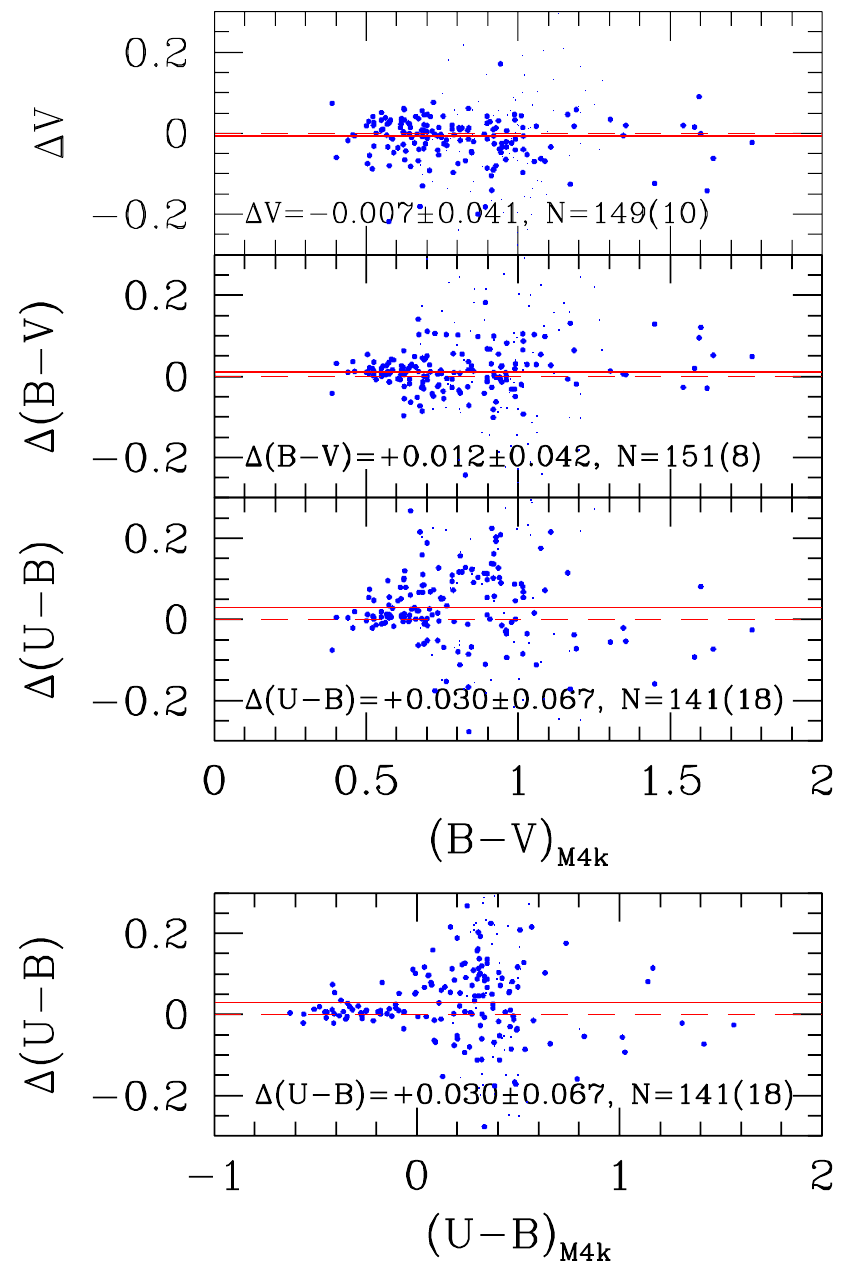}{0.23\textwidth}{(b)} 
          \fig{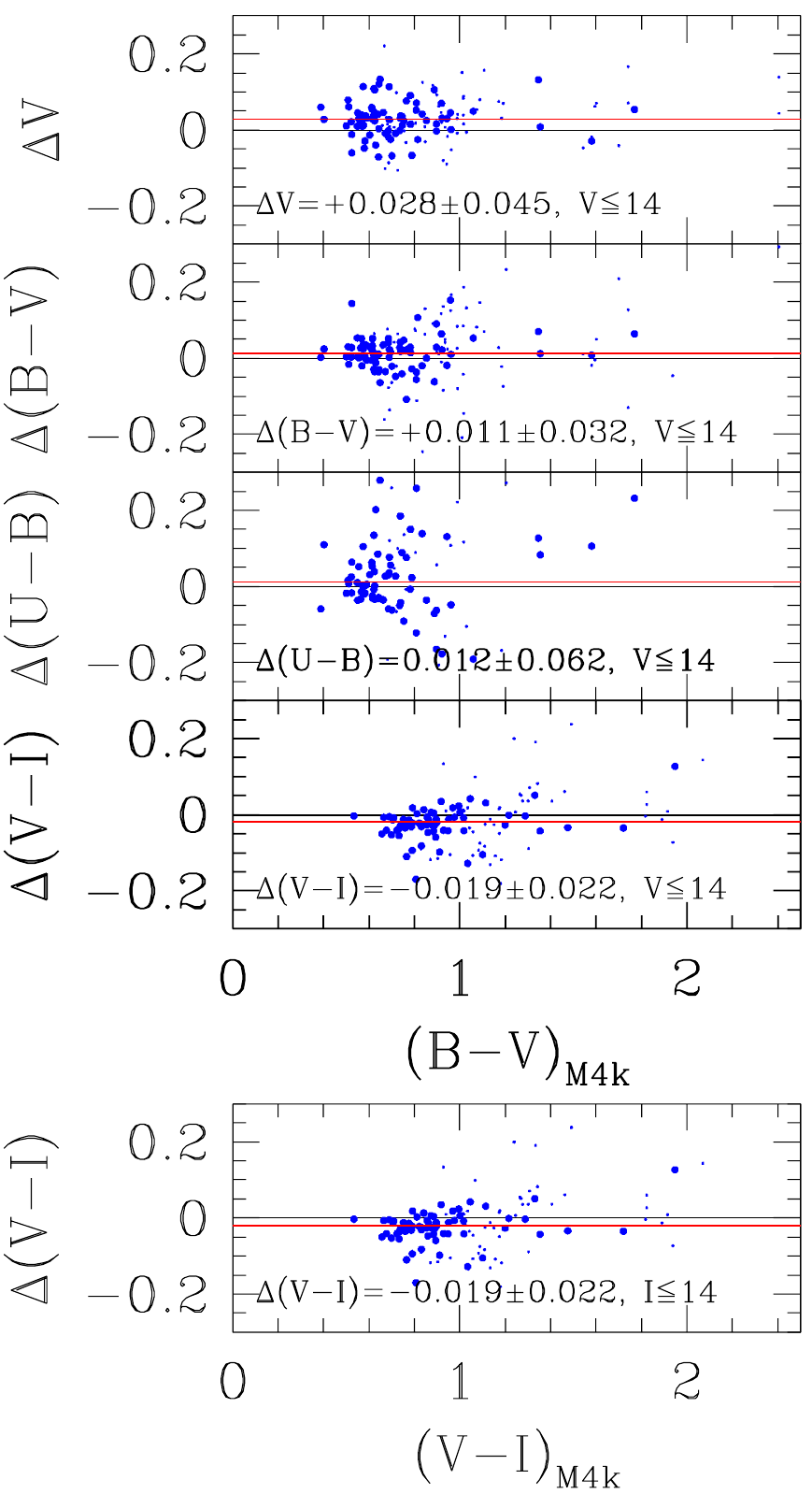}{0.23\textwidth}{(c)}
          \fig{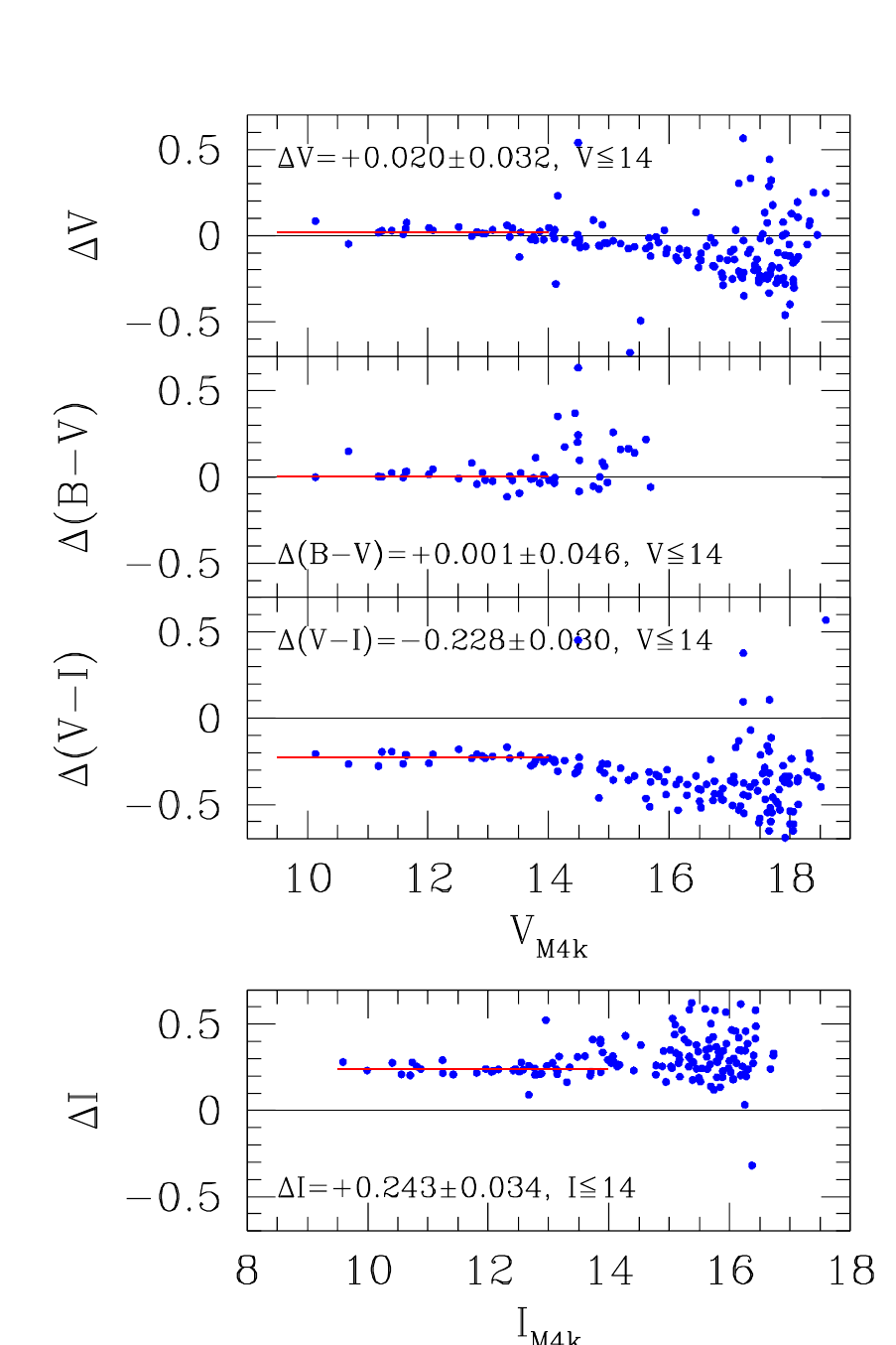}{0.23\textwidth}{(d)}}
\caption{Comparion of photometry. $\Delta$ denotes SNUCam data minus 
others. (a) photoelectric photometry - Magenta dots: \citet{jm55,jh56,hj56,h56},
black square: \citet{gv89}, black cross: \citet{i69}, blue asterisk: 
\citet{js83}, green open square: \citet{usno61}, red circle: \citet{kl83},
(b) CCD photometry by \citet{mjd95}, (c) CCD photometry by \citet{sl95}, and
(d) CCD photometry by \citet{nbes95}. Large and small dots in (b), (c), and 
(d) represent bright stars used in statistics and fainter stars, respectively.
\label{ph_comp}}
\end{figure*}

\floattable
\begin{deluxetable*}{llccccccccccc}
\tablecolumns{11}
\tabletypesize{\scriptsize}
\tablecaption{Comparison with Photoelectric and CCD Photometry \label{tab_comp}}
\tablewidth{0pt}
\tablehead{
\colhead{Catalog} & \colhead{Author} & \colhead{$\Delta V$} & \colhead{n(n$_{\rm ex}$)\tablenotemark{a}} & 
\colhead{$\Delta (V-I)$} & \colhead{n(n$_{\rm ex}$)\tablenotemark{a}} & \colhead{$\Delta (B-V)$} &
\colhead{n(n$_{\rm ex}$)\tablenotemark{a}} & \colhead{$\Delta (U-B)$} & \colhead{n(n$_{\rm ex}$)\tablenotemark{a}} & \colhead{range} }

\startdata
       & \citet{usno61} & +0.082 $\pm$ 0.011 & 8 (1) & \nodata & \nodata & \nodata & \nodata & \nodata & \nodata & $V \leq 14$ \\
            & \citet{i69} & +0.054 $\pm$ 0.034 & 9 (1) & \nodata & \nodata & \nodata & \nodata & \nodata & \nodata & $V < 14 $ \\
       & \citet{kl83}   & +0.086 $\pm$ 0.042 & 17 (5) & \nodata & \nodata & \nodata & \nodata & \nodata & \nodata & $V \leq 14$ \\
CFH12K & \citet{js83}   & +0.044 $\pm$ 0.027 & 63 (20) & \nodata & \nodata & \nodata & \nodata & \nodata & \nodata & $V \leq 14$ \\
       & \citet{mjd95} & +0.016 $\pm$ 0.058 & 429 (35) & \nodata & \nodata & \nodata & \nodata & \nodata & \nodata & $V \leq 15.5$ \\
       & \citet{sl95} & +0.068 $\pm$ 0.052 & 50 (2) & +0.014 $\pm$ 0.051 & 32 (1) & \nodata & \nodata & \nodata & \nodata & $V \leq 14$ \\ 
       & \citet{nbes95} & +0.057 $\pm$ 0.038 & 14 (1) & -0.195 $\pm$ 0.053 & 8 (0) & \nodata & \nodata & \nodata & \nodata & $V \leq 14$ \\ \hline
            & \citet{usno61} & +0.079 $\pm$ 0.035 & 19 (1) & \nodata & \nodata & +0.010 $\pm$ 0.048 & 20 (0) & -0.018 $\pm$ 0.031 & 16 (3) & $V < 14$ \\
            & \citet{i69} & +0.066 $\pm$ 0.028 & 24 (5) & \nodata & \nodata & +0.021 $\pm$ 0.019 & 21 (8) & +0.002 $\pm$ 0.049 & 23 (5) & $V < 14 $ \\
            & \citet{kl83} & +0.090 $\pm$ 0.049 & 33 (4) & \nodata & \nodata & +0.031 $\pm$ 0.034 & 32 (5) & +0.002 $\pm$ 0.056 & 32 (5) & $V < 14$ \\
Maidanak 2k & \citet{js83} & +0.056 $\pm$ 0.032 & 57 (9) & \nodata & \nodata & +0.013 $\pm$ 0.028 & 52 (14) & +0.016 $\pm$ 0.033 & 53 (11) & $V < 14$ \\
            & \citet{gv89} & +0.053 $\pm$ 0.033 & 4 (0) & -0.013 $\pm$ 0.014 & 14 (0) & +0.031 $\pm$ 0.039 & 4 (0) & -0.015 $\pm$ 0.009 & 4 (0) & $V < 12$ \\
            & \citet{sl95} & +0.066 $\pm$ 0.037 & 62 (3) & -0.002 $\pm$ 0.022 & 57 (5) & +0.008 $\pm$ 0.027 & 61 (3) & -0.020 $\pm$ 0.041 & 54 (10) & $V \leq 14$ \\
            & \citet{mjd95} & +0.032 $\pm$ 0.045 & 176 (16) & \nodata & \nodata & +0.006 $\pm$ 0.047 & 177 (16) & +0.027 $\pm$ 0.080 & 162 (31) & $V \leq 15.5$ \\
            & \citet{nbes95} & +0.067 $\pm$ 0.033 & 25 (1) & -0.205 $\pm$ 0.030 & 24 (0) & +0.019 $\pm$ 0.048 & 25 (1) & \nodata & \nodata & $V \leq 14$ \\ \hline
            & Johnson \& Hiltner\tablenotemark{b} & +0.014 $\pm$ 0.009 & 3 (0) & \nodata & \nodata & +0.015 $\pm$ 0.009 & 3 (0) & +0.005 $\pm$ 0.052 & 3 (0) & $V < 14$ \\
            & \citet{usno61} & +0.043 $\pm$ 0.022 & 19 (3) & \nodata & \nodata & +0.014 $\pm$ 0.042 & 21 (1) & -0.018 $\pm$ 0.055 & 19 (1) & $V < 14$ \\
            & \citet{i69} & +0.034 $\pm$ 0.030 & 19 (2) & \nodata & \nodata & +0.017 $\pm$ 0.015 & 16 (5) & -0.010 $\pm$ 0.029 & 17 (4) & $V < 14$ \\
            & \citet{kl83} & +0.049 $\pm$ 0.049 & 30 (3) & \nodata & \nodata & +0.033 $\pm$ 0.029 & 32 (1) & -0.005 $\pm$ 0.056 & 27 (6) & $V < 14$ \\
SNUCam & \citet{js83} & +0.014 $\pm$ 0.031 & 52 (8) & \nodata & \nodata & +0.015 $\pm$ 0.026 & 49 (11) & +0.032 $\pm$ 0.036 & 47 (12) & $V < 14$ \\
            & \citet{gv89} & +0.040 $\pm$ 0.032 & 3 (0) & -0.027 $\pm$ 0.011 & 13 (2) & +0.007 $\pm$ 0.008 & 3 (0) & -0.022 $\pm$ 0.081 & 3 (0) & $V < 14$ \\
            & \citet{sl95} & +0.029 $\pm$ 0.045 & 79 (0) & -0.019 $\pm$ 0.022 & 59 (13)& +0.011 $\pm$ 0.032 & 76 (3) & +0.013 $\pm$ 0.062 & 66 (11) & $V < 14$ \\
            & \citet{mjd95} & -0.006 $\pm$ 0.046 & 184 (17) & \nodata & \nodata & +0.014 $\pm$ 0.051 & 183 (18) & +0.031 $\pm$ 0.090 & 177 (24) & $V < 15.5$ \\
            & \citet{nbes95} & +0.020 $\pm$ 0.032 & 25 (1) & -0.228 $\pm$ 0.030 & 24 (0) & +0.002 $\pm$ 0.023 & 21 (5) & \nodata & \nodata & $V < 14$ \\
\enddata

\tablenotetext{a}{The number of stars excluded in the comparison is indicated in parenthesis}
\tablenotetext{b}{photoelectric data from \citet{jm55,jh56,hj56,h56}}
\end{deluxetable*}

Three sets of photometric data were compared with existing photoelectric photometry
and modern CCD photometry. The results are summarized in Table \ref{tab_comp}
and a few of them are shown in Figure \ref{ph_comp}. 
For the data comparison we used a successive exclusion scheme for data 
which deviated from the mean by more than $2.0 \sigma$ for the photoelectric
photometry and $2.5 \sigma$ for the CCD photometry. The reason for
using a different threshold is that photoelectric photometry is more vulnerable
to a sudden change in the sky conditions.
Photoelectric photometric data were mostly brighter than CFH12K data or Maidanak
2k data by 0.04 -- 0.08, and than SNUCam data by about 0.01 --
0.05 mag in $V$ with a scatter of about 0.02 -- 0.05 mag. Although only
three stars were in common with Johnson and/or Hiltner \citep{jm55,jh56,hj56,h56}
they were well consistent with SNUCam data. On the other hand, the photoelectric
data of \citet{js83} were well consistent with the SNUCam data photometric
zero-points, but many data were excluded in the statistics because of large
deviations.

Figure \ref{ph_comp} (b), (c), and (d) show the differences between the SNUCam
data and CCD photometry by various authors. As \citet{sl95} transformed their
CCD data to the standard system using photoelectric photometric data from
\citet{usno61} for $UBV$ and \citet{gv89} for $(V-I)$, the differences are very
similar to those of the photoelectric photometry. The SNUCam
data are well consistent with the CCD data of \citet{mjd95}. 
But the large scatter in the comparison is related to the brightness range 
used in the comparison, and is probably caused by the relatively 
short exposure time used by \citet{mjd95}.\footnote{Although they did not 
mention the exposure time explicitly, the scatter increases rapidly for $V 
\geq 14$ mag in the comparison with our CFH12K data or SNUCam data.} On the other hand,
the comparison between the SNUCam data and those of \citet{nbes95} shows
a very strange pattern. Although the difference in $V$ is very close to zero 
for bright stars, it is systematically fainter for faint stars ($V \geq 15$ 
mag). In addition, the $I$ magnitude shows an offset by about 0.25 mag for 
bright stars ($I \leq 14$ mag), but the difference increases for fainter stars.
Such an offset is related to the difference between the Cousins and Johnson
$I$ magnitude systems as they transformed the $(V-I)$ of \citet{gv89} to
Johnson's $(V-I)$ using the relation given by \citet{msb79}. The curved feature
in Figure \ref{ph_comp} (d) 
may be related either to the non-linear response of the CCD chip they used 
(Kodak KAF-4200) at the faint regime or to improper sky subtraction.

\subsection{{\it Spitzer} Mid-Inrared Observations and Data Reduction}

\subsubsection{Spitzer Observations of IC 1805 \label{sst_obs}}

Deep optical photometry of IC 1805 reveals that H$\alpha$
emission stars are distributed across the whole FOV of the CFH12K observations (see Figure
\ref{YSOdist}). To confirm whether the distribution of PMS stars with H$\alpha$
emission is real or not, we decided to reduce the {\it Spitzer} MIR images.
The Spitzer mapping observations were performed under program ID 20052
(PI: S. Wolff) in 9 $\times$ 9 mosaics. Each pointing was imaged in the high dynamic 
range mode (exposure time: 0.4 s and 10.4 s). The mapping of IC 1805 was 
performed on 2006 September 20. We refer to the region as ``SST/CM''.
The Astronomical Observation Requests (AORs) 
utilized for this map was number 13846016. For complete photometry of stars 
in the CFH12K FOV in $3.6 \mu m$ and $4.5 \mu m$, we also downloaded and
reduced the GLIMPSE360 data (AOR: 38753280, 38763264, 38769408, 38799104, 38798592, 
38784512, PI: B. A. Whitney).

\begin{figure*}
\gridline{\fig{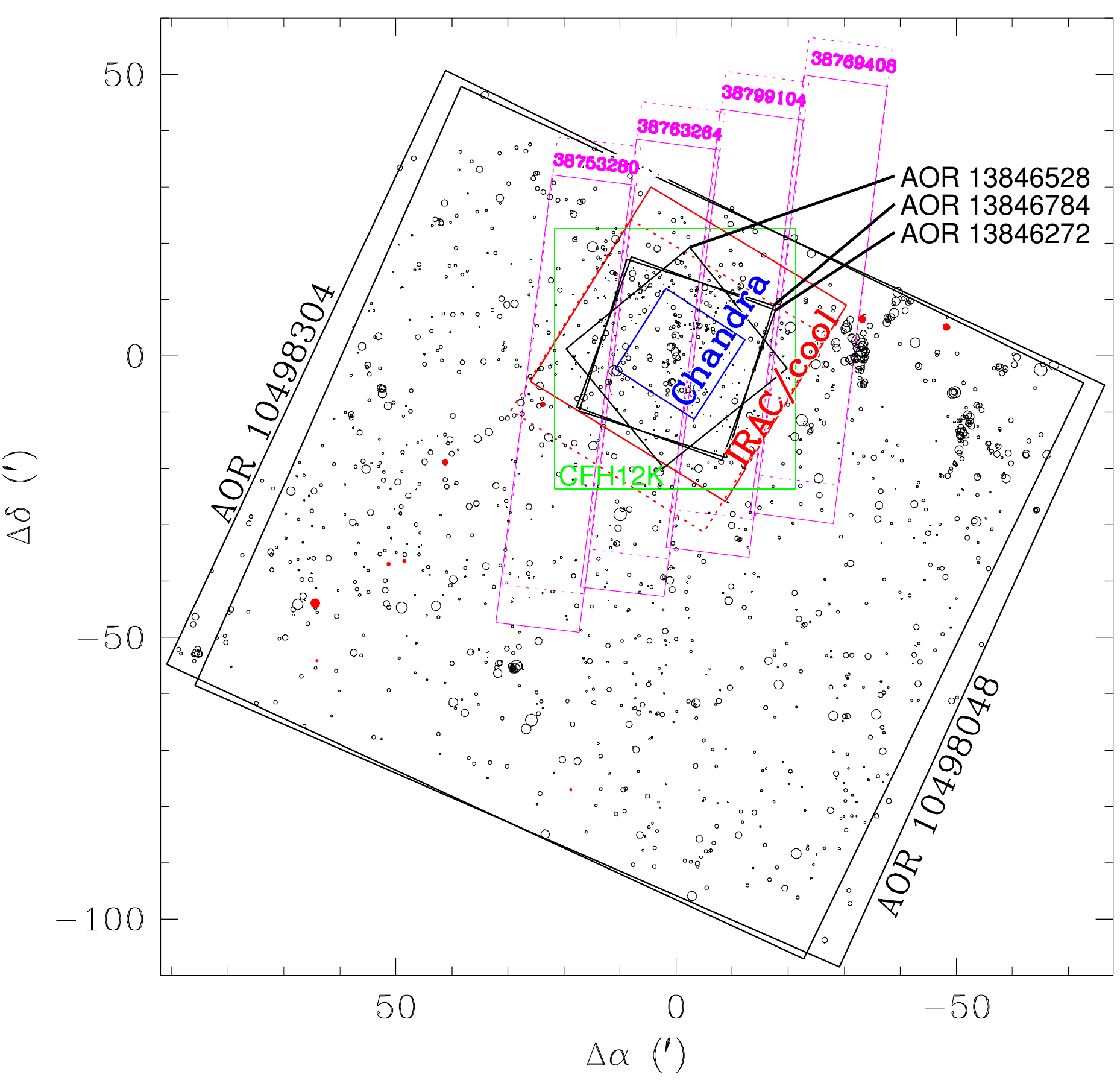}{0.47\textwidth}{(a)}
          \fig{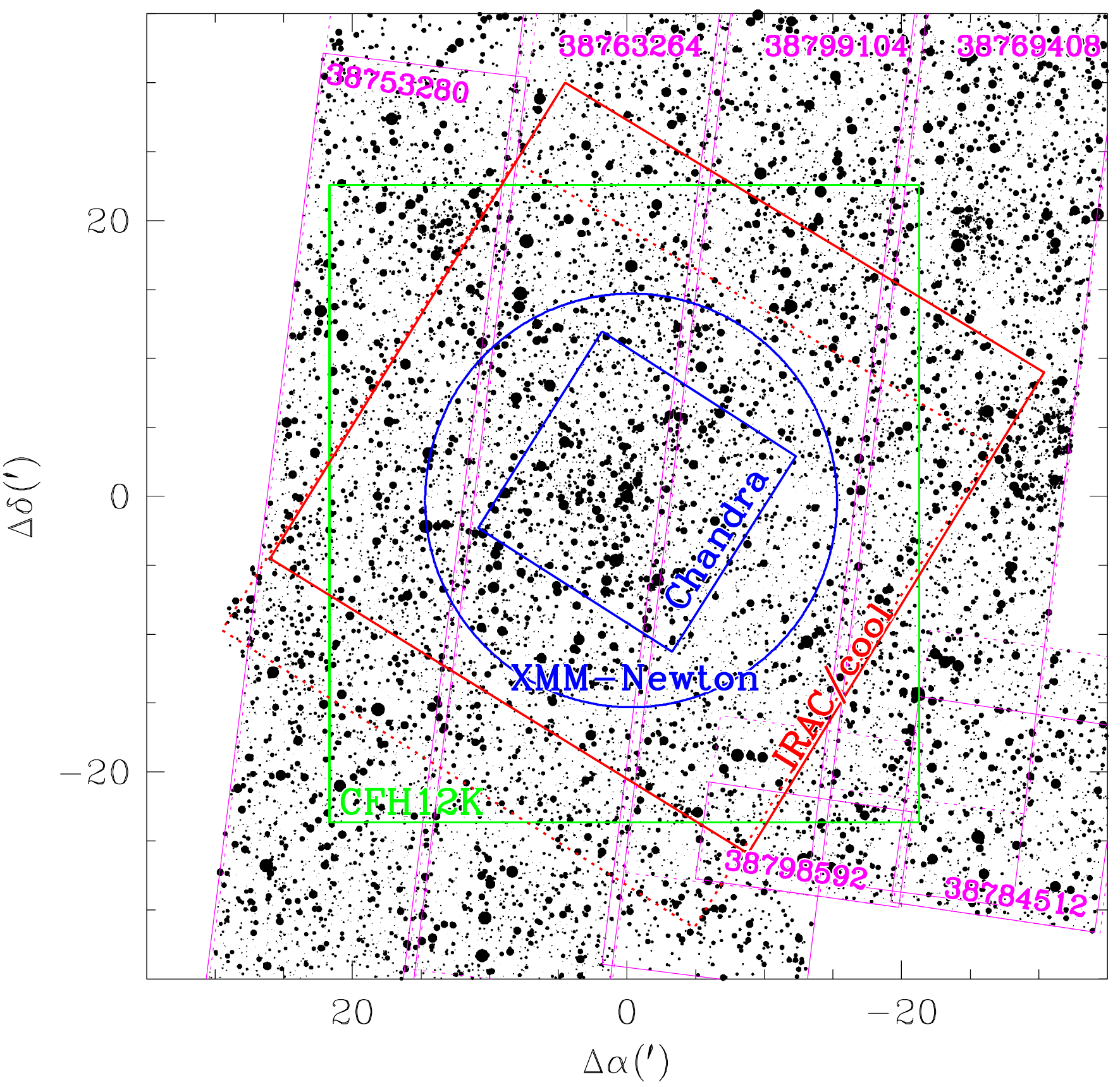}{0.47\textwidth}{(b)} }
\caption{Fields of view of the {\it Spitzer} MIR, {\it Chandra} 
and {\it XMM-Newton}, and CFH12K optical observations. (a) Black, red, magenta,
green, and blue squares indicate the FOVs of {\it Spitzer}/MIPS 24$\mu m$
observations, {\it Spitzer}/IRAC cool mission observations, {\it Spitzer}/IRAC
GLIMPSE360 survey, CFH12K, and {\it Chandra} X-ray observations, respectively.
The solid and dotted lines for the {\it Spitzer}/IRAC FOVs denotes the FOVs of 
IRAC 3.6$\mu m$ (and 5.8$\mu m$) and 4.5$\mu m$ (and 8.0$\mu m$), respectively. 
The size of the circles is proportional to the 24 $\mu m$ brightness. Red dots
represent extended sources from PSF fitting. (b) Central portion of (a).
A blue circle denotes the FOV of the XMM-Newton observations. 
The size of dots is proportional to the brightness in 3.6$\mu m$ or 4.5 $\mu m$.
The number in each strip is the AOR of the GLIMPSE360 images.
\label{all_fov}}
\end{figure*}

MIPS scans of IC 1805 were
obtained on 2005 August 31 and 2005 September 2 (PID 3234, PI: J. S. Greeves)
at the fast scan rate (exposure time: 2.62 s). Twenty five scans of 1.00 
length, with 300 offsets, were used. The observed area is much larger than
the FOV of CFH12K. The AORs utilized for the MIPS mapping
were numbers 10498304 and 10498048. Three MIPS Phot images (AOR: 13846272, 
13846528, 13846784, PI: S. Wolff) were also used. The post-BCD (basic 
calibrated and mosaicked) images were downloaded from the Spitzer heritage 
archive\footnote{\url{http://sha.ipac.caltech.edu/applications/Spitzer/SHA/}}.
The pixel size of the IRAC post-BCD data is $0\farcs6 \times 0\farcs6$, while that
of the MIPS 24 $\mu m$ data is $2\farcs45 \times 2\farcs45$. The data utilized 
pipeline processing software version S18.7.0 for the IRAC (cool mission) images,
S19.1.0 for GLIMPSE360 images, and S18.12.0 for the MIPS 24 $\mu m$ image.

\subsubsection{Photometry}
\floattable
\begin{deluxetable*}{cccccccccccccc@{}c@{}c@{}c@{}c@{}c@{}c@{}cllrl}
\tablecolumns{26}
\tabletypesize{\scriptsize}
\rotate
\tablecaption{Catalog of {\it Spitzer Space Telescope} IRAC and MIPS 24$\mu m$ Sources.\tablenotemark{*} \label{tab_sst}}
\tablewidth{0pt}
\tablehead{
\colhead{Spitzer ID} & \colhead{$\alpha_{\rm J2000}$} & \colhead{$\delta_{\rm J2000}$} & \colhead{[3.6]} &
\colhead{[4.5]} & \colhead{[5.8]} & \colhead{[8.0]} & \colhead{[24]} &
\colhead{$\epsilon_{[3.6]}$} & \colhead{$\epsilon_{[4.5]}$} & \colhead{$\epsilon_{[5.8]}$} &
\colhead{$\epsilon_{[8.0]}$} & \colhead{$\epsilon_{[24]}$} & \multicolumn{5}{c}{N$_{\rm obs}$} &
\colhead{D\tablenotemark{a}} & \colhead{Class\tablenotemark{c}} & \colhead{M\tablenotemark{b}} & \colhead{2MASS ID\tablenotemark{d}} &
\colhead{Optical Counterpart} }

\startdata
\enddata

\tablenotetext{*}{Table  \ref{tab_sst} is presented in its entirety in
the electronic edition of the Astrophysical Journal Supplement Series. A portion is shown here
for guidance regarding its form and content. Units of right ascension are
hours, minutes, and seconds of time, and units of declination are degrees,
arcminutes, and arcseconds.}
\end{deluxetable*}

We used the IRAF version of DAOPHOT to derive PSF-fitting photometry 
for the stars in the field of IC 1805. Because Spitzer IRAC images are undersampled,
PSF fitting yields photometry with relatively poor signal-to-noise. 
For uncrowded fields with little nebulosity, aperture photometry would provide 
photometry with lower noise than PSF-fitting photometry for IRAC data. 
However, portions of the IC 1805 field are crowded, or have highly variable 
and strong nebulosity, or both. We believe that PSF-fitting photometry 
provides more uniform and reliable photometry than aperture photometry, 
admittedly at the expense of having more noise for stars where the backgrounds are 
benign and crowding is not an issue. 
For most stars in the cluster, because we have four independent sets of data, 
the PSF-fitting photometric accuracy is improved by averaging the results from 
the separate AORs. Details for the data reduction can be found in \citet{ssb09}.
We found that
IRAS 02260+6118 (=S022094, YSO class: F) is a point source in 3.6 and 4.5 $\mu
m$ images, but an extended source in 5.8 and 8.0 $\mu m$ images.

\begin{figure*}
\epsscale{0.95}
\plotone{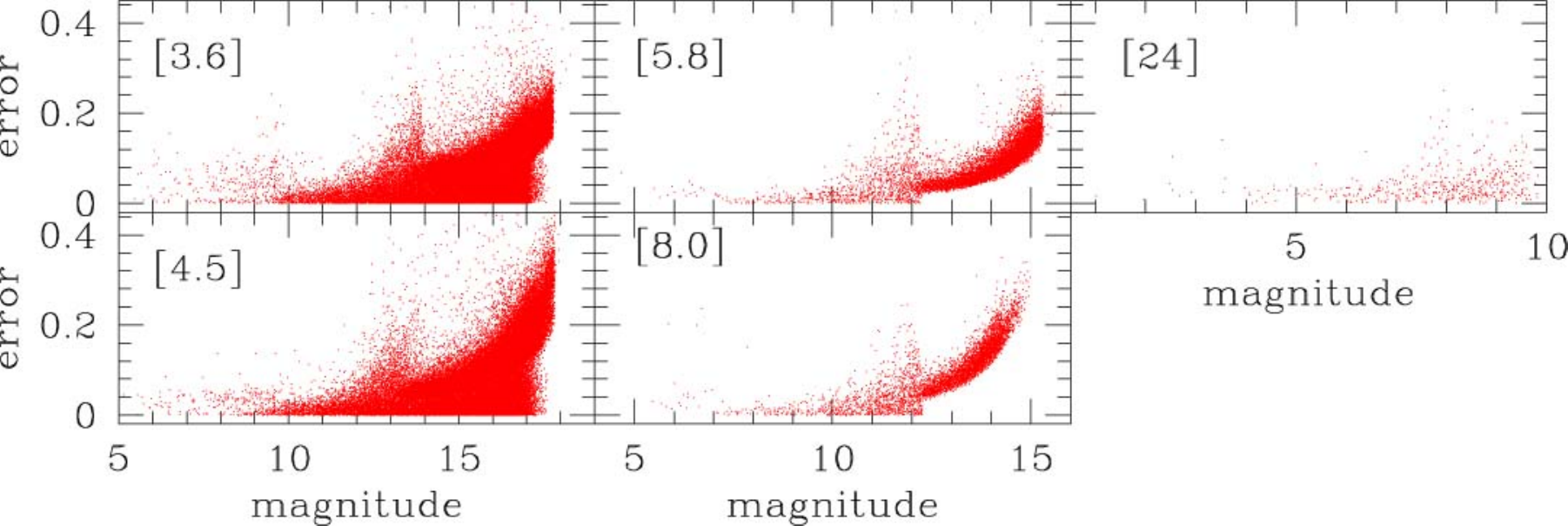}
\caption{Distribution of photometric errors as a function of magnitude. 
\label{sst_error} }
\end{figure*}

The FOVs of the {\it Spitzer}/IRAC and MIPS, CFH12K optical, {\it Chandra} 
and {\it XMM-Newton} X-ray observations are shown in Figure \ref{all_fov}.
The FOV covered by the stars measured is the combination of the FOVs of
AOR 13846016 (cool mission), a full strip of the GLIMPSE360 survey -
AORs 38753280, 38763264, 38769408, 38799104, and a small portion of AORs
38798592, 38784512 as shown in Figure \ref{all_fov}. The weighted mean values
and weighted errors of the magnitudes from multiple observations were 
calculated as in \citet{sl95} (weight = $1 / \epsilon^2$). 
We present the photometric data for four IRAC bands and the MIPS 24 $\mu m$
band for 101,746 objects in Table \ref{tab_sst}. The distribution of photometric
errors is shown in Figure \ref{sst_error}. As the {\it Spitzer}/IRAC images are
undersampled data, the photometric errors are no better than 0.1 mag, even for
bright stars (eg. [3.6] $<$ 10 mag). But as several epochs of data with 
two exposure times per epoch are available, the resulting final error is 
small if the magnitudes from all images are consistent (eg. stars with $\epsilon
\approx 0.0$ at [3.6] $>$ 14 mag). If not, the resulting error will be larger
(eg. stars with $\epsilon > 0.1$ at [3.6] $\approx$ 13.7 mag). The abrupt
increase in photometric errors at [3.6] $\approx$ 14, [4.5] $\approx$ 13,
and [5.8] $\approx$ [8.0] $\approx$ 12 is due partly to a large intrinsic
error from short exposure images and partly to a large difference in
photometry from short and long exposed images. Such a trend
can be seen in Figure \ref{sst_error}. We label the objects in 
Table \ref{tab_sst} as ``S'' + the identification number in the first column. 
The total number and faint limit of objects detected from the photometry are 
100082 stars and 18.3 mag for [3.6], 100989 stars and 18.0 mag for [4.5],
11092 stars and 16.0 mag for [5.8], 5433 stars and 15.0 mag for [8.0], 
and 523 stars and 9.8 mag for [24], respectively. We included in the table, 
the YSO class (see section \ref{ysoclass}), membership 
information (H$\alpha$ or/and X-ray emission - see section 3),
duplicity from the PSF fitting process, 2MASS identification, and any optical 
counterpart.

We compared our data with \citet{wsr11} who published {\it Spitzer}/IRAC and
MIPS data for 974 objects. Among them fourteen objects were listed twice.
We also found no counterpart in our data for 110 of their objects. Most of them 
(83 objects) were outside our FOV shown in Figure \ref{all_fov}. Twenty five 
objects that had no counterpart in our catalog also had no counterpart in 2MASS, 
and were mostly faint ([3.6] $>$ 14 mag). They may therefore be spurious 
detections, such as cosmic ray events. Two objects not in our catalog are 
the bright K2III star BD+60 519 and its neighbor, due to severe saturation 
of BD+60 519. For objects in common with \citet{wsr11}, the differences relative 
to our photometry are +0.011 $\pm$ 0.030 mag (N = 571, 78 excluded), +0.000 $\pm$
0.028 mag (N = 574, 98 excluded), -0.006 $\pm$ 0.051 mag (N = 549, 104 excluded),
+0.022 $\pm$ 0.099 mag (N = 552, 119 excluded), and -0.093 $\pm$ 0.225 mag (N = 34,
5 excluded) in [3.6], [4.5], [5.8], [8.0], and [24], respectively.
The consistency of the photometric zero-points
between the two data sets is good, but the scatter increases for fainter stars.


\subsection{X-Ray Observations}

\subsubsection{{\it Chandra} X-ray Observatory Observations}

The {\it Chandra} X-ray Observatory Observations of IC 1805 (Obs ID: 7033, PI: 
L. Townley) were made on 2006, November 25. The total exposure time was about 79ks.
The properties of 647 X-ray sources were published in \citet{tbg14}, which is
part of ``the Massive Young Star-Forming Complex Study in Infrared and X-Ray
(MYStIX) Project'' \citep{ftb13}. We searched for the optical and MIR counterparts
of these X-ray sources with a matching radius of up to $1\farcs5$. If the candidate
was the closest object within a matching radius of $1\farcs0$, we considered
the object as the optical (MIR) counterpart of the X-ray source and 
assigned the membership ``X''. The second closest object, or the closest object
within $1\farcs5$ from the X-ray source, was considered to be a candidate
X-ray emission object, and assigned the membership ``x''. The membership
information is included in Tables \ref{tab_cfht}, \ref{tab_m2k}, \ref{tab_m4k}, 
and \ref{tab_sst}. Among 647 X-ray sources, 194 objects had no counterpart in our optical source
catalogs and MIR source list within a matching radius of $1\farcs5$ (232 objects
within $1\farcs0$). Twenty-six X-ray sources had only {\it Spitzer} MIR
counterparts.

\begin{figure*}
\epsscale{0.9}
\plotone{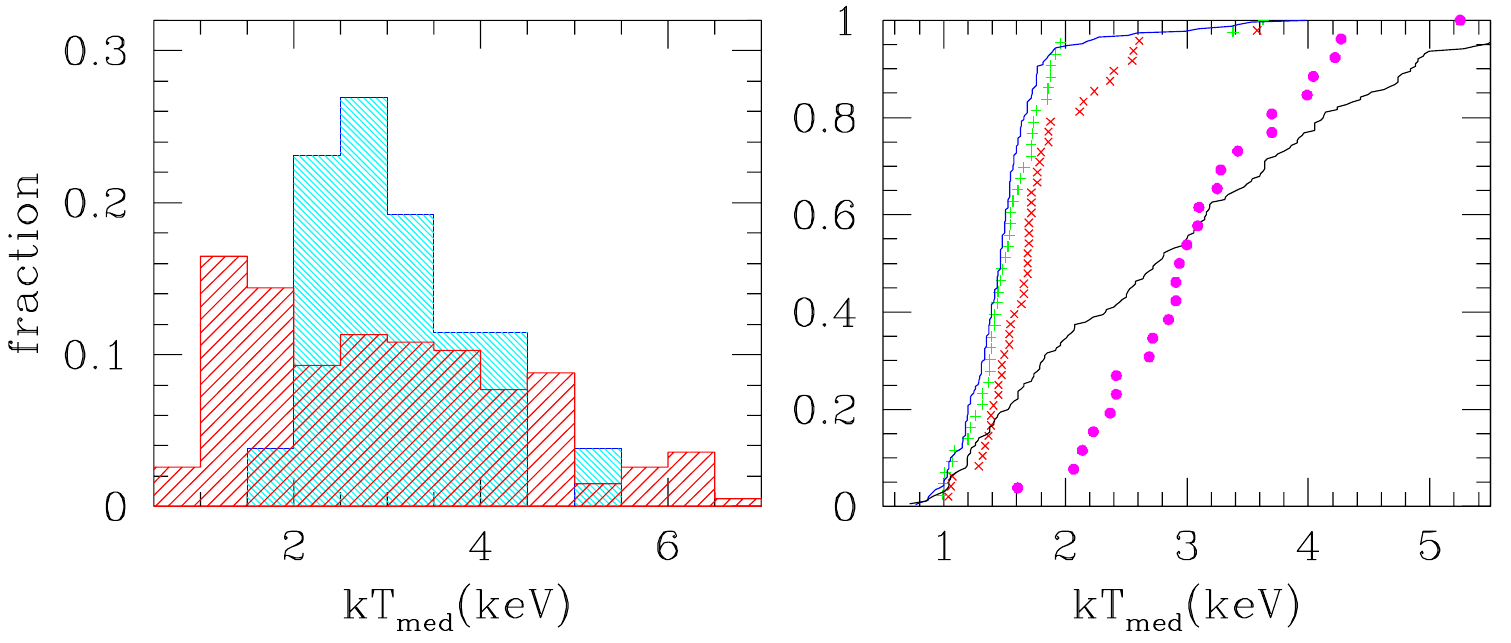}
\caption{ Median X-ray energy distribution of MYStIX sources. (Left) Histogram
of no optical counterpart (cyan) and that of no counterpart in optical as well as
MIR (red). (Right) Cumulative distribution of disk-bearing YSOs (red cross),
Class III (green plus), Class IV (blue solid line), no optical counterpart (magenta dot),
and no counterpart in optical and MIR (black solid line).
\label{mystix} }
\end{figure*}

\citet{tbg14} also released the median X-ray energy of the X-ray sources. The median
energy distribution of the 194 X-ray sources with no optical and MIR counterpart 
showed two peaks with a more than 10\% fraction in a 0.5 keV bin (see left panel of Figure
\ref{mystix}). The median value 
of the median energy was 2.8 ($\pm$ 1.5) keV with the stronger one at 1.0 -- 
2.0 keV, and the second one at 2.5 -- 4.0 keV. On the other hand, the 26
X-ray sources detected only in the MIR {\it Spitzer} observations showed a different
distribution with a single peak near 3.0 ($\pm$ 0.8) keV, which corresponds to
the second peak of the X-ray sources without a counterpart in this study. The median value of the
median X-ray energy of stars with X-ray emission only (i.e.
no emission in H$\alpha$), was 1.5 ($\pm$ 0.6) keV.
However, that of stars with emission in both H$\alpha$ and X-ray 
was slightly harder (1.7 $\pm$ 0.5 keV),
that may be related to the relatively larger column density of the surrounding
circumstellar materials. A similar pattern of X-ray energy distribution can be
found from the median values among different YSO classes (see section 
\ref{ysoclass} for YSO classification). The YSO classes I, 
F, II, t. T, and g, which are definite members of
IC 1805, had slightly higher X-ray temperatures ($<E>_{med}$ = 1.7 $\pm$ 0.7
keV, N = 48) (red crosses in the right panel of Figure \ref{mystix}). While the median
values of Class III and Class IV objects (green plus and blue solid line, respectively,
in Figure \ref{mystix}) were
1.5 ($\pm$ 0.5) keV (N = 43) and 1.5 ($\pm$ 0.4) keV (N = 231), respectively. 

\subsubsection{XMM-Newton Observations}

XMM-Newton observations of IC 1805 were conducted in August 2014 for a single snapshot of 48ks 
duration (ObsID: 0740020101, PI: G. Rauw). The data were reduced using SAS v14 
(see \citet{rn16} for more detail). Source detection was performed 
using the task {\sc edetect\_chain} on both soft (0.4--2.0\,keV) and hard (2.0--
10.0\,keV) band images and for all three EPIC cameras. This task 
first searched for sources using sliding boxes, then applied a PSF fitting to 
yield the best positions and equivalent on-axis count rates. It was run for 
a likelihood of detection of 10, both with, and without considering 
the possibility of extended sources, and simultaneously fitting up to 5 
neighbouring sources, but the results were similar in both cases. A total of 
191 sources were found, nine of them appearing potentially problematic (e.g. 
due to their position in a CCD gap, or in the PSF wings of a brighter source). 
A more detailed study of the X-ray properties of these sources was dealt with in
\citet{rn16}.

We searched for the optical, {\it Spitzer} MIR, and {\it Chandra} X-ray source
counterparts for the XMM-Newton X-ray sources with a matching radius of up to
$6''$ (mostly less than $4''$). Among 191 XMM-Newton sources, 174 optical 
sources (stars or galaxies) were identified as optical counterparts for 167 
XMM-Newton X-ray sources, 182 {\it Spitzer} MIR sources were identified as 
MIR counterparts of 175 XMM-Newton sources. Among 143 XMM-Newton sources 
within the FOV of {\it Chandra} X-ray observation, 130 sources were matched
with one or two {\it Chandra} X-ray sources (141 X-ray sources in MYStIX
catalog). Thirteen XMM-Newton sources had no counterpart in the MYStIX catalog,
and therefore may be spurious detections or X-ray sources with strong 
variability. Four XMM-Newton sources had no counterpart in the optical, near-infrared (NIR)
2MASS, and MIR catalogs.

\section{Membership Selection}

\begin{figure*}
\epsscale{1.0}
\plotone{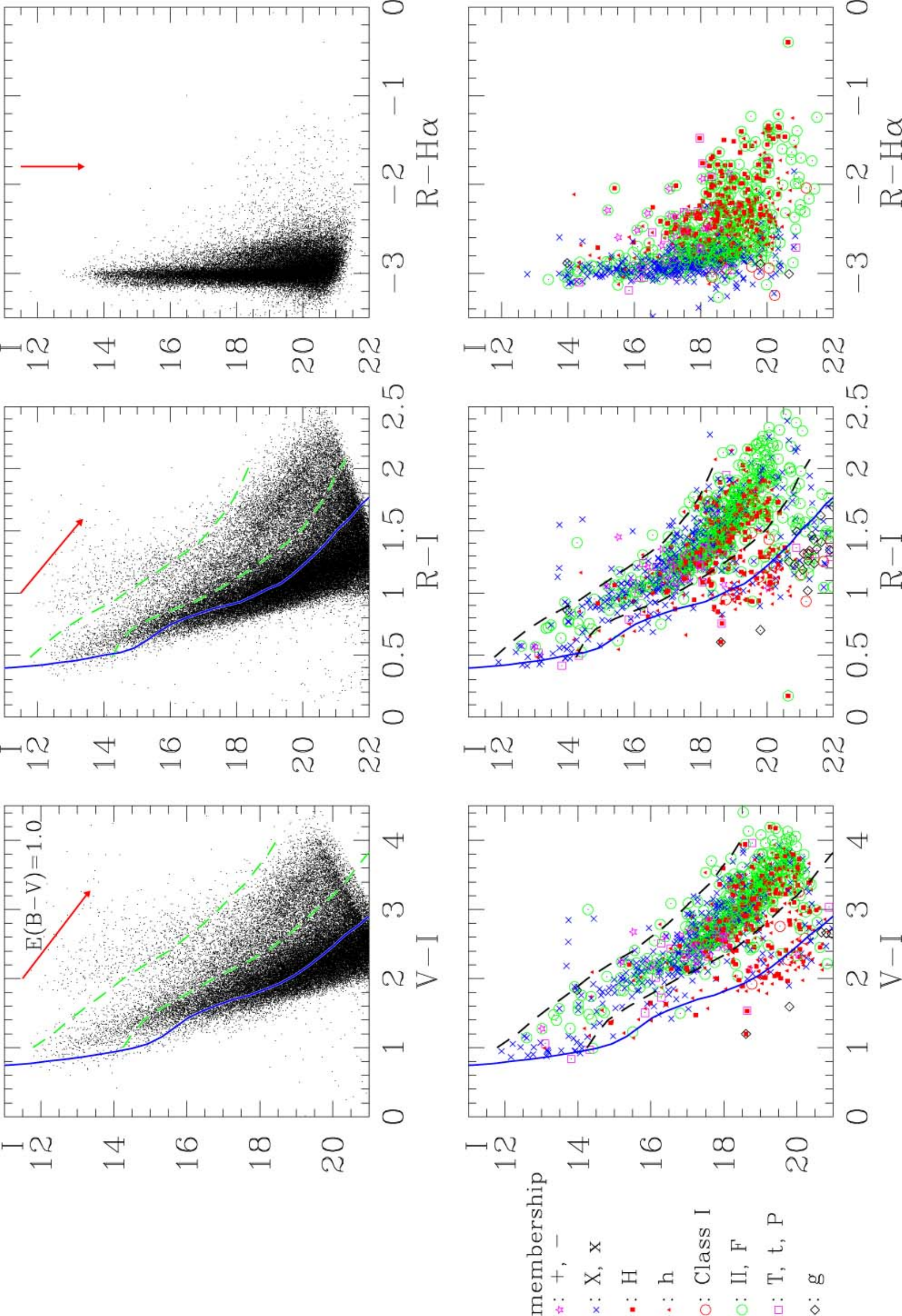}
\caption{Color-magnitude diagrams of IC 1805 from CFH12K observations. Upper
panels: the ($I$, $V-I$) (Left), ($I$, $R-I$) (Center), and ($I$,
$R-$H$\alpha$) (Right) diagrams for all stars, Lower panel: the same
diagrams for those stars assigned as members. The blue solid line represents the ZAMS 
relation \citep{sos} at a distance of 2.4 kpc and $E(B-V)$ = 0.85 mag, and the two dashed lines
in the left two panels are the upper and lower boundary of the PMS stars in IC 1805. The red arrow in
the upper panels is the reddening vector for $E(B-V)$ = 1.0 mag.
The meaning of symbols is presented in the lower left.
\label{opt_cmd}}
\end{figure*}

Membership selection in the study of open clusters is a critical factor in
deriving reliable physical properties of the clusters because, as
most open clusters are in the Galactic plane, we can expect there to be many
field interlopers in the foreground as well as in the background.
We present the color-magnitude diagrams (CMDs) of IC 1805 from CFH12K 
observations in Figure \ref{opt_cmd}. However only a weak enhancement
of stars between the two dashed lines can be seen in the upper panels  of
Figure \ref{opt_cmd}. The locus of PMS stars in the CMDs of young open clusters gives 
several important parameters, such as age, mass distribution, star
formation history, etc. And therefore low-mass membership selection 
is the important first step to precisely determine the locus of PMS stars in Figure \ref{opt_cmd}.
The locus of PMS members is modified and updated based on new membership
selection criteria described below. The color excess ratio
in $R$ and $I$ is somewhat uncertain. The parameterization representation of the
interstellar reddening law \citep{ccm89} predicts $E(R-I) = 0.833 E(B-V)$
and $E(V-I) = 1.592 E(B-V)$ for the $R_V$ obtained in section \ref{reddening_law}
which is not a good fit to the MS band in Figure \ref{opt_cmd}. In addition, the
value of 1.592 is very different from the canonical value 1.25 obtained by \citet{dwc78}.
Alternatively, we determined these values from Figure \ref{mao_cmd} which
give the best fit to the blue MS stars in IC 1805 - $E(R-I) / E(B-V)$ = 0.66 and
$E(V-I) / E(B-V) $ = 1.26.

The membership selection of low-mass members at the PMS stage is very
difficult because most of them are brighter than normal MS stars.
Because classical photometric colors cannot give a reliable membership
selection criterion for low-mass PMS stars in young open clusters,
various useful membership selection criteria have been introduced during the last 20 years,
such as H$\alpha$ photometry \citep{sbl97}, X-ray emission \citep{fms99,sbc04},
and MIR excess emission \citep{gmm08,kag08,ssb09}.
These membership selection criteria
have their own limitations. For a thorough selection of members, several
criteria should be used in conjunction. In this section, we describe 
several membership selection criteria, their merits and their limitations.
The selection criteria for H$\alpha$ emission stars is described in section
\ref{ha_sel}, MIR excess emission stars in section
\ref{ysoclass}, and X-ray emission members in section \ref{xraystar}. 

\subsection{{\rm H}$\alpha$ Emission stars \label{ha_sel}}

\subsubsection{{\rm H}$\alpha$ Emission stars from CFH12K observation \label{cfh_ha_sel}}

\floattable
\begin{figure*}
\epsscale{0.9}
\plotone{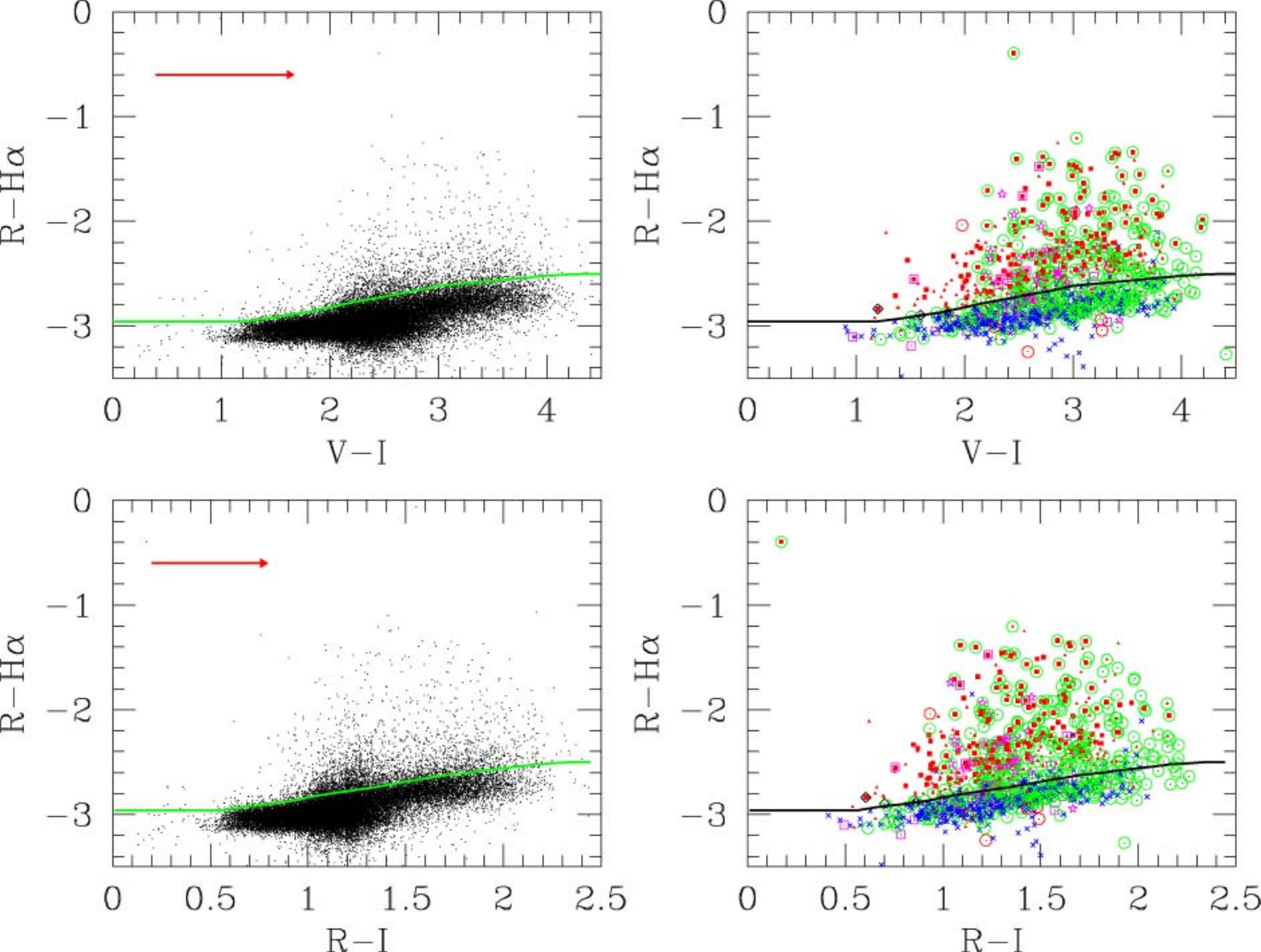}
\caption{The selection criteria for H$\alpha$ emission stars. Left panels:
the $(R-$H$\alpha)$ versus $(V-I)$ diagram (Upper) or the $(R-$H$\alpha)$ 
versus $(R-I)$ diagram (lower) of all stars, Right panels: the same diagrams 
for stars with membership. The green (left panels) or black line (right panels) denotes
the mean line of stars with no appreciable H$\alpha$ emission. The red arrow
in the left panels represent the reddening vector of $E(B-V)$ = 1.0 mag.
The other symbols in the right panels are the same as those in Figure \ref{opt_cmd}.
\label{cfh_ha} }
\end{figure*}

\citet{sbl97} used the H$\alpha$ emission measure index, ($R - $H$\alpha$), as 
a membership criterion for low-mass PMS stars in NGC 2264. We present diagrams 
of ($R-$H$\alpha$) versus ($V - I$) or ($R-$H$\alpha$) versus ($R - I$) in 
Figure \ref{cfh_ha}. The left panels of Figure \ref{cfh_ha} show the distribution
of all stars detected in H$\alpha$. The division of cluster stars and field
stars is not as evident as that in the field of NGC 2264 (see Fig. 5 of
\citealt{sbc08}). This is due to the fact that the less reddened foreground stars are
relatively rare and most field stars (or member stars without any appreciable
H$\alpha$ emission) detected are those in the Perseus spiral arm,
whose reddening is very similar to that of the cluster stars. There is a vertical
scatter at ($V-I$) $\approx$ 2.4 and ($R-I$) $\approx$ 1.2. These objects with
large photometric errors are either faint late-type stars in the Perseus arm, halo 
stars in the FOV, or faint external galaxies. The solid line 
represents the mean line of stars with no appreciable H$\alpha$ emission, 
such as foreground MS stars or weak-line cluster T Tauri stars.  

We used the
same selection criteria for H$\alpha$ emission stars as in \citet{sbc08}, i.e.
$\Delta (R-$H$\alpha) >$ 0.2 as H$\alpha$ emission stars (membership class: H)
and $\Delta (R-$H$\alpha) >$ 0.1 as H$\alpha$ emission candidates (membership
class: h) if their combined photometric error in ($R-$H$\alpha$) was less than
0.07 mag. As the depth of the H$\alpha$ images for the North and South
regions was not the same as that for the Center, we used different faint limits
for selecting H$\alpha$ emission stars - H$\alpha$ = 24 mag for 
the Center ($\Delta \delta = -14.'5$ -- $+13.'5$), 22 mag for the North ($\Delta
\delta \geq +13.5$), and 23.2 mag for the South ($\Delta \delta \leq -14.5$).
In addition, a more stringent criterion was applied for the faint stars ($R >
20.75$) to avoid many spurious detections due to their large intrinsic errors.

\subsubsection{ {\rm H}$\alpha$ Emission Stars from SNUCam Observation \label{m4k_ha_sel}}

\begin{figure*}
\gridline{\fig{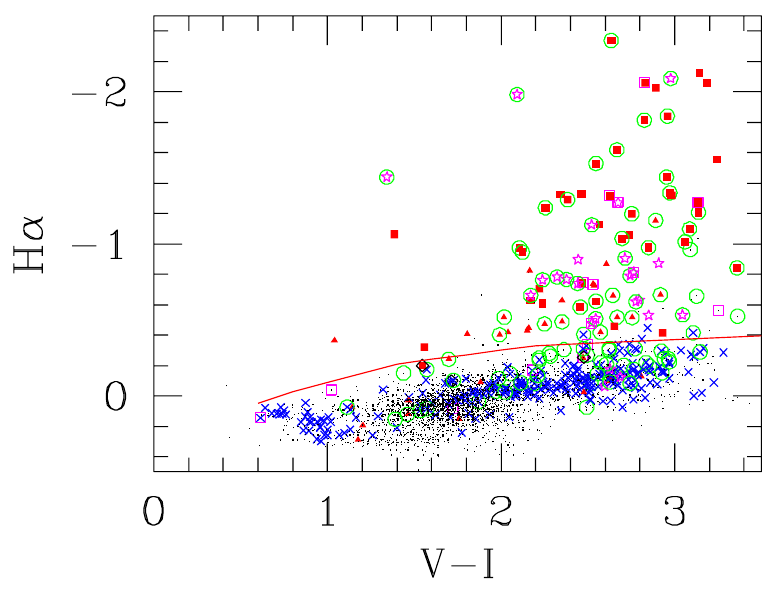}{0.48\textwidth}{(a)}
          \fig{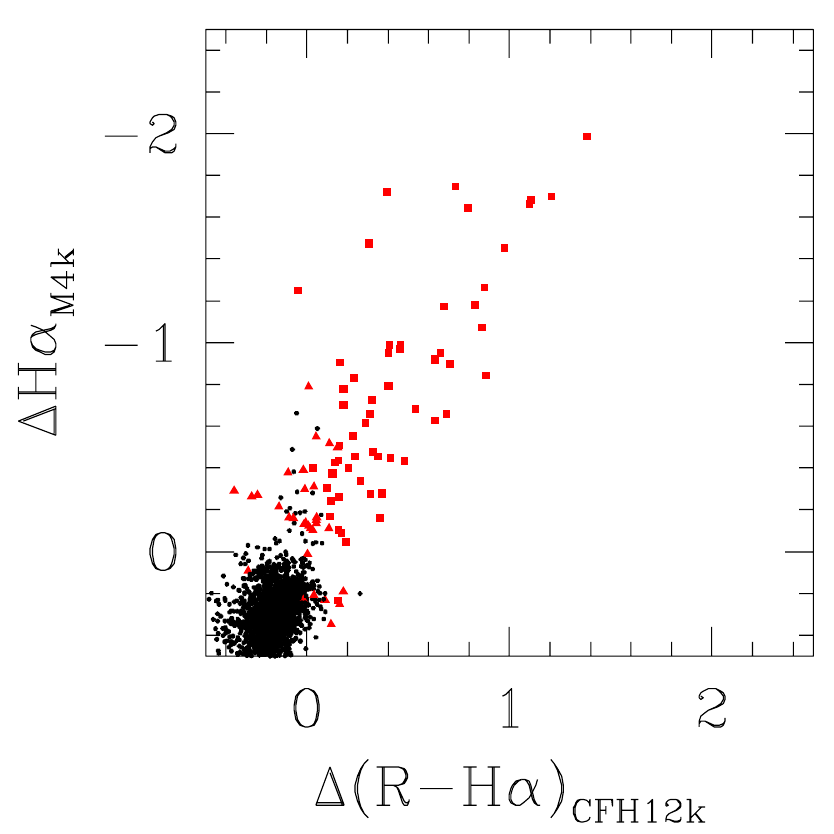}{0.36\textwidth}{(b)}}
\caption{(a) The selection criterion for H$\alpha$ emission stars from the
SNUCam data. The solid line represents the photospheric level of 
unreddened stars. The other symbols are the same as Figure \ref{opt_cmd}.
(b) Comparison of H$\alpha$ emission indices between $\Delta (R-$H$\alpha)$
from the CFH12K and $\Delta$H$\alpha$ from the SNUCam data. Red squares and 
triangles are H$\alpha$ emission stars and candidates, respectively.
\label{m4k_ha} }
\end{figure*}

As the depth of the H$\alpha$ images at the Maidanak Astronomical 
Observatory was much shallower than those obtained with
the CFH12K and the seeing was also relatively poor, the H$\alpha$ emission star selection 
from the M4k data was limited
to the relatively bright stars ($I$ + H$\alpha$ $\leq$ 18 mag) to reduce the
number of spurious detections due to large photometric errors in H$\alpha$.
The selection criterion of H$\alpha$ emission stars is the same as that in
\citet{lsk14a,lsk14b} as shown in Figure \ref{m4k_ha}. 

A total of 45 H$\alpha$ emission stars and 16 H$\alpha$ candidates were selected from
the SNUCam data. 
The H$\alpha$ emission indices $\Delta (R-$H$\alpha)$ and $\Delta$H$\alpha$ are
compared in Figure \ref{m4k_ha} (b). Overall consistency between the two indices
was good, but some stars showed a large difference that may be related to the 
variabilty of the star. In addition, some stars
showed weak emission in one index, but not in the other. As the time difference
between the CFHT observation and the SNUCam observations was 5.5 or 7 years,
the level of stellar activity of some stars could have changed.

As most T Tau-type PMS stars show strong variability, especially in 
H$\alpha$, the ($R-$H$\alpha$) index of some stars also showed variability even 
over a 1-day timescale, therefore the H$\alpha$ membership criteria from one data set
was not always the same as that from another data set. In addition, stars with
strong variability may have a large combined photometric error. As we had 5
sets of independent photometry in H$\alpha$ (3 sets from the CFH12K 
observations and two sets
from the SNUCam data), we applied an additional selection criterion to recover
the membership of such stars, regardless of their combined photometric error
in ($R-$H$\alpha$). If a star was classified as an H$\alpha$ emission star
(either H or h) more than twice from their ($R-$H$\alpha$) index
or H$\alpha$ index used in the SNUCam data, we classified the star as an 
H$\alpha$ emission star. From this procedure we selected 26 stars as
H$\alpha$ emission stars (membership class: H). Among the newly selected
``H'' stars, 20 stars were originally classified as ``h'' from the first classification
scheme. Similarly we selected 94 H$\alpha$ emission candidates (membership class: h) 
where H$\alpha$ emission was detected only once from several H$\alpha$ 
observations.

From these selection procedures a total of 182 H$\alpha$ emission stars and 199 H$\alpha$
emission candidates were selected. Among them, SNUCam data 
contributed wholly for 3 H$\alpha$ emission stars (VSA 113 = MWC 50 = M4k0795,
C41178 = M4k1785, \& C66601 = M4k6084) and 21 H$\alpha$ emission candidates, 
and partly contributed (i.e. detected once) to the selection of
10 H$\alpha$ emission stars.
The H$\alpha$ emission stars in \citet{osp02} were cross-matched with our H$\alpha$
emission stars, and we found that four stars in \citet{osp02} (BRC 7-4,
-7, -8, \& -9) were also classified as H$\alpha$ emission stars from our classification scheme,
and two (BRC 7-1, \& -7) were Class II objects from our MIR data. BRC 7-6 was not classified
as an H$\alpha$ emission star nor a Class II object, but is in the PMS locus. The other three
stars (BRC 7-2, -3, \& -5) are not classified as H$\alpha$ emission stars as well as not
being in the PMS locus. 

We can find H$\alpha$ emission stars over the whole FOV, and the degree of
concentration is rather low. The distribution of H$\alpha$ emission stars is shown in the left
panel of Figure \ref{YSOdist}. The highest
density region of H$\alpha$ emission stars coincides with the brightest part of
the nebula near HD 15629 (O4.5V - \citealt{smw11}). 
As mentioned in section \ref{mimm}, the distribution 
is very similar to that of the intermediate-mass stars of IC 1805.

\subsection{MIR Excess Emission Stars \label{ysoclass}}

\begin{figure*}
\epsscale{1.0}
\plotone{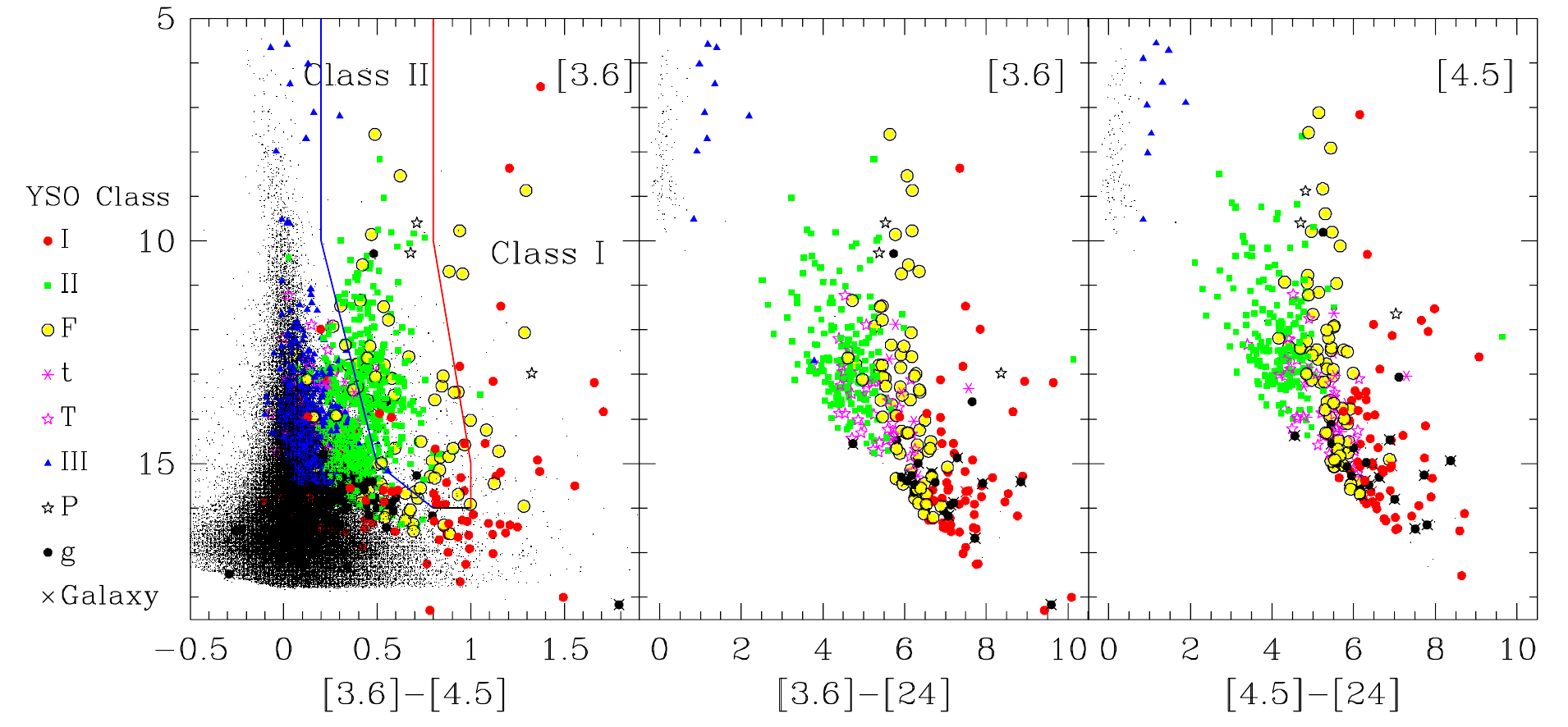}
\caption{Color-magnitude diagrams (CMDs). Red dots, green squares, yellow dots with a black
circle, magenta asterisks, magenta stars, blue triangles, large black dots with a
cross, large black dots, and black stars represent, respectively, Class I, 
Class II, flat spectrum objects, objects with a pre-transition disk, transition 
disks, Class III, visually confirmed galaxies, photometric galaxies, and objects with PAH 
emission. Small dots denote objects with no YSO classification. Blue and
red solid lines in the ([3.6], [3.6]-[4.5]) CMD are dividing
lines used for the classification of Class II and Class I objects
detected only in the [3.6] and [4.5] bands.
\label{sst_cmd}}
\end{figure*}

\begin{figure*}
\epsscale{0.8}
\plotone{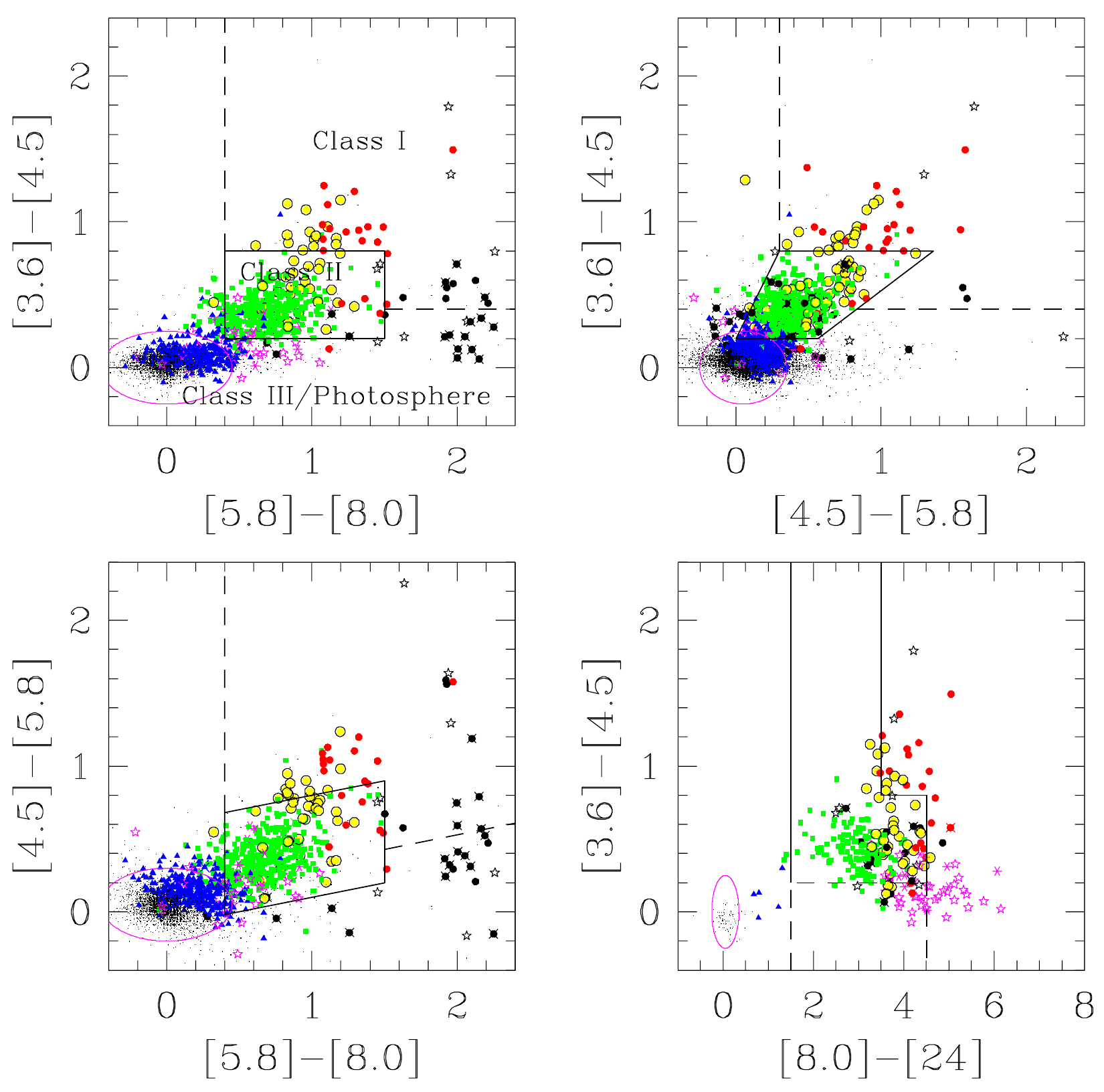}
\caption{Two-color diagrams. Symbols are the same as those in Figure 
\ref{sst_cmd}.
\label{sst_ccd}}
\end{figure*}

Classification of YSOs is a basic step towards the study 
of the properties and evolutionary status of YSOs. The same classification 
scheme described in \citet{ssb09} was employed, i.e. both the use of two-color 
diagrams (TCDs) and the slope of the spectral energy distribution (SED) ($\alpha
\equiv d \log ({\lambda F_\lambda}) / d \log \lambda$) with a proper
weighting scheme. In addition, for the classification of stars outside the SST/CM FOV,
we have to use a simple classification criterion, i.e. a star's
location in the ([3.6], [3.6]-[4.5]) CMD (see Figure
\ref{sst_cmd}). In actual application of the above classification criterion,
the photometric errors in [3.6] and [4.5] were also taken into account. 
A total of 11910
objects with reasonable photometric errors were classified. The number of class
I ($\alpha \geq +0.3$), Flat ($+0.3 > \alpha \geq -0.3$), class II ($-0.3 > \alpha
\geq -1.8$), class III ($-1.8 > \alpha \geq -2.55$), and class IV ($\alpha < -2.55$
- stellar photosphere) objects were 76, 85, 542, 1433, and 9681, respectively.
\citet{ssb09} also introduced $\alpha_{IRAC}$ (SED slope from four IRAC bands)
and $\alpha_{LW}$ (SED slope between 8.0 $\mu m$ and 24 $\mu m$) to classify
stars with (pre-)transition disks. 
If  $\alpha_{IRAC}$ = -0.3 -- -1.8 and $\alpha_{LW} > +0.3$, we classified the object
as a YSO with a pre-transition disk (``t''), and if  $\alpha_{IRAC} < -1.8$ and
$\alpha_{LW} > +0.3$, then a YSO with a transition disk (``T'').
In addition, if the 8.0 $\mu m$ flux of an object was more than 0.3 dex larger than the flux
estimated from the 5.8 $\mu m$ flux and 24 $\mu m$ flux, we assigned the object
as ``P'' (an object with polycyclic aromatic hydrocarbon emission). And if [5.8]-[8.0] $> 1.5$,
we classified the object as ``g'' (an object with MIR colors similar to those of starburst
galaxy). The number of objects with classes P, t, T, and g is 6, 11, 37, and 35,respectively.
The distribution of these objects in the TCDs
is shown in Figure \ref{sst_ccd}. Although we used the MIR CMDs for the classification
of YSOs outside the SST/CM FOV, as YSO classification is largely dependent on the SED slope,
YSO classification outside the SST/CM FOV may be incomplete.

The nature of the spectral classes of objects (classes P and g) is also of interest.
There are 6 objects with a class P. Two objects are stars - S57629 (=C43353) is a normal 
A type star, the other (S50327 = C34233) is an H$\alpha$ emission object below
the PMS locus. Two objects (S48812 = C32227; S91183) are faint extended sources.
\footnote{S91183 is not measured from the CFH12K images and 
so is not listed in Table \ref{tab_cfht}, but a faint elongated 
object can be seen on the CFH12K images. C32227 (=S48812) is a faint object below 
the PMS locus.} Both may be galaxies. Finally, the last two sources are 
S60140 and S97616. The former is the counterpart of two optical sources C46494 
(a star in the PMS locus) and C46539 (a faint extended object below the PMS locus, 
and hence a background galaxy), while the latter is not in the CFH12K FOV.

There were 35 objects with a class g. Thirty objects are extended objects
in the CFH12K images, hence are  galaxies. Two (S47970 = C31194, 
S57498 = C43188) are stars in the PMS locus. S73461 is the counterpart of two
optical sources C63264 (an H$\alpha$ emission object below the PMS locus) 
and C63272 (a star? in the PMS locus). However it is very difficult to judge
whether S73461 is an elongated galaxy or a close optical double. 
The remaining two are not in the CFH12K FOV.
The total number of optically confirmed galaxies is 122.

\begin{figure*}
\epsscale{0.95}
\plotone{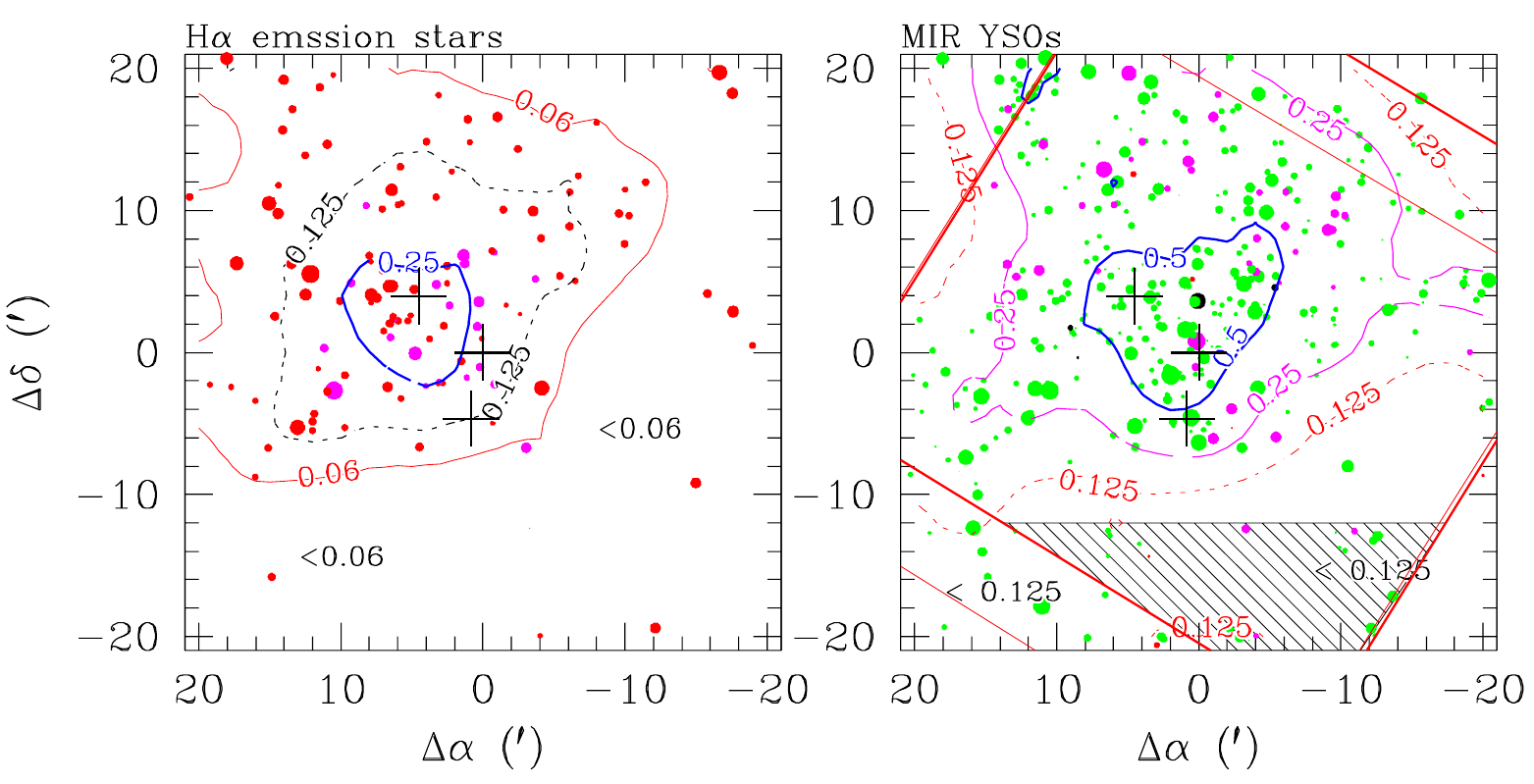}
\caption{The spatial distribution of H$\alpha$ emission stars (Left) and YSOs with
YSO class I, F, II, t, T, g (Right). The position of HD 15558 ($\Delta \alpha$ = 0$\farcm$0,
$\Delta \delta$ = 0$\farcm$0), HD 15570 ($\Delta \alpha$ = 0$\farcm$82,
$\Delta \delta$ =-4$\farcm$66) and HD 15629 ($\Delta \alpha$ = 4$\farcm$54,
$\Delta \delta$ =3$\farcm$94) is marked as ``+''. The contour with numbers represents
the surface density of H$\alpha$ emission stars and YSOs in units of [star
arcmin$^{-2}$ ]. Thick and thin large squares in the right panel represent the FOV of IRAC 3.6 $\mu m$
and 4.5 $\mu m$, respectively. The shaded area in the right panel represents the control field 
selected for the correction of field star contribution to the initial mass function (see section 
\ref{imf}). Dots represent either H$\alpha$ emission stars (Left) or YSOs
(Right) whose size is proportional to the brightness of the objects. The color of the
dots indicates the type of membership - (Left) red: H$\alpha$ emission stars,
magenta: H$\alpha$ emission stars with X-ray emission; (Right) red: Class I,
green: Class II or flat spectrum (F) objects, magenta: objects with (pre-)
transition disks (t or T) or with PAH emission (P), and black: galaxy candidates
(g). \label{YSOdist} }
\end{figure*}

\begin{figure*}
\epsscale{0.95}
\plotone{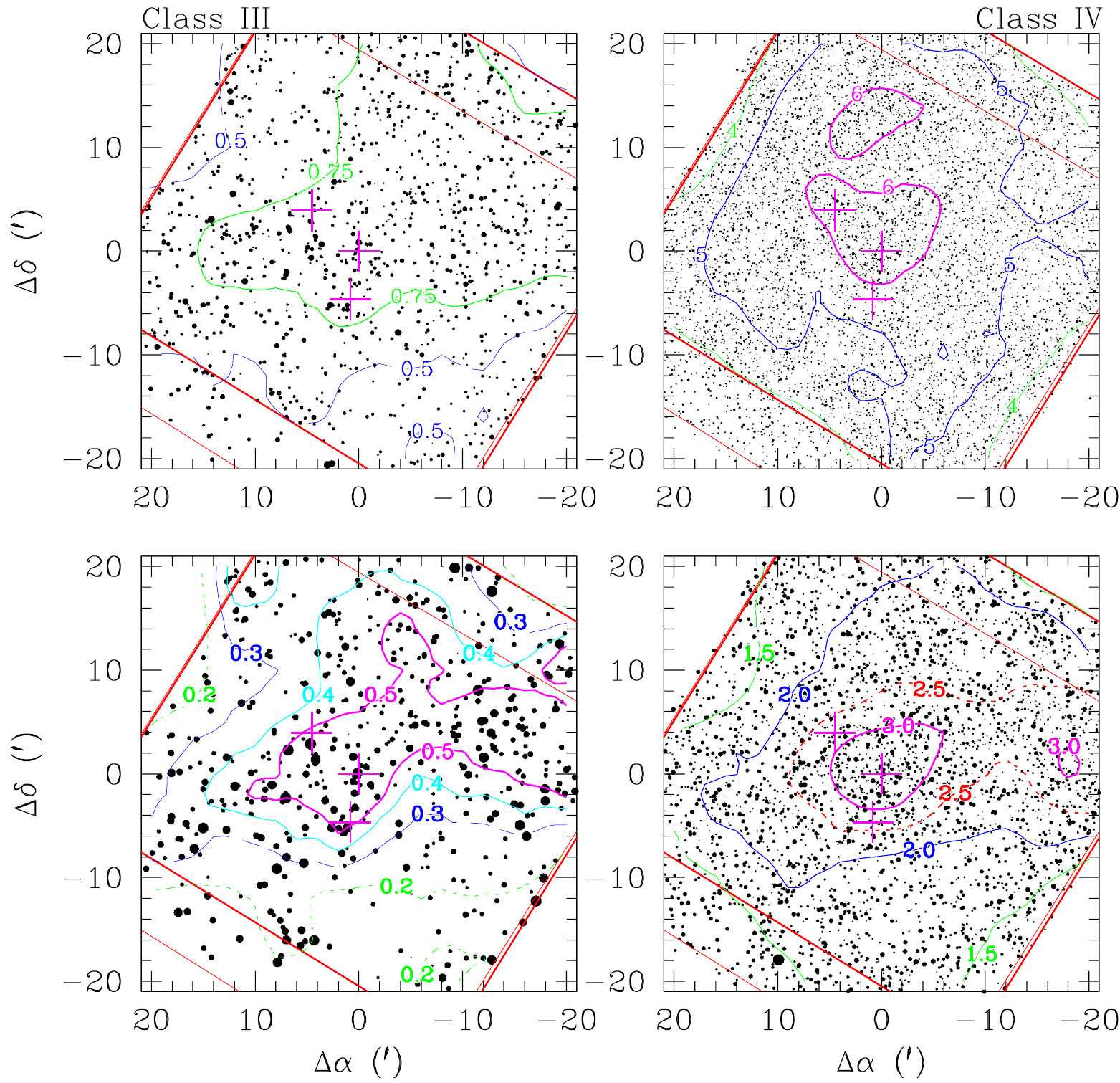}
\caption{The spatial distribution of stars with YSO Class III and IV.
(Upper) all stars with YSO class III (left) and IV (right). 
(Lower) the stars with the given YSO class in the PMS locus.
The numbers on the contour denote
the surface density in units of [star arcmin$^{-2}$]. 
\label{Cl3n4} }
\end{figure*}

Among stars with a YSO classification
from MIR photometric diagrams and SED slopes, YSO classes I, II, F, 
t, T, and stars of YSO class P and g are considered as probable low-mass PMS members of IC 1805.
We present the spatial distribution of YSOs in the right panel of Figure
\ref{YSOdist}.  The highest density region of YSOs is
midway between HD 15558 (O4.5III(f)) and HD 15629. The density of
these objects decreases as the distance from the peak increases. 
The gradient of the surface density is high to the south and south-west,
but low to the north. The lowest density of these objects is south-west
of the cluster center. In addition there is a weak signature of density
enhancement of these objects in the far southern region. The marginal difference
between the two distributions in the north may be caused by the difference
in the photometric depth of the H$\alpha$ observations as well as
the incompletenesses of YSO classification from MIR photometry.

The surface density distribution of Class III and IV objects is shown in Figure
\ref{Cl3n4}. These objects show a weak enhancement near the cluster center.
There is no physical reason for Class III or IV objects above or below the
PMS locus to show any radial variation of the surface density. However
the surface density of these objects in the PMS locus may show a radial
variation, and hence we have drawn their surface density in the lower
panels of Figure \ref{Cl3n4}. This fact implies that the disk lifetime of some
PMS stars  may be very small or the strong ultraviolet (UV) radiation from hot massive
stars in the cluster center may affect the disk lifetime \citep{ssb09}.

\subsection{X-ray Emission Stars \label{xraystar}}

Strong X-ray emission is one of the more prominent properties of PMS stars 
and therefore can be used as a membership criterion for the PMS stars in young open clusters. 
However, as X-ray emission from late-type stars persists for a long time \citep{ssb08},
and in addition, as the activity level of X-ray emission from PMS stars covers a wide range,
i.e., $\log {L_X / L_{bol}}$ = -5 -- -3 \citep{fgg03}, we should expect there to be some
foreground or background interlopers with X-ray emission. With the above caveats
we tentatively identify the optical counterparts of the X-ray emission sources as cluster members,
and then check their distribution in the optical CMDs. From the overall distribution of
H$\alpha$ emission stars, MIR excess emission stars, and X-ray emission stars,
the locus of PMS stars in the CMDs is finally derived as was done for the young open cluster
NGC 2264 \citep{sbc04}. The optical counterparts of X-ray sources in the PMS locus are
considered to be members of IC 1805.

The lower panels of Figure \ref{opt_cmd} show the optical CMDs of the stars with 
PMS membership (stars with H$\alpha$ emission, X-ray
emission, and/or PMS stars with YSO class I, F, II, t, T, P, and g). Most of these stars with
PMS membership are well located between the two dashed lines, and therefore
the two dashed lines represent the PMS locus of IC 1805. However some stars
with PMS membership are on, or near the reddened ZAMS line. Some of them were
selected as PMS members both from H$\alpha$ photometry and MIR photometry.
These stars are most probably PMS stars with nearly edge-on disks. Other stars
selected from only one membership criterion (H$\alpha$ photometry or MIR
photometry) may not be real PMS members, but spurious detections due to their
intrinsic large photometric errors. Some X-ray emission stars near the reddened
ZAMS line are either X-ray active stars among field stars in the Perseus
spiral arm, or background galaxies. In addition, a few X-ray emission stars above the upper
limit of the PMS locus, are foreground active late-type stars in the local arm.
There is a YSO class F (C01044 = S022094) which is far brighter than
the other PMS members in the PMS locus. This object is the
optical counterpart of IRAS 02260+6118, and could be one of the youngest
objects near the border between W3 and W4 (see also \citealt{pcp14}). 

\subsection{Massive and Intermediate-Mass Members \label{mimm}}

\begin{figure*}
\epsscale{0.8}
\plotone{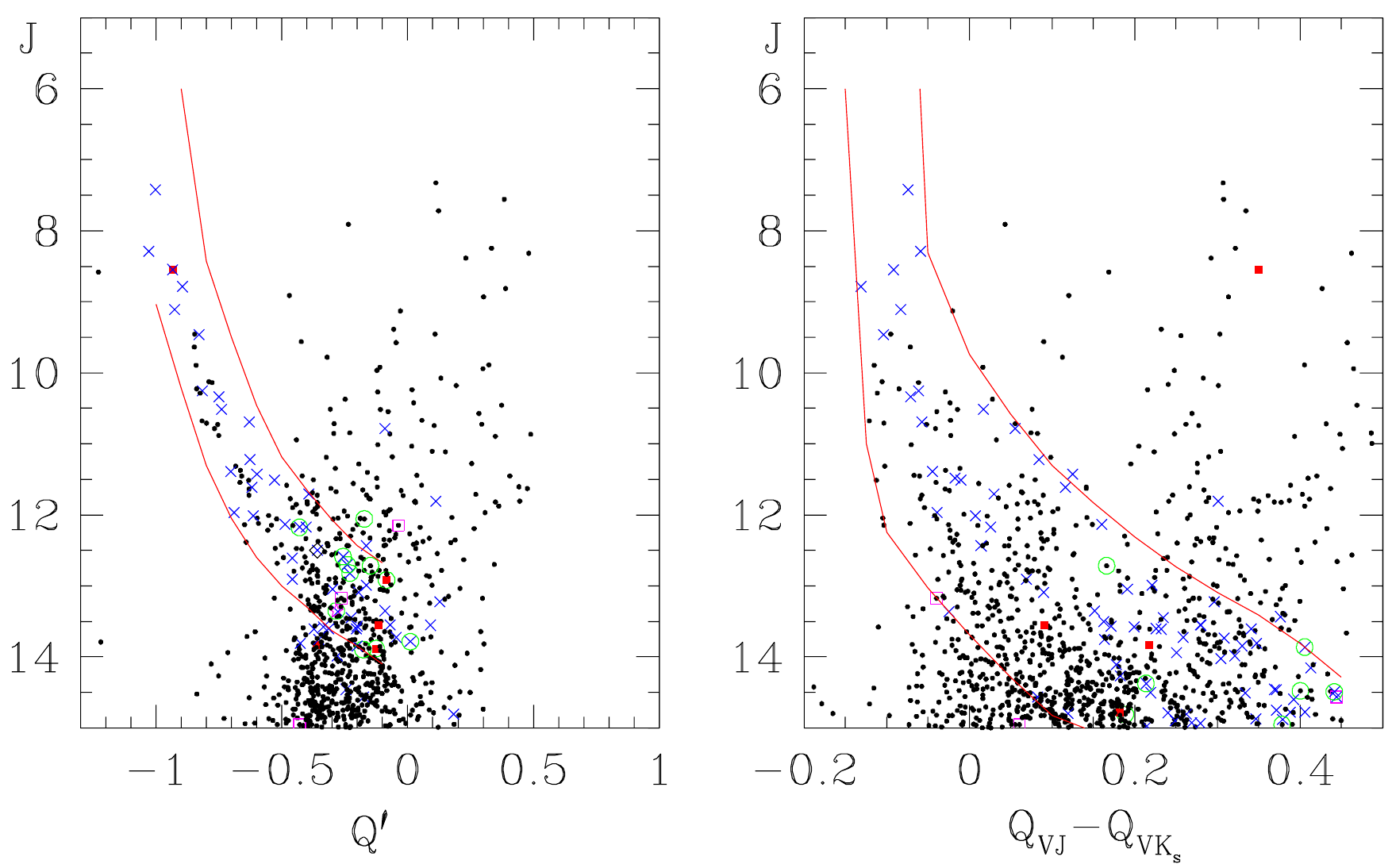}
\caption{($J, ~Q'$) diagram (Left) and ($J, ~Q_{VJ} - Q_{VK_s}$) diagram (Right)
of stars in the Maidanak 2k data. The modified Johnson's $Q$ [$Q' \equiv (U-B)
- 0.72 (B-V) - 0.025 E(B-V)^2$] and the reddening-free indices 
$Q_{V\lambda}$ are defined in \citet{sos}. The sold lines represent the upper
and lower limit of massive members and candidate intermediate-mass members
of IC 1805.  The other symbols are the same as Figure \ref{opt_cmd}.
\label{dmq2} }
\end{figure*}

\begin{figure*}
\epsscale{1.0}
\plotone{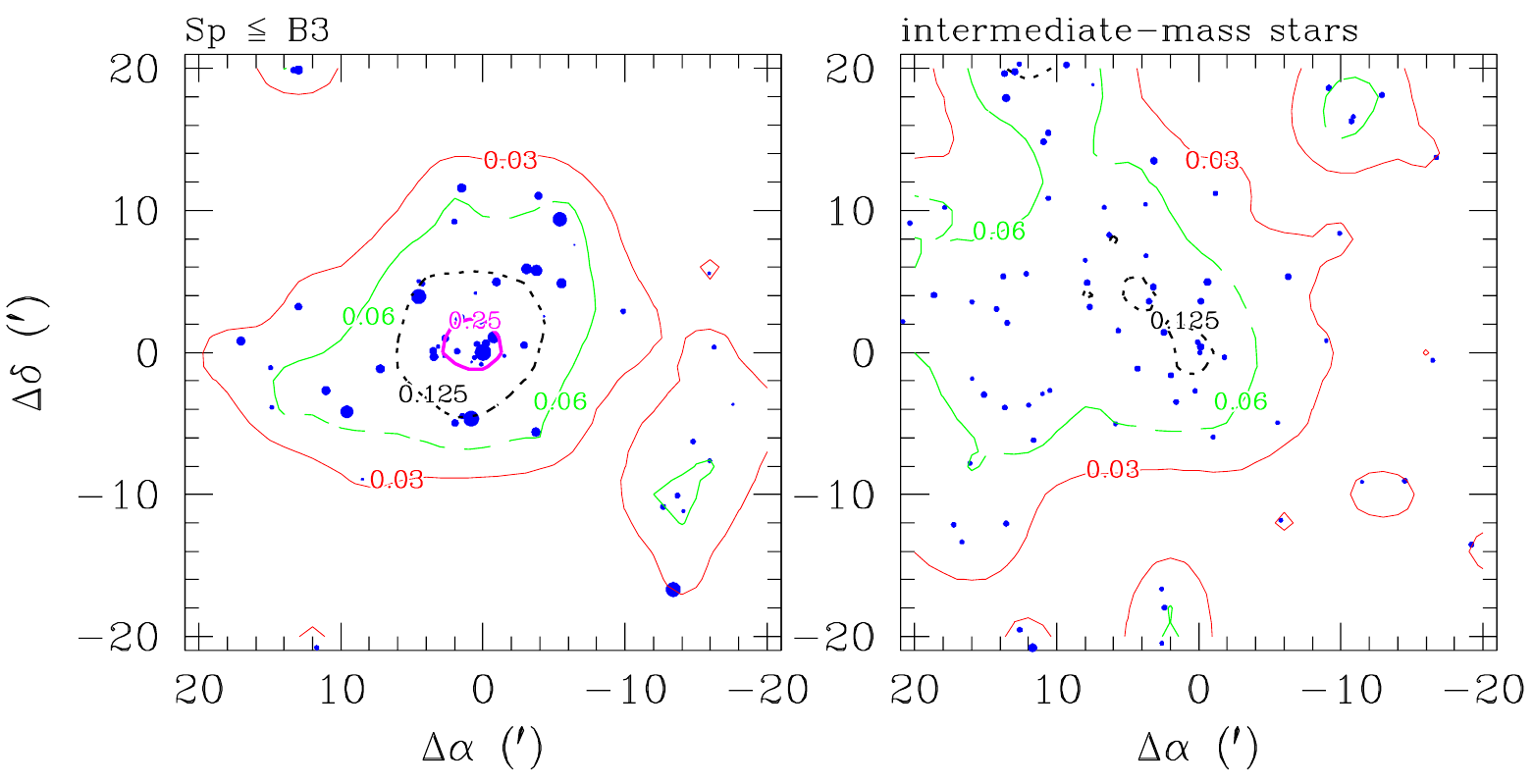}
\caption{Spatial distribution of massive members (left) and intermediate-mass
members (right). Blue dots represent the selected members of IC 1805. 
The size of dots is proportional to the brightness of the star, and the number 
on the contour denotes the surface density in units of [star arcmin$^2$].
\label{massive}}
\end{figure*}

O- and early B-type (Sp $\lesssim$ 
B5) stars are generally found in young stellar systems. Such massive members of 
young clusters or OB associations can be easily selected from the ($U-B, ~ B-V$) TCD.
The number of O- and early B-type
stars in the CFH12K FOV is 8 and 71, respectively. These stars are all considered to be massive members
of IC 1805. However optical photometry alone cannot discriminate members of IC 1805 
from those of the Cas OB6 association.

Selection of late B- to F-type members of young open clusters is very difficult. The disks around these
stars are relatively short-lived \citep{ssb09}, therefore H$\alpha$ or MIR
photometry is useless except for Herbig Ae/Be stars. In addition, late B- to F-type stars are
quiet in X-rays because they have no surface convective zone, nor any strong stellar
wind. Only a fraction of them are detected from X-ray observations, and in these rare cases,
the X-ray emission is considered to originate from a
low-mass companion that is in the T Tauri stage \citep{dms16}. Spectral classification
may be the only reliable membership criterion for these stars in young open
clusters. Unfortunately the spectral types of only a limited number of stars 
in IC 1805 are known. In view of such limitations, we had to select most of
the intermediate-mass members ($V \leq 14.5$ mag) of IC 1805 using photometric
data alone. 

First we selected probable member candidates in the ($J, ~Q'$) and 
($J, ~Q_{VJ}-Q_{VK_s}$) diagrams as shown in Figure \ref{dmq2}. The reddening-
independent indices $Q_{V\lambda}$ were defined in \citet{sos}, and one of them
($Q_{VI}$) was originally introduced in \citet{sb04} to determine the distance
to the starburst type young massive cluster NGC 3603. The object at ($J, ~Q'$)
$\approx$ (8.58, -1.23) is a very red object BIRS 119 \citep{e80} (=C15111
=M2k0678). The abnormal $Q'$ value is due to its extreme ($B-V$).
Using the two selection criteria mentioned above,
157 stars including O- and early B-type stars in the whole observed field
were selected. We then checked their position in all
available TCDs and CMDs in optical and NIR pass bands including the
reddening-corrected CMDs [($V, ~V-I$), ($V,~B-V$), ($V,~U-B$), ($B-V,~V-I$),
($U-B,~B-V$), ($V, ~V-J$), ($V, ~V-H$), ($V, ~V-K_s$), ($V_0,~(V-I)_0$),
($V_0,~(B-V)_0$), ($V_0,~(U-B)_0$), ($Q_{VI},~Q'$), ($Q_{VJ},~Q'$),
($Q_{VH},~Q'$), and ($Q_{VK_s},~Q'$)]. We also checked the reddening of
each star estimated from the ($U-B,~B-V$) TCD, with that from  the reddening map (see
section \ref{reddening}), and that from its spectral type if its spectral type were known.
As we did not observe the whole CFH12K FOV in $UBV$, we had to use 
previous investigators' photoelectric and CCD photometric $UBV$ data which
were therefore inevitably inhomogeneous, and had large errors
for fainter stars ($V \gtrsim 14$ mag). The quality of ($U-B$) is the most
critical factor in the selection of intermediate-mass members. Using this
procedure we selected 50 intermediate-mass members from photometric data alone.
In addition, amongst the stars plotted between the two lines in Figure \ref{dmq2},
six X-ray emission stars, an H$\alpha$ emission star, and an X-ray emission
star with H$\alpha$ emission  were also classified
as intermediate-mass members of IC 1805. 

The spatial distribution of massive members
and intermediate-mass members selected in this section is shown in Figure
\ref{massive}. Most massive members are concentrated at the center. Several
members are between the bright central nebula and the faint north-south
nebula along the western edge of W4. And a few are at the edge of BRC 5.
These stars are considered as the second generation
stars triggered by the strong radiation field from the massive O-type
stars in the IC 1805 center \citep{osp02,pcp14}. 
However the distribution of intermediate-mass stars is more distributed and extended 
to the northeast connecting the central cluster and BRC 5. But we could not 
find any enhancement of intermediate-mass stars in the southwest. The distribution  is very similar
to that of H$\alpha$ emission stars in Figure \ref{YSOdist}.

\section{Reddening and distance}

\subsection{Two-Color Diagrams and Reddening \label{reddening}}
 
\begin{figure*}
\gridline{\fig{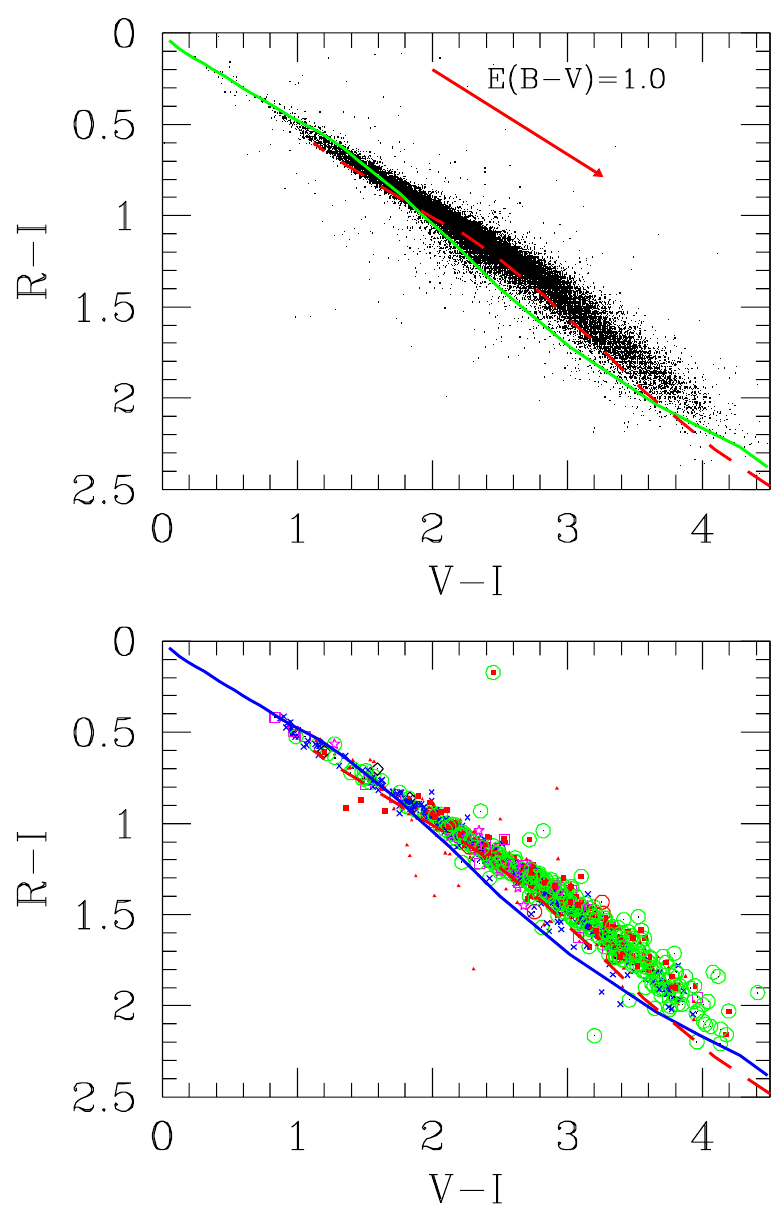}{0.30\textwidth}{(a)}
          \fig{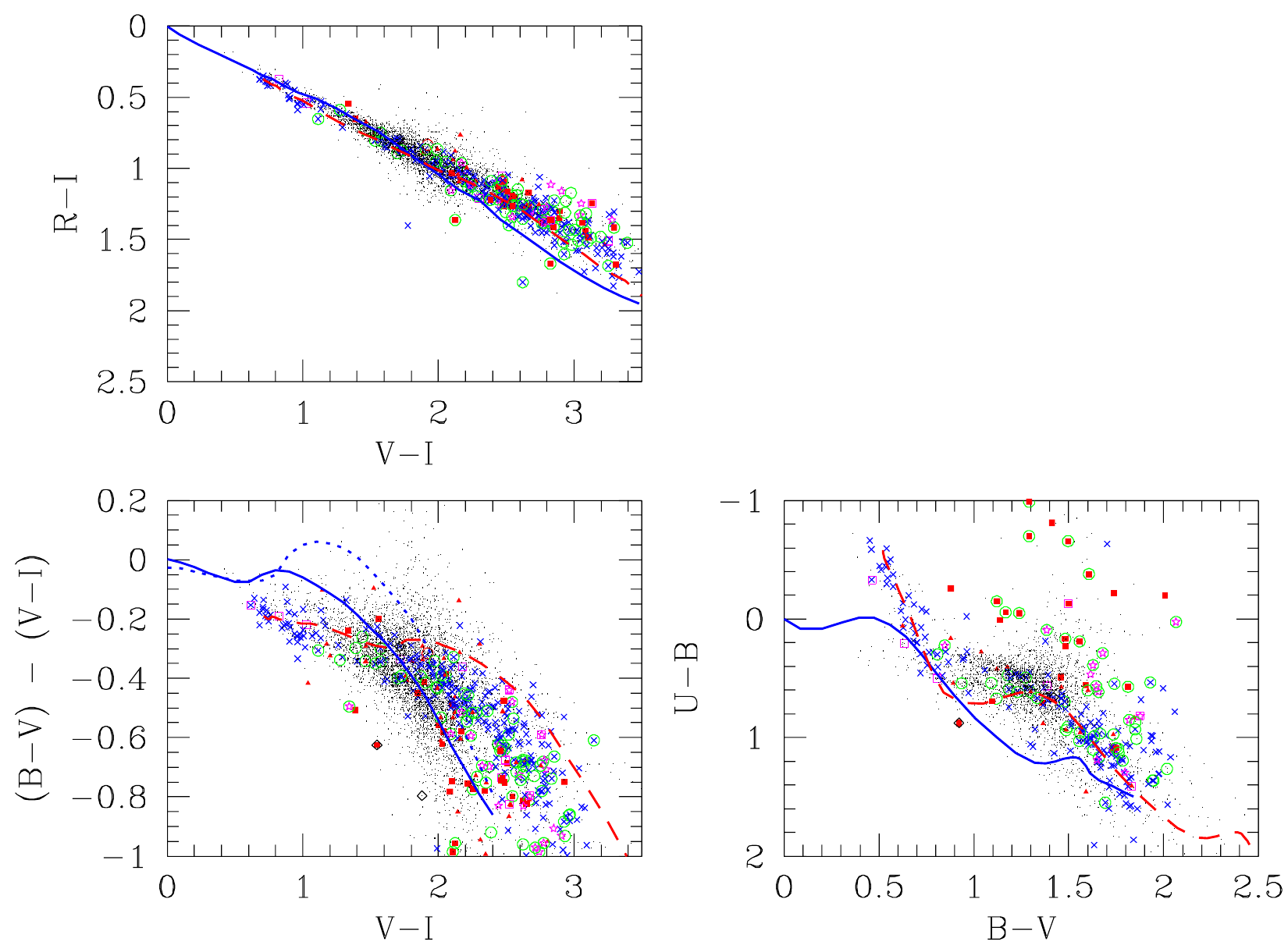}{0.60\textwidth}{(b)} }
\caption{Two-color diagrams. (a) the ($R-I,~V-I$) TCDs from CFH12K observations.
The upper and lower panels show the TCD of all stars and that of stars with
PMS membership. (b) three TCDs from Maidanak 2k and 4k observations. The blue
solid and red dashed lines represent, respectively, the intrinsic and reddened
color-color relations of MS stars. The median reddening of the early-type members
of IC 1805 $E(B-V)$ = 0.85 mag is applied in the diagrams. The dotted line in the lower
center is the intrinsic color-color relation of giant stars. The other symbols
are the same as in Figure \ref{opt_cmd}. \label{figccd}}
\end{figure*}

The TCDs of stars in the observed region are shown
in Figure \ref{figccd}. As the reddening vector in the ($R-I,~V-I$) TCD is
very similar to the intrinsic color-color relation of MS stars, the field stars
in the foreground or in the Perseus spiral arm and cluster stars show a similar
distribution in the diagram, and so the diagram cannot be used for any membership
criterion. The situation is slightly improved in the ($B-V, ~ V-I$) TCD, but
the loci of cluster PMS stars and that of field MS or giant stars largely
overlap each other.

However, the ($U-B,~ B-V$) diagram is the basic diagram for estimating the
reddening of early-type stars without ambiguity, at least for the stars
earlier than B5. The reddening $E(B-V)$ of 87 early-type stars in IC 1805 
and 4 early type stars outside the CFH12K FOV is
determined from the ($U-B,~ B-V$) diagram. The range of $E(B-V)$ is
between 0.72 and 1.23 mag, and the mean value of $E(B-V)$ is 0.88 ($\pm$ 0.10)
mag (median value is 0.85 mag), which are very similar to that obtained by
\citet{gv89,mjd95}. Although \citet{sl95} applied a slightly different reddening law,
they obtained a similar range and mean value. However, \citet{js83,hgb06} derived
a somewhat smaller range and mean value. The latter authors derived the reddening
from SED fitting to the O-type stars in the Cas OB6 association. Their $E(B-V)$ is
mostly consistent with ours except for the O7Vz star BD +60 513 and the most 
evolved massive star HD 15570. While the $E(B-V)$ for the former is smaller than ours,
that of the latter is larger.

\begin{figure}
\epsscale{0.7}
\gridline{\fig{IC1805IRAC_ebv_small.pdf}{0.53\textwidth}{(a)}
          \fig{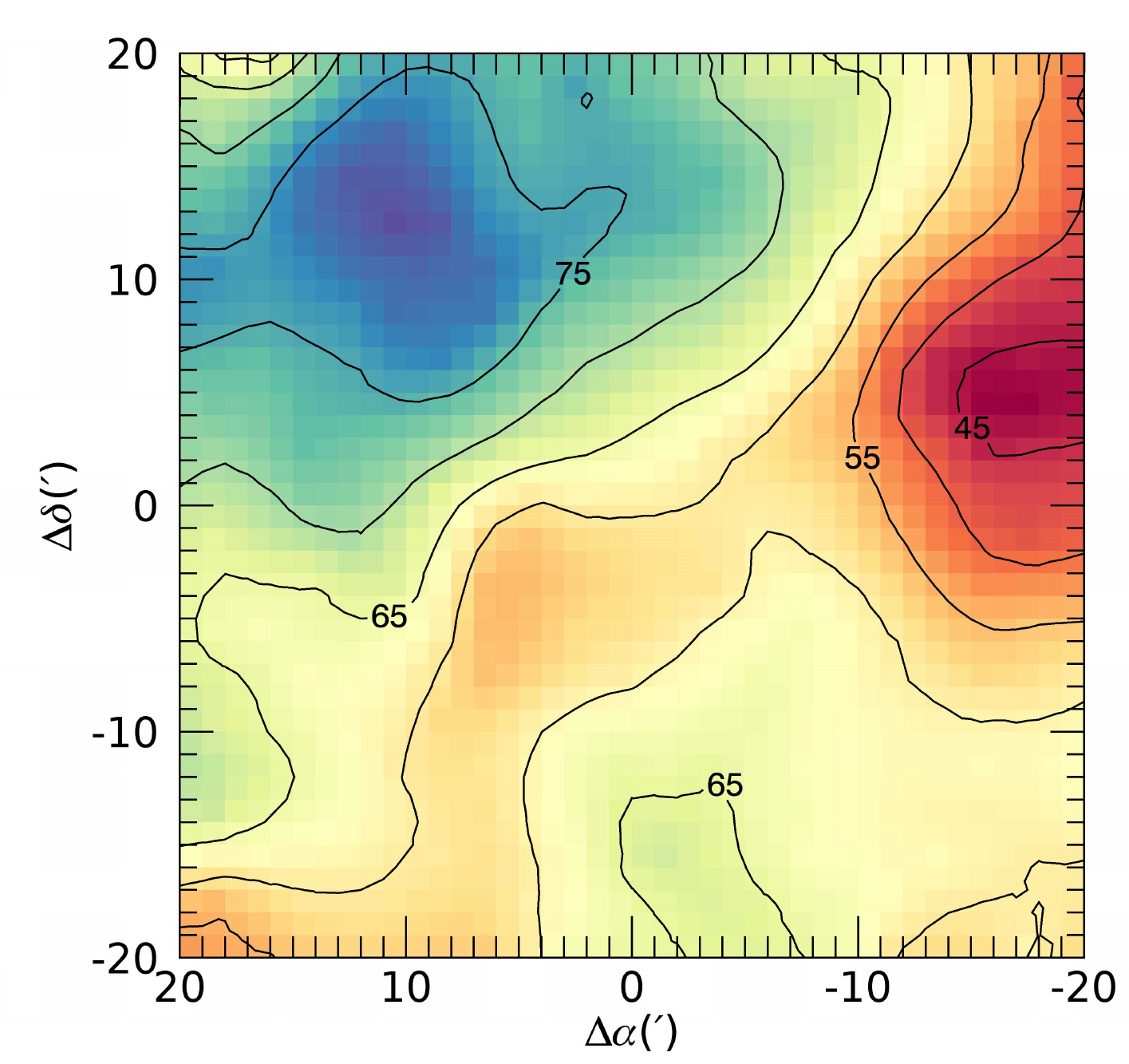}{0.47\textwidth}{(b)} }
\caption{(a) The reddening map of IC 1805 superimposed on the color-composite
MIR image (color encoding - red: 8.0 $\mu m$, green: 4.5 $\mu m$,
blue: 3.6 $\mu m$). The lines represent 
the iso-reddening contours smoothed with the scale length of $1.'5$.
The line type and thickness represent different amounts of reddening $E(B-V)$
as shown in the figure. The circles indicate the early-type stars used in the
reddening determination. The size of the circles is proportional to the brightness
of the stars. The color of the dots is related to the membership of the star -
red: H$\alpha$ emission star, blue: X-ray emission star, and yellow: normal 
early-type star. (b) Surface density of field stars ($I$ = 17 -- 20.5 mag \& below 
the PMS locus). The numbers on the contour denote
the surface density in units of [star arcmin$^{-2}$]. \label{ebvmap} }
\end{figure}

The spatial variation of reddening, i.e. the reddening map, is derived from 
the 91 early-type stars in and around the observed FOV, and is shown in
Figure \ref{ebvmap} (a), which is superimposed on the color-composite
MIR image. The reddening is in general larger in the west (close to
the active SFR W3), which implies that the PAH emission nebula in the west
is in front of IC 1805. However there seems to be no close correlation
between the variation in the reddening and the emission nebula at the center.
This fact implies that the bright emission nebula
is illuminated by the strong UV radiation from hot massive stars at the cluster 
center, but is probably at the immediate background of the main cluster.
The reddening in the south-east of
IC 1805 is slightly larger, implying that the nebula is partly associated with
the cluster stars in this region. The smallest reddening occurs at ($\Delta
\alpha \approx 6',~\Delta \delta \approx 10'$), where the PAH and 
CO emission \citep{chs00} is relatively absent. Figure \ref{ebvmap} (b) shows
the surface density of field MS stars below the PMS locus with $I$ = 17 -- 20.5 mag.
The surface density map also supports the radial structure of this region - 
the field stars are densely populated in the region where PAH emission is
absent. The density is lowest in the western region. These facts also support
the clouds associated with W3 being in front of IC 1805. The PAH emitting nebula 
just behind the central cluster effectively blocks the light from the background, 
therefore the surface density of field stars is relatively low along the nebula that 
extends from northwest to southeast. If the surface density of field
stars along the line-of-sight is homogeneous in the observed FOV, about two thirds of them are in front of
the Perseus arm, about 15\% are between the PAH nebula in the west 
(probably at the same distance as W3) and IC 1805, 
and about 25\% are in the background of IC 1805. The reddening map
will be used to estimate the reddening of the low-mass PMS stars in IC 1805.

Although the ($U-B$) values of the faint stars were not very good because of the smaller aperture
of the telescope used, there are non-negligible numbers
of UV-bright stars in the ($U-B,~ B-V$) diagram. This implies
that many low-mass PMS stars in IC 1805 are still actively accreting.
A similar situation can be found in other young open clusters in the Perseus
spiral arm \citep{lsk14a,lsk14b}.

\subsection{Reddening Law \label{reddening_law}}

\begin{figure*}
\epsscale{1.0}
\plotone{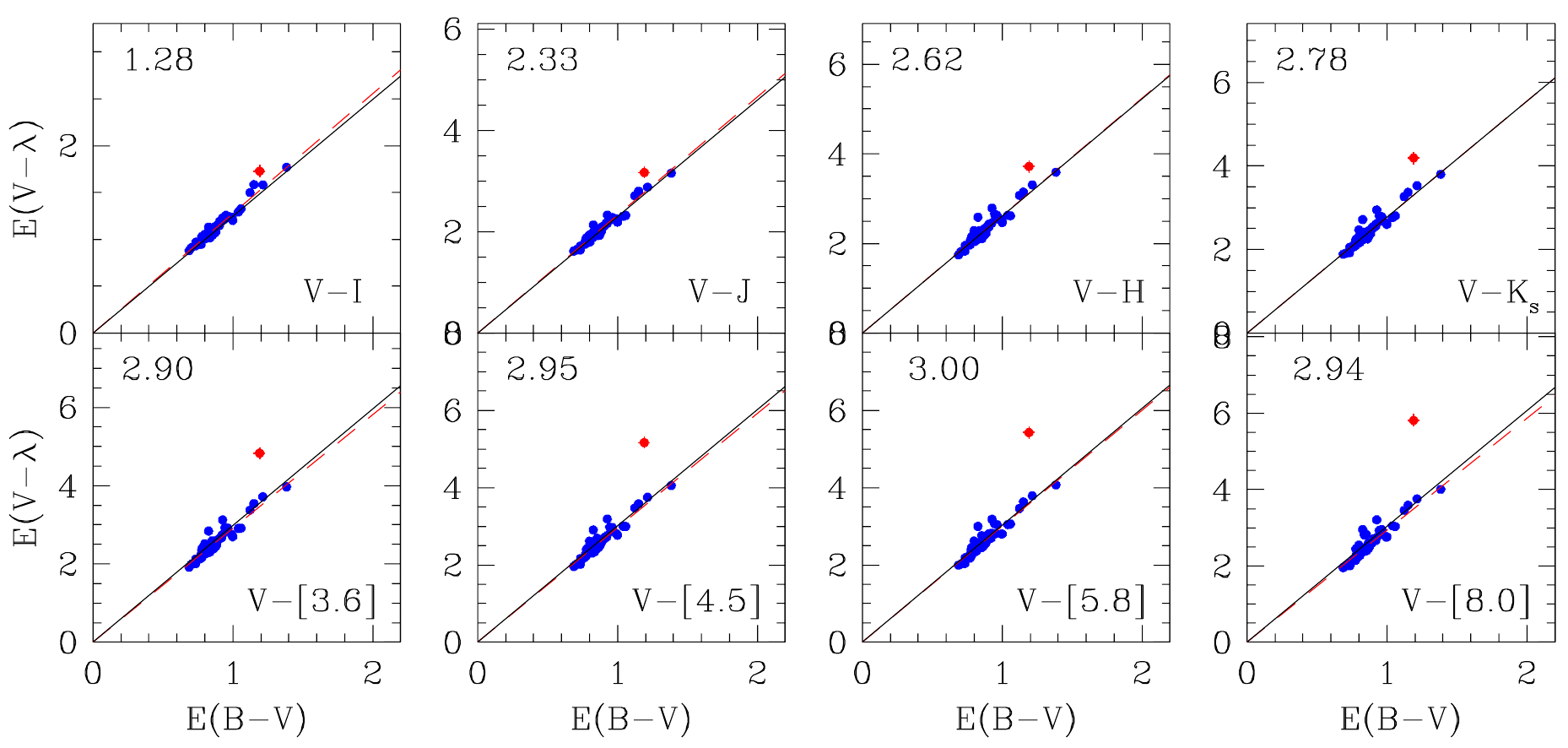}
\caption{The reddening law of IC 1805. The solid line is the $ E(V - \lambda) /
E(B - V) $ ratio for $R_V = 3.05$, while the red dashed line represents the mean color
excess ratio from normal early type stars in the CFH12K FOV. The number in each 
panel denotes the mean value of the color excess ratio for the given color. 
The H$\alpha$ emission star VSA 113 (O9.5Ve  or Be - red diamond) shows an abnormal 
value due to the excess emission from a circumstellar disk.
\label{redlaw} }
\end{figure*}

The interstellar reddening law is one of the fundamental parameters involved
in determining the distance to astronomical objects, and is known to be
different from one line-of-sight to another in the Galaxy \citep{fm09,sb14}.
In addition, \citet{fm09} acknowledged that there is no universal NIR
extinction law. \citet{f78,gv89} presented a method to determine the 
total-to-selective extinction ratio $R_V$ using color excess ratios of optical
and NIR colors, and \citet{ssb13} extended this relation to the MIR
{\it Spitzer} colors. These relations have been successfully used to determine
the $R_V$ of several young open clusters. The color excess ratios of several
young open clusters have been well fitted to a single line with a normal $R_V$ 
\citep{ksb10,lsk11,ssb13,lsk14a,lsk14b,lsb15,lsh15}. However the color excess
ratios of some extremely young open clusters are best fitted by a combination of two lines
with different slopes, which means that two different media with different
extinction properties exist in the line-of-sight, i.e. an abnormal reddening
law for the intracluster medium with a normal $R_V$ for the foreground medium
[e.g. NGC 1931 \citep{lsb15}, Westerlund 2 \citep{hps15}, or Tr 14 and Tr 16
\citep{hsb12}].

Figure \ref{redlaw} shows the color excess diagrams for IC 1805 that we used
to determine the total-to-selective extinction ratio $R_V$.
We excluded the H$\alpha$ emission star VSA 113 (O9.5Ve or Be) from the fits as its 
colors were affected by emission from its circumstellar disk.
The color excess ratios are all well fitted to a single line, which implies that 
(1) the dust size distribution of the foreground medium and intracluster medium are very similar and 
(2) a fairly normal $R_V$ in the direction of IC 1805 is obtained from the
64 O and early B-type stars (Sp $\leq$ B4V), $R_V = 3.052 \pm 0.058$.
From optical and NIR photometry and polarimetry,
\citet{gv89} arrived at the same conclusion for the properties of the
dust in IC 1805 and in the foreground ($R_V = 3.1 \pm 0.1$). 
Recently, \citet{mmb07} deduced at the same dust properties from
CCD polarimetry of IC 1805. \citet{hc93} also had arrived at the same conclusion
from extinction curve fitting from NIR to UV wavelengths, but
obtained a slightly smaller $R_V$ of about 2.9.

Previous $R_V$ determinations for the cluster fall into two groups. One group obtained a nearly
normal $R_V$ as in the current work. \citet{hgb06} obtained $R_V$ = 2.94 -- 3.13 from SED fitting; 
\citet{kl83} obtained $R_V = 3.06 \pm 0.06$ for the central region; \citet{sl95},
$R_V = 2.9$ from the spectral type versus $M_V$ (Sp - M$_V$) relation.
While another group of authors obtained somewhat larger values. \citet{j68},
$R_V = 5.7$ for the Cas OB6 region from the Sp - M$_V$ relation;
\citet{kl83}, $R_V = 3.82 \pm 0.5$ for the peripheral region; \citet{i69},
$R_V = 3.8 \pm 0.5$ using radio and H$\alpha$ emission measures;
\citet{pun03}, $R_V = 3.56 \pm 0.29$ from various color excess diagrams, color-color relations, 
TCDs and CMDs.

\subsection{The Distance of IC 1805 and Color-Magnitude Diagrams \label{dist_cmd} }
  
\begin{figure*}
\epsscale{1.0}
\plotone{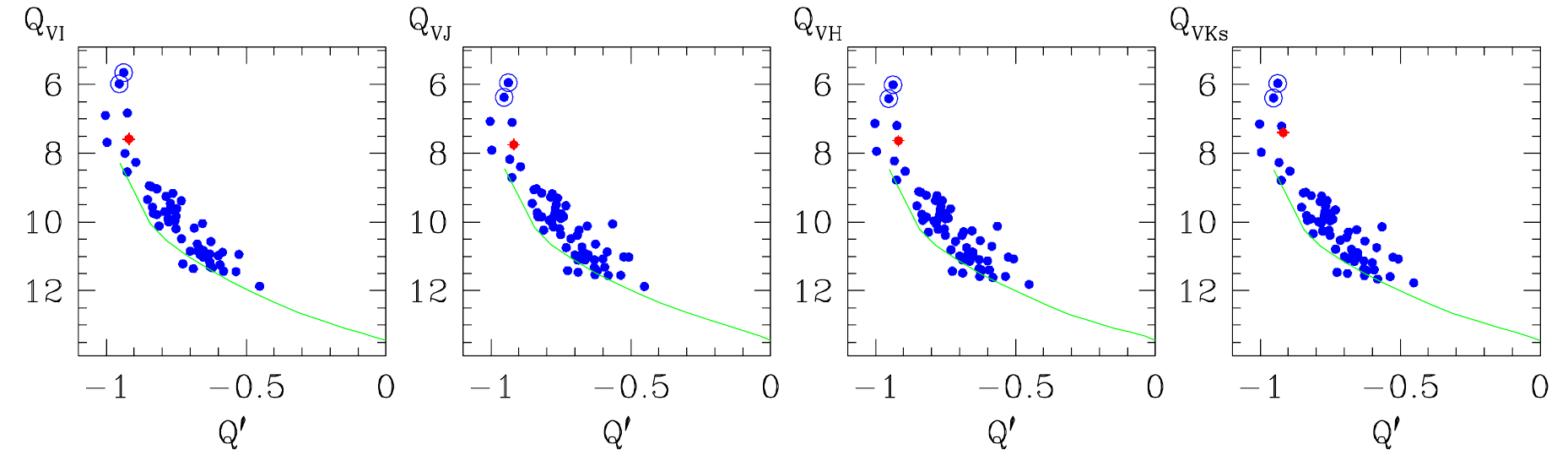}
\caption{The reddening-independent index $Q_{V \lambda}$ versus a modified
Johnson Q ($Q'$) diagram of IC 1805. Dots represent early type stars used
in Figure \ref{redlaw}. Dots with an open circle denote two evolved stars,
HDS 15558 and HD 15570. The thick solid lines in each panel represents the ZAMS
line at a distance modulus of 11.9 mag. \label{dm_q} }
\end{figure*}

As O stars in IC 1805 are used for the calibration of the Sp-M$_V$
relation \citep{ca71}, the distance to the cluster is very important.
However, because the distance to an astronomical object is strongly dependent
on the adopted 
$R_V$ and the adopted or derived $R_V$ in the direction of IC1805 varied from 2.9 to 5.7, the derived
distance to IC 1805 ranged from 0.76 kpc \citep{j68} to 2.4 kpc
\citep{kl83,sl95}. The distance of IC 1805 from most photometric studies have been based
on ZAMS fitting or the Sp-M$_V$ relation, and converges around 2.3 -- 2.4 kpc
\citep{kl83,js83,mjd95,sl95}. In support, \citet{gs92} derived the distance
of the surrounding Cas OB6 association to be 2.4 kpc. In the absence of more recent determinations 
investigators have therefore assumed or adopted the distance determined by \citet{mjd95}.
But a recent challenge to the distance of IC 1805 has emerged from radio
astrometry of H$_2$O or methanol masers in massive star forming regions.
Very Long Baseline Interferometry (VLBI) astrometry of a methanol maser \citep{xrz06}
or H$_2$O maser \citep{hbm06} in the nearby HII region W3 has given a consistent distance of 2.0 $\pm$ 0.05
kpc for W3(OH). These astrometric results will be discussed further in section 6.1.

\begin{figure*}
\epsscale{0.95}
\plotone{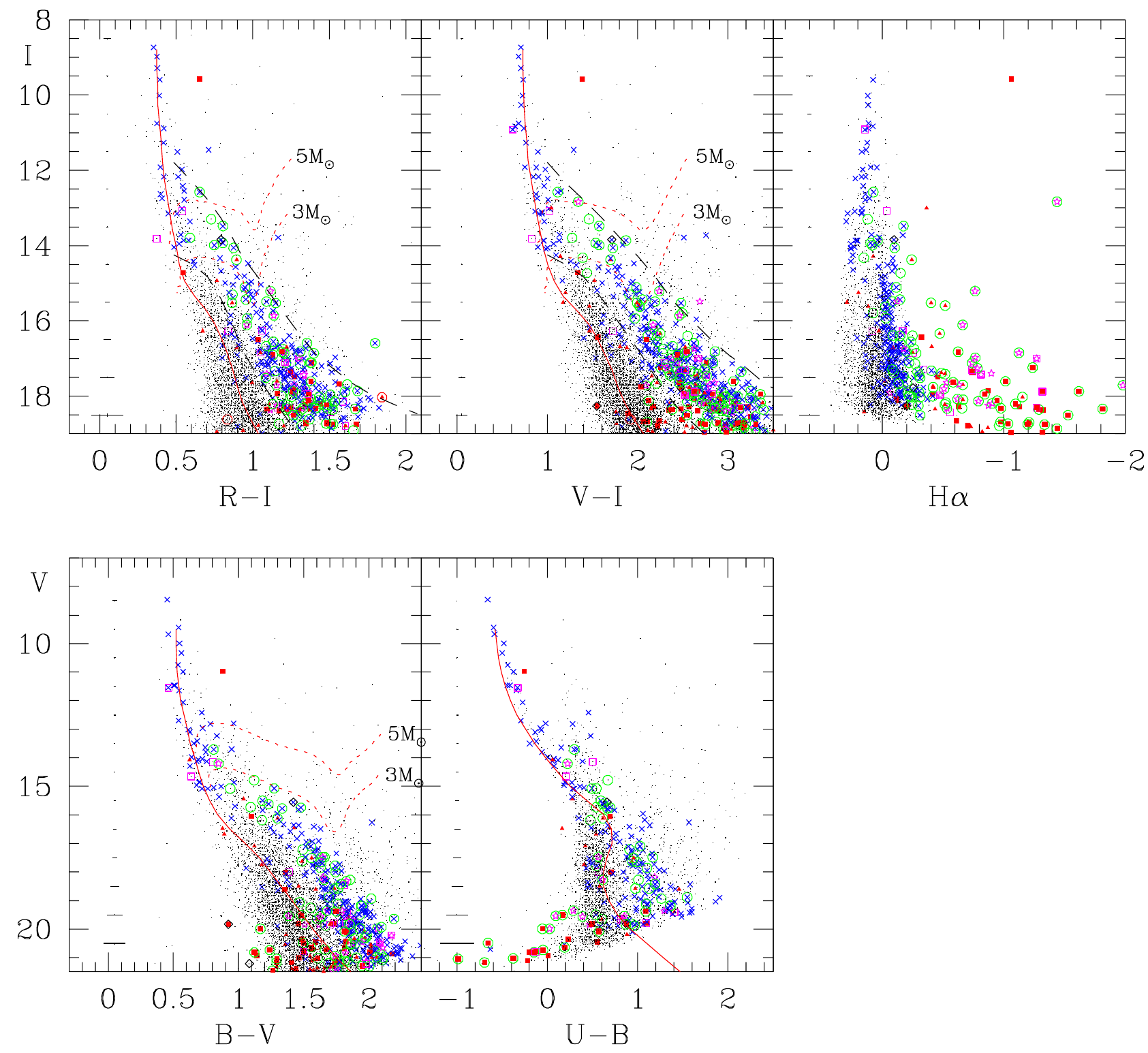}
\caption{Color-magnitude diagram of IC 1805 based on the data obtained with
the AZT-22 1.5m telescope at Maidanak Astronomical Observatory. The solid line in each
panel represents the reddened ZAMS with $E(B-V)$ = 0.85 and $V_0 - M_V =$ 11.9
mag, while the two dashed lines denote the upper and lower boundary of the PMS locus
of IC 1805. Mean photometric errors are shown in the left of each panel.
The other symbols are the same as in Figure \ref{opt_cmd}. The two dotted
lines in the CMDs are the PMS evolutionary tracks of 3 M$_\odot$ and 5 M$_\odot$ stars,
respectively, from \citet{sdf00}.
\label{mao_cmd} }
\end{figure*}

We have independently derived the distance to IC 1805 using a modified ZAMS fitting technique.
The CMDs of the reddening-free index $Q_{V\lambda}$ and a modified Johnson
$Q$ ($Q'$) were used as shown in Figure \ref{dm_q}. The definition of these
indices is presented in \citet{sos} (repeated in \citealt{lsb15}). When we
fit to the ZAMS, we should take the lower ridge line of the MS band to avoid
the effects of evolution during the MS stage, contamination by
systems of multiple stars or chemically peculiar stars, and/or scatter due
to photometric errors. Our derived distance modulus of IC 1805 is 
11.9 ($\pm$ 0.2) mag (equivalently
$d$ = 2.4 ($\pm$ 0.2) kpc), which is well consistent with previous 
determinations, but about 400 pc more distant than the nearby SFR W3(OH).
The error quoted here is an assumed error.\footnote{We have selected
proper motion and parallactic members of 34 nearby open clusters using
the {\it Gaia} DR1 TGAS data to check the reliability of the ZAMS relation,
and found that the error in distance modulus
is 0.21 $\pm$ 0.10 mag due to the scatter of individual parallaxes among members.}
In addition, ZAMS fitting is affected by the photometric errors.
However, as can be seen in Figure \ref{dm_q} the ZAMS well describes
the lower part of the cluster stars in all four reddening-free CMDs.

The CMDs of IC 1805 are presented in Figure \ref{opt_cmd} and \ref{mao_cmd}.
The ZAMS with the median reddening and the adopted distance of IC 1805 is over-plotted.
From the CMDs in Figure \ref{opt_cmd} we can barely detect the existence
of cluster stars, but in Figure \ref{mao_cmd} we can easily recognize
the well-developed sequence of early-type members to the left of each CMD.
The reddened ZAMS well follows the early-type MS stars in each CMD.
The locus of low-mass PMS stars in IC 1805 is marked in two CMDs whose color
is less affected by the UV excess due to mass accretion activities.
In the ($V,~ U-B$) CMD, early type members
are clearly separated from field stars which are distributed vertically at
($U-B$) $\approx$ 0.3 -- 0.6 mag.
However, the separation between cluster stars and field stars is not conspicuous
in most CMDs in Figure \ref{mao_cmd}. As mentioned in section \ref{mimm},
late B and A-type stars in IC 1805 (masses of PMS stars between 3 M$_\odot$
and 5 M$_\odot$
stars in Figure \ref{mao_cmd}) overlap with field stars in the CMD,
and cannot be reliably separated from field stars with any combination of
optical and/or IR colors. The ($I, ~$H$\alpha$) CMD which could be used as
an age indicator of young open clusters as claimed by \citet{dms16} is also
presented. Due to the small number of X-ray emitting B- and A-type stars,
the hooked feature at H$\alpha \simeq$ 0.2 is less pronounced. However the length
of the hooked feature is shorter than that of NGC 6231 and the feature is well separated
from the vertical distribution of low-mass PMS stars at H$\alpha \approx$ 0.0. 
These features indicate that IC 1805 is younger than NGC 6231 (age = 4.0 -- 7.0
Myr for massive stars).
All 8 O-type stars in IC 1805 are X-ray emitters, but the fraction of
X-ray emitters is about half for early B-type stars (Sp $\leq$ B4). 

\subsection{Radius of IC 1805}

\begin{figure*}
\epsscale{0.9}
\plotone{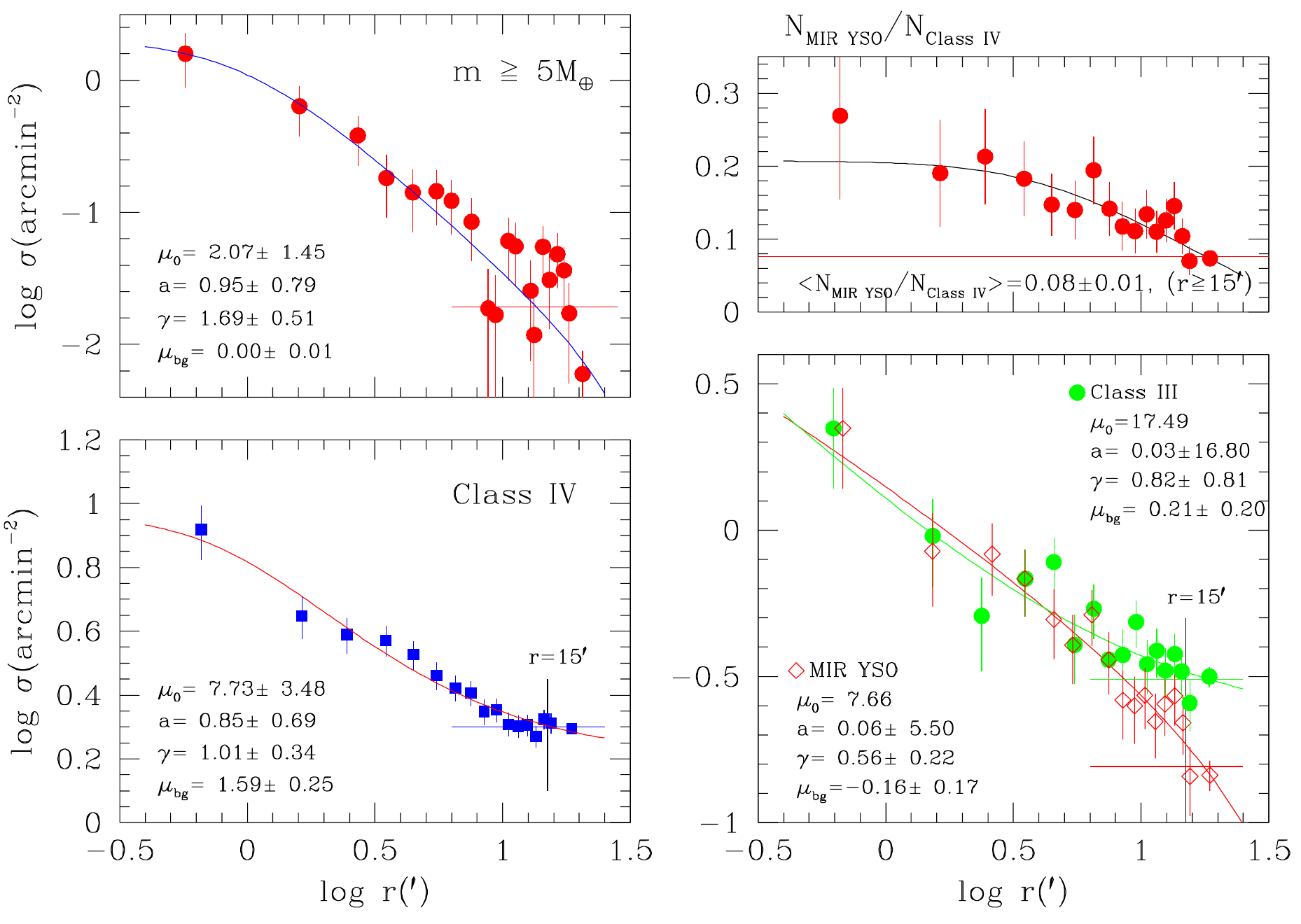}
\caption{The surface density profile of massive stars  ($m \geq 5 M_\odot$ - Upper
left panel), Class IV (Lower left panel), radial variation of the ratio between MIR YSOs and
Class IV objects (Upper right) and of MIR excess PMS members and Class III objects
(Lower right panel). The solid line represents the best-fit to the EFF model \citep{eff87},
and the fitting parameters are shown in each panel. The horizontal line in each panel represents
the average surface density of each object ($r \geq 12'$ for massive stars, and $r \geq 15'$ for
the others).  The error bars are derived by assuming Poisson statistics.
\label{figrho} }
\end{figure*}

\begin{deluxetable*}{c||cccc|ccc}
\tablecolumns{8}
\tabletypesize{\scriptsize}
\tablecaption{Fitting Parameters for the Surface Density Profile \label{profile}}
\tablewidth{0pt}
\tablehead{
\colhead{Object} & \colhead{$\mu_0$ [arcmin$^{-2}$]} & \colhead{a ($'$) } & 
\colhead{ $\gamma$ } &  \colhead{ $\mu_{bg}$ [arcmin$^{-2}$]} & 
\colhead{ $\rho_0$ [arcmin$^{-2}$]} & \colhead{r$_c$ ($'$)} & \colhead{ $\rho_{bg}$ [arcmin$^{-2}$]} }

\startdata
& \multicolumn{4}{c}{EFF model} & \multicolumn{3}{c}{King Model} \\
massive star & 2.069 $\pm$ 1.453 & 0.950 $\pm$ 0.787 & 1.688 $\pm$ 0.510 & -0.004 $\pm$ 0.010 &
1.686 $\pm$ 0.844 & 1.377 $\pm$ 0.434 & 0.001 $\pm$ 0.004 \\
MIR YSO & (7.658 $\pm$ 396.7) & (0.059 $\pm$ 5.497) & 0.561 $\pm$ 0.220 & -0.158 $\pm$ 0.172 &
0.930 $\pm$ 0.189 & 4.510 $\pm$ 0.943 & 0.101 $\pm$ 0.021 \\
Class III & (17.485 $\pm$ 7028) & (0.034 $\pm$ 16.80) & 0.825 $\pm$ 0.814 & 0.212 $\pm$ 0.198 &
0.618 $\pm$ 0.236 & 4.098 $\pm$ 1.726 & 0.282 $\pm$ 0.031 \\
Class IV & 7.733 $\pm$ 3.484 & 0.847 $\pm$ 0.689 & 1.011 $\pm$ 0.340 & 1.589 $\pm$ 0.246 &
4.014 $\pm$ 0.670 & 3.095 $\pm$ 0.508 & 1.838 $\pm$ 0.052 \\
N$_{MIR ~YSO}$ / N$_{Class~IV}$ & 1.342 $\pm$ 47.83 & 4.056 $\pm$ 10.98 & 0.068 $\pm$ 2.629 & -1.134 $\pm$ 47.81 &
0.172 $\pm$ 0.034 & 13.39 $\pm$ 5.235 & 0.016 $\pm$ 0.043 \\
\enddata
\end{deluxetable*}

The radius of a cluster is one of the important parameters in the study of 
cluster systems. However IC 1805 is a very sparse cluster with no strong
central concentration, and so it is not easy to define the radius of the
cluster. Although, as shown in Figure \ref{YSOdist}, \ref{Cl3n4}, and
\ref{massive}, massive stars are concentrated at the center, the spatial
distributions of H$\alpha$ emission, MIR excess stars, and intermediate-mass
stars are extended toward the northeast direction and show an abrupt
decrease to the southwest. Therefore, the radius
or spatial extent of IC 1805 is not well represented by the radial
distribution of one type of stars. Despite such a limitation, we tried to
determine the radius of IC 1805 from the radial distribution of
member stars. Before calculating the radial density profile of one type
of object, we should find the center of the cluster. The apparent center of IC 1805 
derived from the surface density distribution of massive stars in Figure \ref{massive} 
is around the brightest star HD 15558 (O4.5IIIe), however the spatial distribution 
of H$\alpha$ emission stars or MIR YSOs in Figure \ref{YSOdist} indicates
the center of these objects is to the immediate north of HD 15558 or somewhere
between HD 15558 and HD 15629 (O4.5V). As we did for the young open cluster
NGC 6231 \citep{ssb13}, we calculated the mass-weighted mean value of
($\Delta \alpha$, $\Delta \delta$) of stars with $m \geq 5 M_\odot$.
The resultant center is (-0$\farcm$057, +0$\farcm$867) north of HD 15558 ($\alpha_{J2000}
= 2^h ~32^m ~42.^s06$, $\delta_{J2000} = +61^\circ ~28' 2\farcs8$).

To determine the radius of IC 1805, we calculated the surface density profiles 
of the massive stars ($m \geq 5 M_\odot$), Class IV, Class III, and MIR excess PMS stars,
and these profiles are shown in Figure \ref{figrho}. The profile for massive stars
was calculated for the whole CFH12K FOV, but profiles for the others were calculated
for the SST/CM FOV because of the completeness of membership selection.
In order to estimate the radial extension of IC 1805, we fitted the profile to
the EFF model ($\mu (r) = \mu_0 [ 1 + (r/a)^2 ]^{-\gamma /2} + \mu_{bg}$,
\citealt{eff87}), and have given the fitting results in each panel of Figure \ref{figrho}
and Table \ref{profile}. The fitting was performed with the IDL routine MPFIT.
The fitting results relatively well represent the observed radial profile of massive stars
and Class IV stars, however those for Class III stars or MIR YSOs have
a large error due to an abrupt increase at the very center of IC 1805 ($r \lesssim
1'$). We also tried to fit the profiles with the King model ($\rho (r) = \rho_0
 / [ 1 + (r/r_c)^2 ] + \rho_{bg} $, \citealt{k62}), and presented the results
in Table \ref{profile}. In contrast to the EFF model, the King model does not well
describe the profile, especially near the center. The surface density of
MIR YSOs and the ratio between MIR YSOs and Class IV stars decrease
abruptly at $r \approx 15'$. The surface density
profiles as well as the fitting results to the EFF model show that the radius of
IC 1805 is about $15'$ (= 10.5 pc at d = 2.4 kpc). This value is about 1.7 times
larger than the radius obtained by \citet{psp17} who estimated the radius
from the radial density profile of their selected YSO members. 

The surface density of massive stars is high enough
to derive some information on their radial distribution. The core radius
$r_c$, which is defined as the radius where the surface density 
reaches half of the central value, is about $1.'07$ (equivalently
0.75 pc). This value is very similar to that of the massive stars
in NGC 6231 ($r_c = 0.88 \pm 0.02$ pc). However the full radius of
IC 1805 is about 1.75 times larger than that of NGC 6231 ($r \approx$ 6.0pc).

\section{AGE and THE INITIAL MASS FUNCTION OF IC 1805}

\begin{figure}
\epsscale{0.6}
\plotone{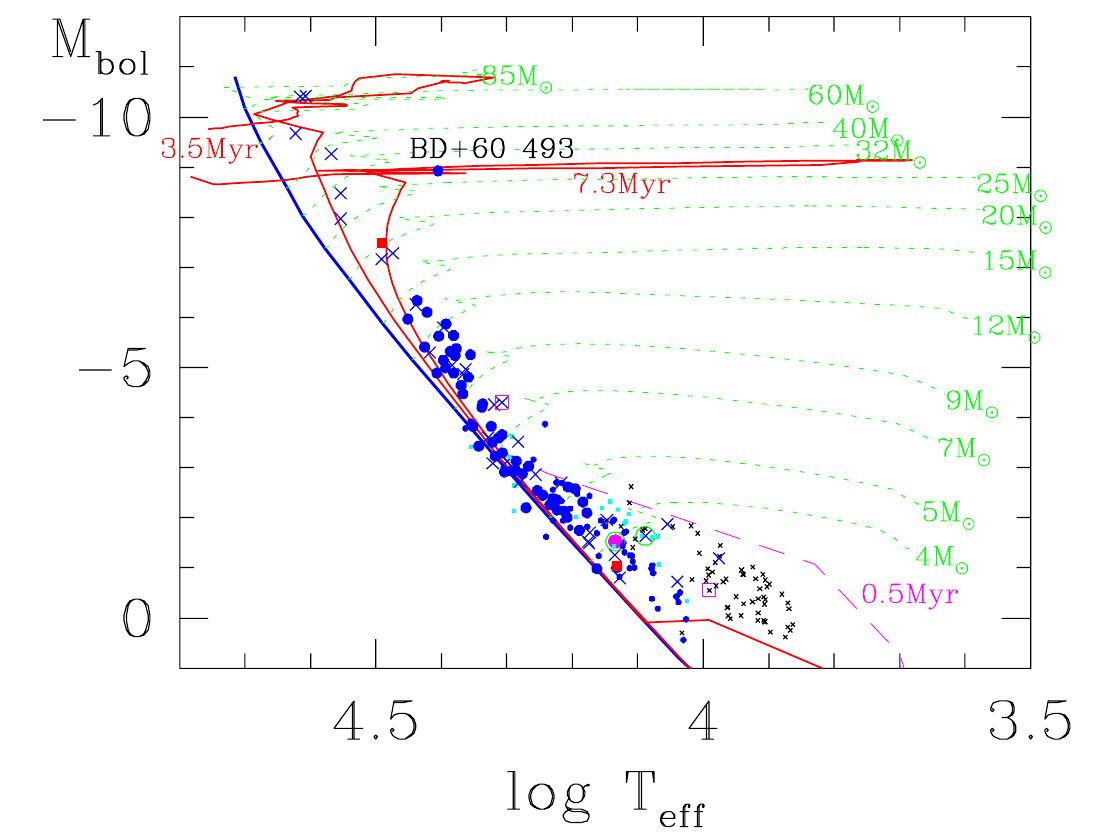}
\caption{The Hertzsprung-Russell diagram of bright stars in IC 1805. 
Large crosses, red squares, large magenta dots, large blue dots, small blue
dots, small cyan squares, and black crosses represent,
respectively, X-ray emission stars or candidates, H$\alpha$ emission stars or
candidates, H$\alpha$ emission stars with X-ray emission, early type stars,
early type candidates, stars with uncertain membership, and stars with no
membership information. Stars with YSO class I, F, II, t, T, P, and g
are superposed with an additional symbol as in Figure \ref{opt_cmd}.
The thick solid line is the ZAMS of \citet{geneva12}, while thin solid lines
are the isochrones for age 3.5 Myr and 7.3 Myr interpolated from the stellar
evolutionary tracks of \citet{geneva12} and the PMS evolution tracks of
SDF00. The thin dashed line in the lower right is the isochrone of age
0.5 Myr interpolated from the PMS evolution tracks of SDF00.
The dotted lines with mass to the right are the stellar evolution tracks
of \citet{geneva12}. \label{hrd_hm} }
\end{figure}

The mass and age of a star can be derived from the Hertzsprung-Russell diagram (HRD)
with the help of stellar evolution models and PMS evolution tracks.
To construct the HRD of a stellar system we have to employ various calibrations
in order to properly locate the stars in the HRD. The various calibrations required are summarized in
\citet{sos}. For massive O-type stars, the adopted spectral type is very important
for estimating the effective temperature and bolometric magnitude \citep{sos}.
Although minor differences in spectral type were mentioned in \citet{rn16},
we adopt the spectral types from \citet{smw11}.
Currently, two stellar evolution models of massive stars with 
stellar rotation are used in the mass and age estimate of massive stars,
and these are compared in \citet{ssb13}. For consistency with the mass and age scale of
massive stars with previous studies of our group \citep{hsb12,ssb13,lsk14a,
lsk14b,lsb15,hps15}, the age and mass of massive stars are determined
using the stellar evolution models of \citet{geneva12}.

For a long time the PMS evolution tracks of \citet{sdf00} (hereafter SDF00) were used in
the age and mass estimate of low-mass PMS stars. Recently \citet{bhac15} (hereafter BHAC15)
published new PMS evolution tracks for masses less than 1.4 M$_\odot$.
We compare the masses and ages of low-mass PMS stars from  these two PMS
evolution tracks.
 
\subsection{The Hertzsprung-Russell Diagram and Age of IC 1805 \label{hrd}}

We constructed the HRD of IC 1805 using the calibrations described above.
The HRD is shown in Figure \ref{hrd_hm} with several isochrones interpolated from
stellar and PMS star evolution tracks. The brightest stars in the cluster
are evolving away from the MS. The most evolved star HD 15570 (O4.5If+) is
considered to be at the transition stage between a normal Of star and a WN star
\citep{rn16}. The optically brightest star HD 15558 (O4.5III(f)) is an SB2
system with a primary that has possibly a very large minimum mass 
\citep{dbrm06} [see \citep{rn16} for more recent result].
The age of stars at the MS turn-on in the lower part of the HRD seems to be much
younger than that of the massive O-type stars, but as mentioned in the previous
sections, their membership is very uncertain. More discussion on the age
distribution of low-mass PMS stars will be dealt with in detail below.

One of main issues in studying IC 1805 is
the star formation history and its relation to the star formation activity
in the active SFR W3. \citet{gv89} noticed a large scatter of early-B type
stars in the reddening-corrected CMD, and interpreted it as an old population
of IC 1805 (age : about a few 10 Myr) prior to the formation of most massive stars in the cluster.
However, from the size of the HII region, \citet{dts97} estimated the age of the superbubble 
to be between 6.4 and 9.6 Myr. 

From Figure \ref{hrd_hm}, most O-type
stars in IC 1805 are well fitted to the isochrone of age 3.5 Myr.
There are two evolved early-type stars in the observed FOV - BD +60 493 (B0.5Ia 
- \citet{i70,sh99}) and BD +60 498 (O9.7II-III - \citet{smw11})\footnote{
Recently \citet{rn16} claimed that the luminosity class of the star is V rather
than II-III from the absorption of He II $\lambda$4686 and the ratio
of Si IV $\lambda$4088 to He I $\lambda$4143. If the luminosity class of the
star is MS, BD +60 498 is probably a member of IC 1805 rather than a member of
Cas OB6 association.}. These two stars can be thought of
as members of the Cas OB6 association scattered around the W3-W4-W5 region.
If we fit these two stars, the age of the best-fit isochrone is 7.3 Myr,
which is well matched to the expansion age of the superbubble \citep{dts97}.
Although we can see a large scatter of early B-type stars in the HRD as noticed
by \citet{gv89}, and if we assume the scatter as the result of stellar evolution,
we could find at least one or two evolved stars with a luminosity class of Iab.
But we cannot find any evolved counterpart of these early B type stars in or
around the observed FOV. \footnote{The G7Ib star VSA 199 (= M2k3581
= M4k4207) could be a possible
member of this age group, but the absolute magnitude of the star 
($M_V$ = -2.6) is somewhat fainter than that of luminosity class Ib.
The star is too faint and therefore too old to be an evolved counterpart
of these bright B type stars.} If their scatter is a result of the star formation
history in IC 1805, their spatial distribution or kinematic properties may
preserve some information of that. But we could not find any differences
between the two groups (see section \ref{propermotion} for details).

\begin{figure*}[b!]
\epsscale{0.9}
\plotone{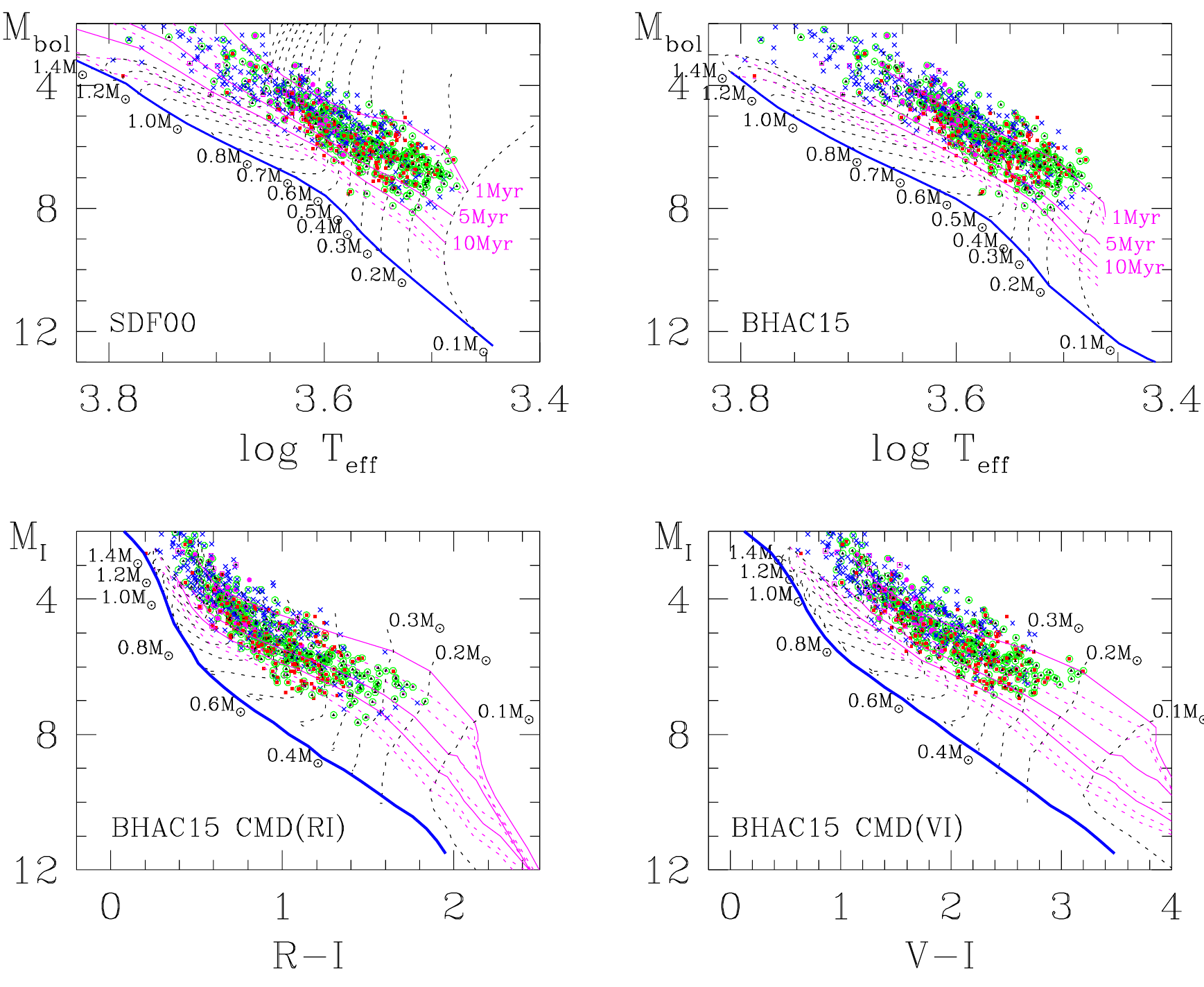}
\caption{Hertzsprung-Russell diagrams (Upper panels) and the ($M_I, ~ R-I$)
or ($M_I, ~V-I$) color-magnitude diagrams (Lower panels) of PMS stars in IC 1805.
The thick blue solid line in each diagram is the ZAMS relation.
The magenta solid lines represent the isochrones of age 1, 5, and 10 Myr interpolated from
the PMS evolution tracks, while the magenta dashed lines are the isochrones of age 2, 3, 7 (upper)
or 8 (lower),  15, and 20 Myr. The dotted lines with mass to the left or right
are the PMS evolution tracks for the mass. The other symbols are the same
as in Figure \ref{opt_cmd}. The upper-left panel is based
on the PMS evolution models by SDF00, and the other panels are based
on the recent PMS evolution models by BHAC15.
\label{hrd_pms} }
\end{figure*}

\begin{figure*}
\epsscale{0.85}
\plotone{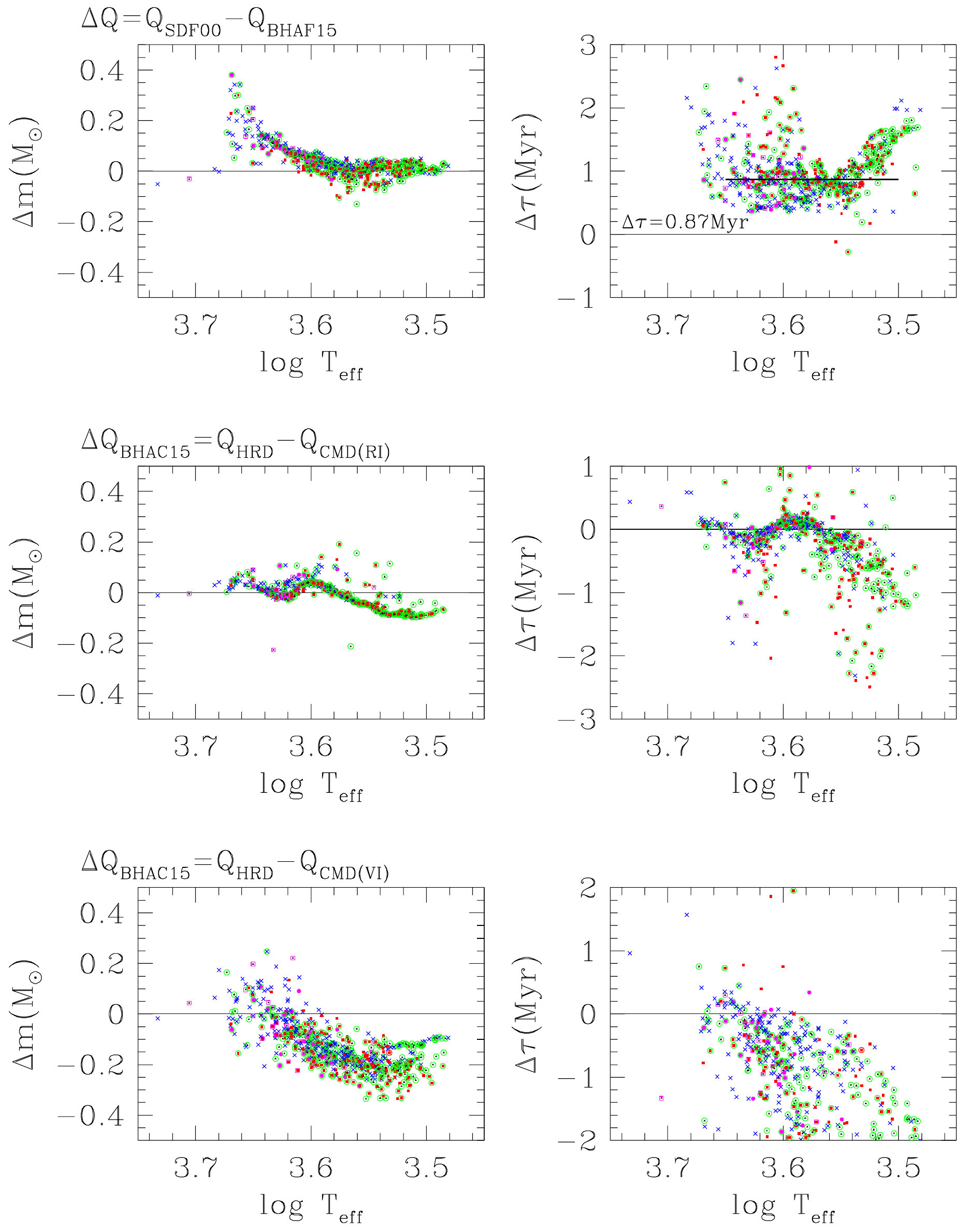}
\caption{Comparison of mass and age from each diagram in Figure \ref{hrd_pms}.
The meaning of $\Delta$ is explained in the top of the left panels. The upper panels
compare the mass (left) or age (right) from the HRDs in the upper panels of
Figure \ref{hrd_pms}. The middle and lower panels show the difference of
mass or age from the HRD (upper right panel of Figure \ref{hrd_pms}) and 
that from the CMDs (lower panels of Figure \ref{hrd_pms}) of BHAC15.
\label{pms_cmp} }
\end{figure*}

Low-mass PMS stars  in young open clusters give valuable information on
the star formation history of the clusters because the mass and age of
PMS stars can be determined from the PMS evolution tracks.
The HRD of low-mass PMS members is shown in
Figure \ref{hrd_pms}. In the figure we compared two PMS evolution models
- SDF00 in the upper left panel and BHAC15 in the other panels.
Most PMS stars in IC 1805 are well enclosed between
the two isochrones with ages 1 Myr and 5 Myr for SDF00. However, many
of them are brighter than the 1 Myr-isochrone of BHAC15, but their
distribution well follows the isochrone of age 1 Myr. In addition, BHAC15
published the absolute magnitudes in $VRIJHKLM$, and we showed 
the distribution of PMS stars in the CMDs
in the lower panels. The distribution of PMS stars in the ($M_I , ~V-I$) diagram
follows relatively well the isochrone of age 1 Myr, but that in the ($M_I ,~R-I$)
diagram does not well match the isochrone. This fact implies that $M_R$
magnitude of BHAC15 is not well matched to the $M_R$ magnitude
of real PMS stars. 

\begin{figure*}[t]
\epsscale{0.9}
\plotone{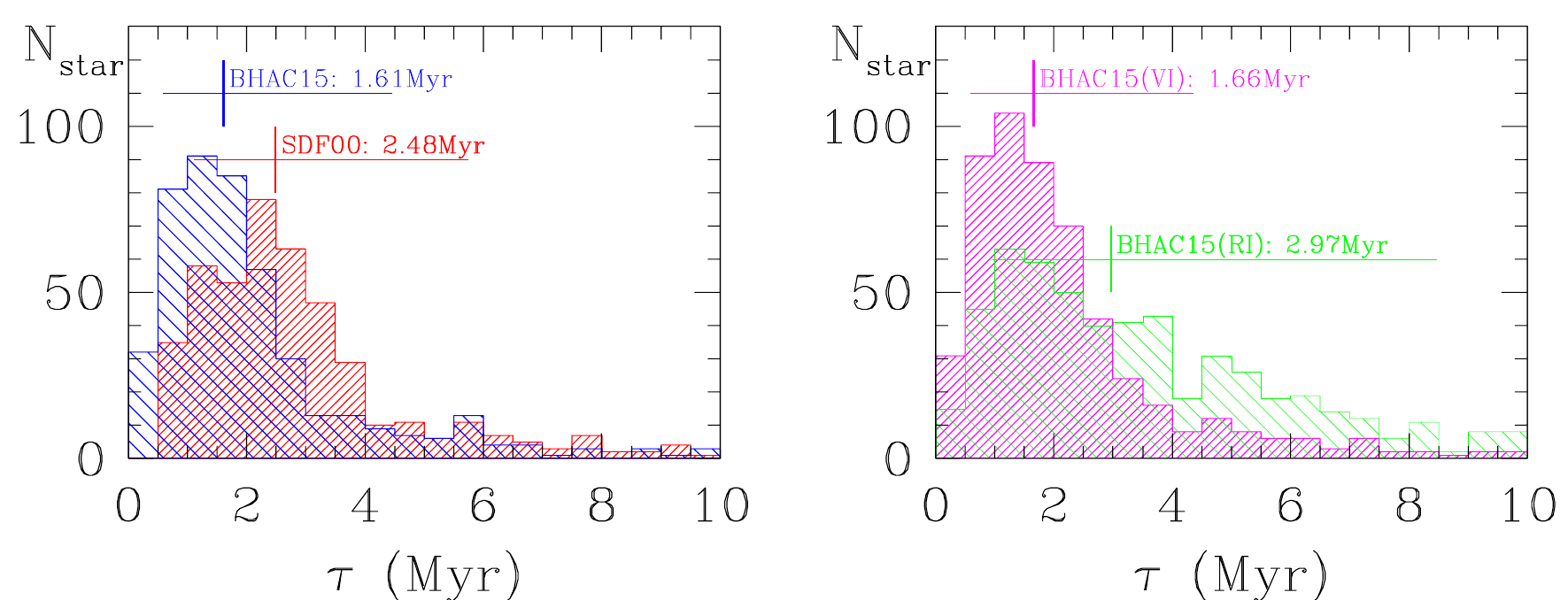}
\caption{Age distribution from PMS models. The vertical bar represents 
the median age of stars with masses bewteen 0.3 and 1.0 $M_\odot$, and
horizontal bar shows the range of 10 and 90 percentile of the distribution.
\label{pms_age} }
\end{figure*}

We estimated the mass and age of individual PMS stars by
interpolating the PMS evolution tracks, and compared the age and mass
estimated from each diagram in Figure \ref{hrd_pms}.  Figure \ref{pms_cmp}
shows the difference in mass and age from each diagram. As many PMS
evolution models show a mass-age relation (see \citealt{sbl97,sbc04} for details),
we compared the mass and age of 485 stars with $m$ = 0.3 -- 1.0 $M_\odot$
from SDF00. The masses from the two
PMS evolution models are well consistent with each other for $\log T_{eff} \lesssim
3.6$. The difference increases for hotter stars, but this may be due to the mass limit
of BHAC15 ($m \leq 1.4 M_\odot$). As expected from Figure \ref{hrd_pms},
the age of low-mass PMS stars from BHAC15 is systematically younger
than that from \citet{sdf00} by about 0.87 Myr. As the SDF00 isochrones of younger 
age ($\lesssim$ 5 Myr) do not follow well the distribution of low-mass PMS 
stars in the HRD, the difference increases for
low-mass stars. The middle and lower panels of Figure \ref{pms_cmp} compare
the mass and age from the HRD and two CMDs based on the PMS evolution
models of BHAC15, which show the internal consistency of mass and
age from various diagrams. Although there is some scatter, the mass and
age from the HRD and the ($M_I, ~V-I$) CMD are in general
consistent with each other. The small difference in mass implies that the temperature
scale of BHAC15 and that of \citet{sos} are well consistent with each other.
We checked the relation between temperature and $(V-I)$ using the same
stars as BHAC15, and found that the relation is well consistent 
within the observational errors. However, the differences are very large between
the physical parameters from the HRD and the ($M_I, ~R-I$) CMD. The difference
in mass is rather systematic, but the difference in age is very large and not
systematic. Therefore, although reliable masses and ages of PMS stars
can be obtained from the HRD or the ($M_I, ~V-I$) CMD,
it is advisable to not use the ($M_I, ~R-I$) CMD.

Figure \ref{pms_age} shows the distribution of age from each diagram. 
The median age from SDF00 is 2.48 Myr with 10 and 90 percentiles of
1.11 Myr and 5.75 Myr, respectively. The median age is about 1 Myr younger
than the age of the most massive stars in IC 1805. The age spread from the age
distribution of PMS stars is about 4.6 Myr according to the definition by
\citet{sb10}. This value is well consistent to the age spread of NGC 2264
obtained by \citet{sb10,lsk16}. The median age from the HRD and  the PMS
evolution model  by BHAC15 is 1.61 Myr with an age spread of about
3.9 Myr. The median age and age spread from the ($M_I,~V-I$) CMD is very
similar to those from the HRD. However those from
the ($M_I, ~ R-I$) CMD are far different from the others
- the median age and age spread are about 3.0 Myr and 7.5 Myr, respectively.

\citet{ssb13} obtained an age spread of about 3 Myr for massive stars and 6 Myr
spread for low-mass PMS stars in the massive young open cluster NGC 6231.
However, we could find no noticeable age spread among the massive stars
in IC 1805 ($\lesssim$ 1.5 Myr).

\subsection{The Initial Mass Function  \label{imf}}

\begin{figure}[b!]
\epsscale{0.9}
\plotone{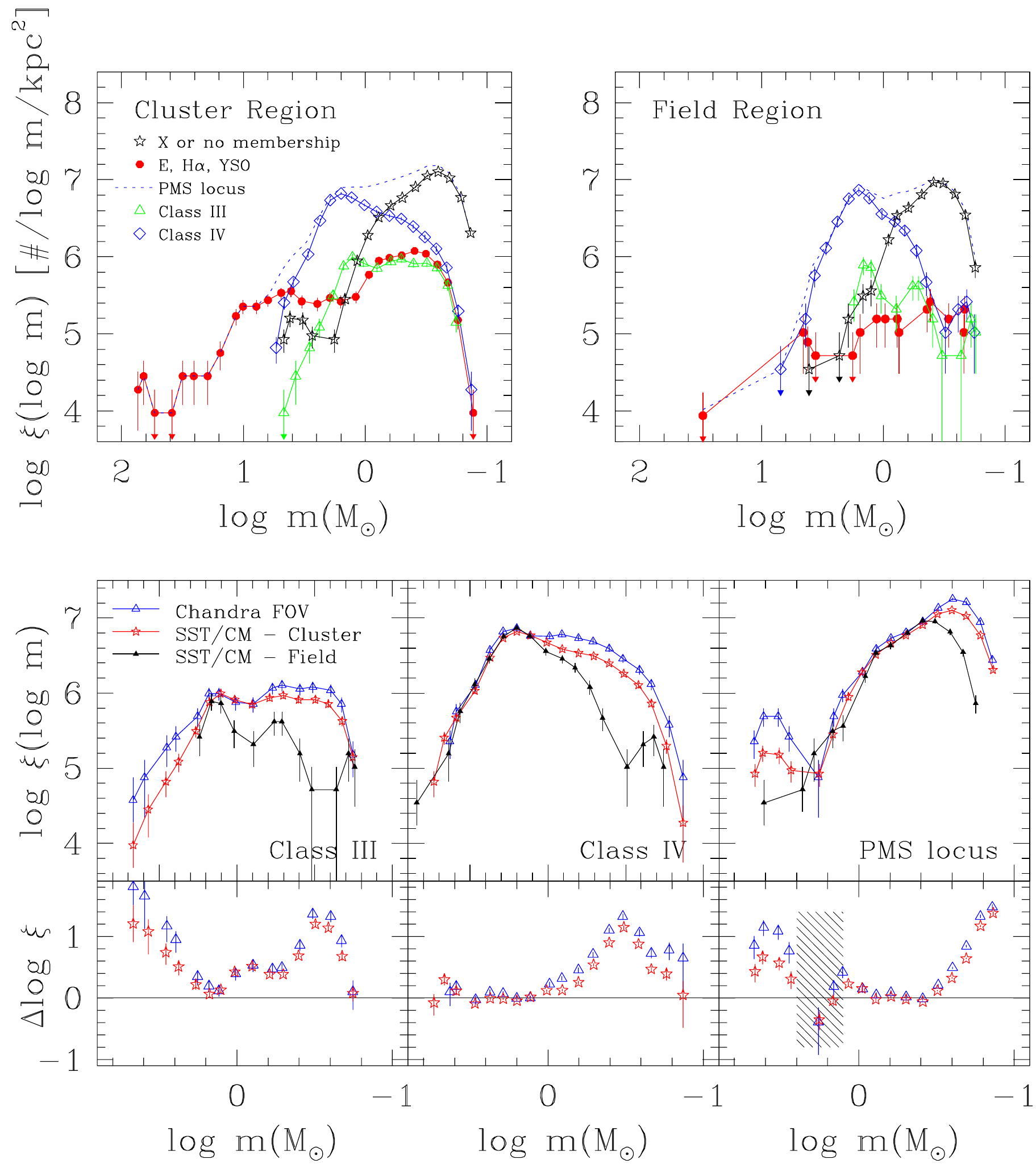}
\caption{(Upper panel) The mass spectrum of each component of the cluster region (left) and 
that of the field region (right). Different components show different contributions to the
total mass spectrum (dashed line). (Middle panel) Mass spectrum of three regions - Chandra FOV, 
SST/CM cluster region, and SST/CM field region - for Class III (left), for Class IV (center), and
for stars in the PMS locus (X-ray emission stars or stars with no membership criterion) (right).
(Lower panel) The difference in the mass spectrum relative to that of the field region. The shaded region
in the right panel represents the over-subtracted range.
\label{mass_sp} }
\end{figure}

The mass of an individual star can be estimated from the HRD.
It is implicitly assumed that all stars are single stars even though the multiplicity of a few
massive stars are known \citep{rdb04,hgb06,dbrm06,rn16}. The effect of
binarity on the shape of the IMF has been discussed in \citet{sb10}.
The mass estimate of massive stars from the HRD
is not easy because a small difference in age gives a very different position in
the HRD, and it is therefore impossible to use a single isochrone or 
mass-luminosity relation. In addition, the complex evolutionary tracks of
massive stars make this matter even more difficult.
We used the same method of estimating the mass of massive stars ($m \geq
20 M_\odot$) as described in \citet{ssb13} based on the stellar evolution tracks of \citet{geneva12}.
The mass of intermediate-mass MS stars was estimated using the mass-luminosity 
relation of the isochrone of age 3.5 Myr. The mass and age of PMS stars were estimated
by interpolating the PMS evolution tracks. 
For consistency with the mass scale of our previous works, and due to the lack
of PMS evolution tracks for $m > 1.4 M_\odot$ in BHAC15,  we estimated 
the mass of PMS stars using the PMS evolution models of SDF00. 
And then the number of stars in a logarithmic mass interval of $\Delta \log m$ = 0.2
was calculated. 

\begin{figure}
\epsscale{0.5}
\plotone{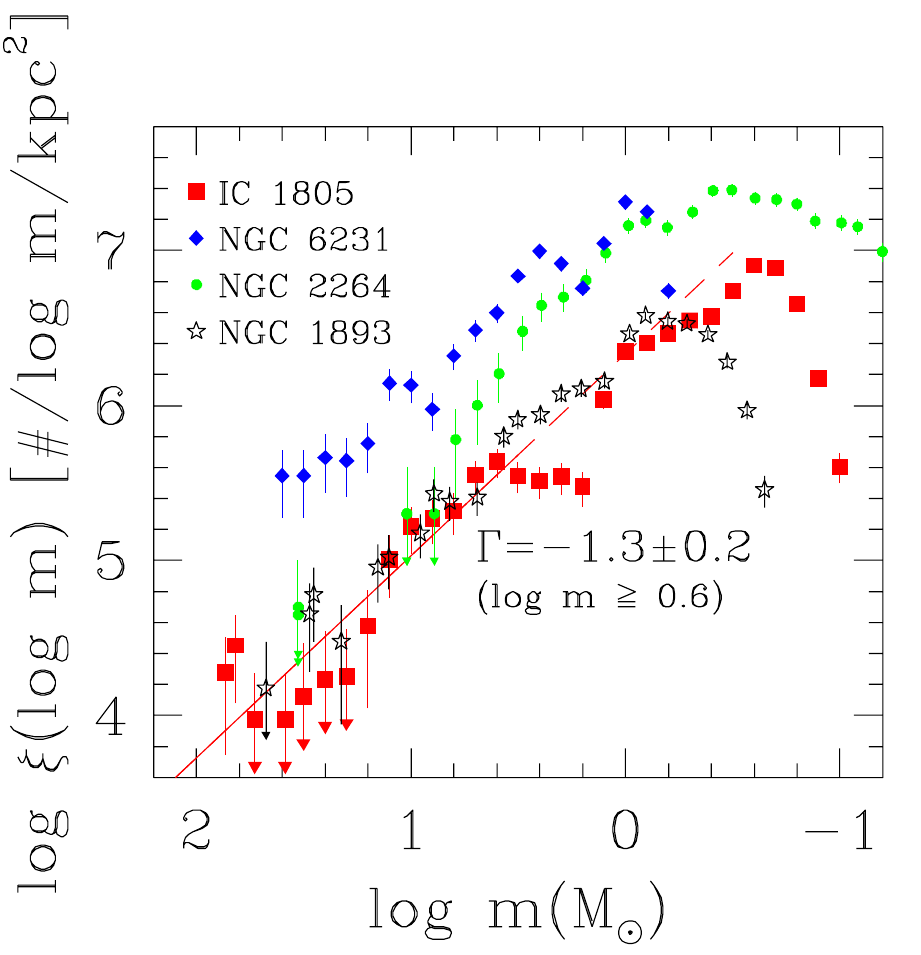}
\caption{The initial mass function of IC 1805 corrected for the contribution of
Cas OB6 association and field stars. The over-subtraction of field contribution
is evident for $\log m$ = 0.1 -- 0.6 as shown in Figure \ref{mass_sp}. The IMF of 
NGC 6231 (diamond), NGC 2264 (dot), and NGC 1893 (star mark)
are also shown for comparison.
\label{figimf} }
\end{figure}

However we should consider two factors when we derive the IMF of IC 1805. The first 
is to select the region where the membership selection is homogeneous. Although
X-ray observation gives the highest membership selection probability \citep{sbc04},
{\it Chandra} or {\it XMM-Newton} observations are restricted to the cluster center.
In addition, H$\alpha$ photometry is shallow in the extreme sourthern and northern regions. 
Despite its lower selection probability, the selection of MIR YSOs is homogeneous
at least in the SST/CM FOV. Although we observed a much larger area, the IMF will
be determined only for the SST/CM FOV.  The next issue is to subtract the contribution of 
Cas OB6 association and field interlopers. To estimate the number of field interlopers
in the PMS locus of IC 1805, we checked the CMD of the nearby old open cluster
Tombaugh 4 \citep{scj10}. The cluster region is slightly more reddened ($E(B-V)
\approx 1.1$), and predicts more high-mass PMS stars ($m \approx 3 \sim 5 
M_\odot$) than the number of stars in the PMS locus.
In addition, the depth of their photometry was too shallow to estimate the contribution
of faint red foreground stars. Finally we decided to search for a control field within
the observed region. We checked the surface density variation of massive stars,
Class III, IV stars, and MIR YOSs, and found that it reached a saturation level between
$r \simeq$ 12$'$ -- 15$'$\footnote{Recently \citet{psp17} derived about 9$'$
for the radius of IC 1805.}. After various trials the southern region of SST/CM FOV
($\Delta \delta \leq 12'$) was selected as the control field (the hatched area in Figure
\ref{YSOdist}). The presumed Cas OB6 association member BD+61 493 is in
the selected field region, and presumably the contribution of the Cas OB6 association
can be subtracted. Although the density of H$\alpha$ emission stars or MIR excess
stars is very low, some number of cluster members could still occur within the control field.
In  that case, over-subtraction of the field contribution is inevitable.

The first step is to check the mass spectrum of each component - members 
(early-type members, H$\alpha$ emission stars, and MIR YSOs), Class III stars,
Class IV stars, stars with X-ray emission or no membership criterion,
and all stars in the PMS locus. The mass spectrum of each component
in the cluster region and field region is shown in the upper panels of Figure \ref{mass_sp}.
In the figure we can easily see that each component has a different contribution
to the total mass spectrum (dotted line) - cluster members dominate in the massive
part, while Class IV objects  and stars with no membership criterion or X-ray emission occupy
the greater part at intermediate-mass and at the low-mass regime, respectively.
The surface density of member stars is very low, but non-negligible in the field region.
BD +61 493 is the only star with $m \geq 10 M_\odot$ in the field region.
In the middle panels of Figure \ref{mass_sp} we compared the surface density
of three regions for three components. The surface density is, in general,
highest in the {\it Chandra} FOV and lowest in the field region. However the difference
in surface density is a strong function of mass as shown in the lower panels.
The surface density of stars in the PMS locus with no membership criterion
or X-ray emission is higher than that of the cluster region and even that of {\it Chandra}
FOV in the mass range of $\log m$ = 0.1 -- 0.6, which is due to higher contribution of
background MS stars as can be seen in the right panel of Figure \ref{ebvmap} 
(the surface density of field MS stars is lowest in the center). The shaded region
in the lower-right panel represents the mass range of over-subtraction.

The IMF of IC 1805 is derived by subtracting the contribution of the Cas OB6 association
and field stars (SST/CM - Field) from the mass spectrum of the cluster region
(SST/CM - Cluster). The net number of cluster stars for each component
(members, Class III, Class IV, stars in the PMS locus with no membership or
X-ray emission) is calculated for the given mass range, and
then we calculate the IMF of IC 1805. The calculation is performed for the interval
of $\Delta \log m$ = 0.1 for $\log m <$ 1.5 if the net number of cluster
stars is larger than 0. The final IMF of IC 1805 is presented in Figure \ref{figimf}.
The IMFs of NGC 1893, NGC 2264, and NGC 6231 are also shown for comparison.
The IMF of IC 1805 shows a large fluctuation for massive stars ($\log m \gtrsim$
1.0), a peak at $\log m \approx$ -0.6 -- -0.7, and then declines rapidly due to
the incompleteness of the photometry. The over-subtraction of the field contribution
is evident in the mass range $\log m$ = 0.1 -- 0.5. The slope of the IMF of
IC 1805 is $\Gamma = -1.3 \pm 0.2$ for $\log m \geq 0.6$, which is very 
similar to that of NGC 1893 ($\Gamma = -1.3 \pm 0.1$, \citealt{lsk14b}) in the Perseus arm,
but slightly steeper than that of NGC 6231 ($\Gamma = -1.1 \pm 0.1$, \citealt{ssb13}),
or shallower than the nearby young open cluster NGC 2264 ($\Gamma = -1.7 \pm 0.1$, \citealt{sb10}).

\begin{figure}
\epsscale{0.5}
\plotone{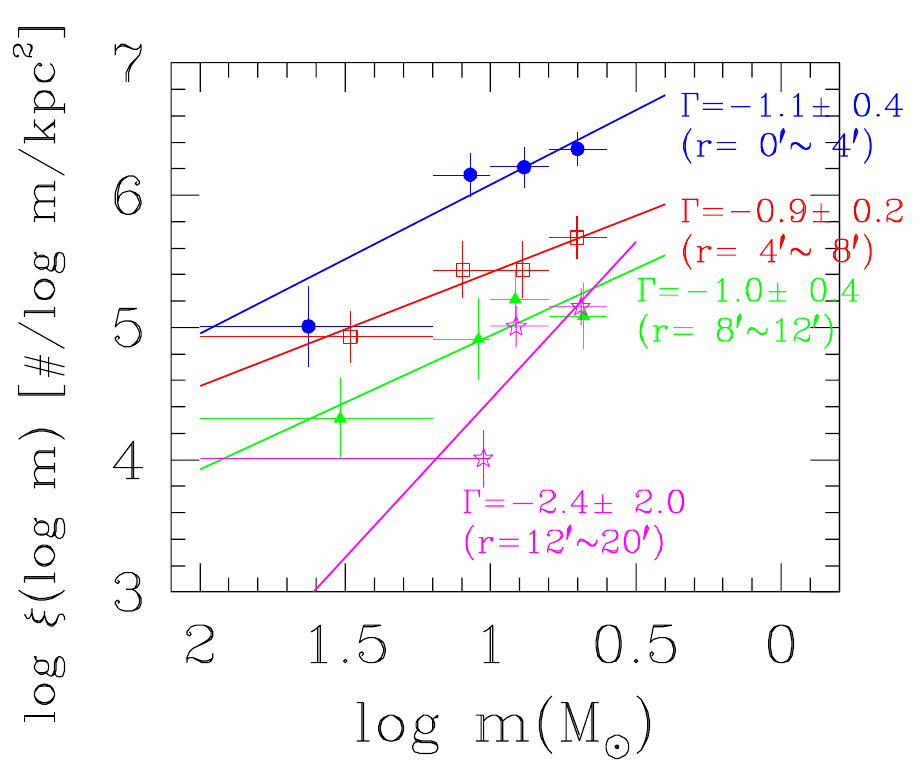}
\caption{Radial variation of the IMF of massive stars. The slopes
of the IMF of the inner 3 annuli have nearly the same value, but
it increases abruptly at $r > 12'$.
\label{figvimf} }
\end{figure}

Massive stars in young open clusters are concentrated to the center.
The origin of such mass segregation is still uncertain. \citet{ssb13}
prefer a primordial origin (see also \citealt{lkj16,kjdf16}), however many theoreticians favor the
dynamical origin under the subvirial condition (e.g. \citealt{mvpz07}).
As mentioned in the introduction the surface density of stars in 
IC 1805 is very low, and so it is sometimes called a ``stellar aggregate'' or
``association''. To address whether mass segregation is prevailing
in IC 1805 or not, we checked the radial variation of the IMF
of the massive stars as shown in Figure \ref{figvimf}. The slopes of
the IMF for $r \lesssim 12'$ (equivalently about 8.4 pc at $d$ = 2.4 kpc) are
nearly the same ($\Gamma \approx -1.0 \pm 0.1$).
No O-type star can be found outside this radius. The only O-type
equivalent massive star outside the radius is BD +60 493 (B0.5Ia) which
is considered a member of the Cas OB6 association.
This fact also implies that the radius of IC 1805 is not much larger than 12$'$.
And mass segregation of massive stars is not evident, at least for $r
< 12'$ in IC 1805. 

\subsection{Total Mass of IC 1805}

The total mass of a cluster is an important parameter.  \citet{wkb10} estimated
the total mass of IC 1805 as 10885$_{-5528}^{+11137} M_\odot$. The mass was
originally estimated to be 14400 $M_\odot$ by \citet{wsd07} by assuming that
the stars in the mass range 6 -- 12 $M_\odot$ constitute 5.5\% of total mass
and the number of early B type stars in IC 1805 is 99. However, the
total mass of member stars selected in section 5 is about 1800 $M_\odot$. 

The mass of IC 1805 can be estimated by the direct integration of the IMF.
Before integrating the IMF we should slightly modify the IMF in Figure \ref{figimf}:
we replaced the IMF in the mass range $\log m$ = 0.1 -- 0.5 by the dashed line 
and corrected the IMF for $\log m \leq$ -0.7 using the IMF of NGC 2264 by 
assuming that the difference in the IMF between a given mass and $\log m$ =
-0.7 is the same amount. This calculation gives 2110 M$_\odot$ , which is
definitely a lower limit because we cannot take into account the contribution
of multiple systems among member stars. We also calculated the total mass
of stars with masses larger than 5 $M_\odot$ and 7 $M_\odot$, and obtained
515.3 $M_\odot$ and 595.3 $M_\odot$, respectively.

To derive the total mass of IC 1805, we simulated a model cluster with the given IMF
mentioned above and binarity. To be a more realistic model cluster, multiplicity
should be taken into account. But it is virtually impossible to do so because
we do not have enough information on the frequency distribution of
multiples and the mass ratio distribution and can only take the binary
fraction distribution and its mass ratio distribution summarized in \citet{dk13}.
From several simulations we calculated the ratios between the total cluster mass
and the total primary mass ($M_{total}/M_p$), the mass of the primaries larger than 5 $M_\odot$
[$M_{total} / M_p (\geq 5 M_\odot)$], and that larger than 7 $M_\odot$
[$M_{total} / M_p (\geq 7 M_\odot)$]. The ratios are 1.344, 4.310, and 5.128, respectively.
The mass of IC 1805 (the cluster region selected in section \ref{imf}) is estimated to be 2690 $\pm$ 190 $M_\odot$.
The upper limit of the cluster mass 
also can be derived using the total mass of all member stars in the cluster region with masses larger than 
5 $M_\odot$ and 7 $M_\odot$ and the ratios above,
and obtained 3710 $\pm$ 30 $M_\odot$. And the number of
O-type stars ($m \geq 15 M_\odot$) is also estimated to be 6.9 $\pm$ 1.0,
which is in agreement with the number of observed O stars (see introduction).

This total mass of IC 1805 is far lower than the cluster mass estimated by 
\citet{wsd07} and \citet{wkb10}. Were the cluster mass of IC 1805 similar to that estimated
by \citet{wkb10}, we would expect to find about 29 O-type primaries ($m \geq
15 M_\odot$) (+16 O-type secondaries) rather than the actual content of 8 or 9. 
Their large estimated cluster mass is probably caused by the inclusion of early B-type stars 
belonging to the Cas OB6 association.

\section{DISCUSSION}

\subsection{Distance of W3 and W4 \label{shape}}

Most investigators implicitly assumed that the three active SFRs W3, W4, and W5 in
the Cas OB6 association are at the same distance
(e.g. \citealt{mto08}). The reason for this assumption is
that star formation in W3 is considered to have been triggered by the massive
young open cluster IC 1805 because this region of the Galaxy
has for a long time been thought of as the site of triggered sequential
star formation. However, there is no direct evidence of
triggered star formation in W3 by W4. The young open cluster in W3, IC 1795, is nearly the same
age as IC 1805 \citep{owkw05}, and the three young SFRs in W3
(W3 Main, W3(OH), and W3 North) show far different
populations \citep{ft08}, indicating that different star formation
mechanisms operated in the different SFRs.
The distance of IC 1805, derived from the ZAMS fitting, gives 
2.4 ($\pm$ 0.2) kpc, which is consistent with the distance obtained by
previous optical investigators as mentioned in section \ref{dist_cmd}.
For a long time, the distance from ZAMS fitting to clusters
provided the most accurate distances and was the most important step 
in the distance ladder out to the distance of external galaxies in the local group,
However,
this assumption has been challenged by the emergence of $\mu$-arc-second ($\mu$as)
accuracy astrometry from radio VLBI observations
\citep{rh14}.  \citet{xrz06} and \citet{hbm06} measured the parallax of
a methanol maser and H$_2$O maser in W3(OH), respectively, and 
obtained a consistent distance of 2.0 kpc. More recently \citet{mhi11}
obtained an even smaller distance of 1.67$_{-0.17}^{+0.21}$ kpc based on
methanol maser emission from an ultracompact H II region in W3(OH)
from a shorter baseline observation.
Although the internal error of radio VLBI astrometry is very small
(less than 1 $\mu$as), there could be several sources of systematic
external errors - variability of maser sources (very few maser spots 
persist for more than a year), spatial motion of maser spots, variation
of the centroid position of astrometric references due to variability
or jet ejection from AGNs, up to a few $\mu$as error due
to the zenith delay correction, and some systematic sensitivity
variation due to the angular offset of the astrometric reference.
And therefore unknown external errors could be much larger
than the published internal errors. 
Most recently, the astrometric satellite {\it Gaia} released the first result
from the data collected during the first 14 months. \citet{gaia} compared
the {\it Gaia} astrometric data with those from radio VLBI astrometry
for a representative sample, and
showed large differences for some objects (see their Table C.1).
To check the reliability of the ZAMS relation, we have selected
proper motion and parallactic members of 34 nearby open clusters ($d
\lesssim$ 1 kpc) using the {\it Gaia} DR1 TGAS data, and found that 
their parallactic distance is well consistent with the ZAMS-fitting distance
within the scatter of parallaxes (about 0.2 mag) among the selected members. 

Although the reddening-free indices used in this paper are relatively immune to
variations in abundance, it is better to check the effect of
abundance differences on the ZAMS relation. According to \citet{glb14} the abundance
gradient from $\delta$ Cepheid variables, which represent the young
population in the Galactic disk, is [Fe/H] = 0.49 ($\pm$ 0.03) - 0.051
($\pm$ 0.003) $R_{GC}$ (kpc). The abundance at IC 1805 ($R_{GC}
\approx$ 10.3 kpc)  is estimated to be about -0.035 dex. And the gradient between
$R_{GC} \approx$ 7 -- 10 kpc is much shallower than the average slope
(see the lower panel of Fig. 4 of \citealt{glb14}), therefore
the abundance near IC 1805 is probably close to the Solar value,
therefore there should be no difference in abundance, hence no impact on distance determination.
However, the abundance of the stars in IC 1805
may be lower than the value estimated above [e.g. [Fe/H]
$\approx$ -0.13 if we adopt the abundance gradient from
red giant stars ($d {\rm [Fe/H]} / d R_{GC}$ = -0.07 for $\tau <$
1 Gyr, \citealt{acm16})], the bolometric magnitude difference of
the ZAMS relation of massive stars between [Fe/H] = 0.00 and -0.13
is estimated to be about 0.14 ($\pm$ 0.02) mag  at a given
temperature from the stellar evolution models of \citet{geg13}.
The abundance difference from the Solar metallicity, if true, may
reduce slightly the difference between the distance of IC
1805 from the ZAMS fitting and that of W3(OH) from radio
astrometry.

Recently, \citet{bbb16} determined a distance of 1.7 $\pm$ 0.2 kpc 
to the early-type eclipsing binary DN Cas in the Cas OB6 association
and classified DN Cas as a B0V+B1V system.
However, \citet{h56} classified the star as a O8Vvar which would make a 
difference to the deduced distance. 
The spectra in Figure 2 of \citet{bbb16},
as well as our unpublished high resolution echelle
spectra obtained with the Bohyun-san Observatory Echelle Spectrograph
(BOES - \citealt{boes}), show many Helium lines, such as He II $\lambda\lambda$
4200, 4541, 4686, He I $\lambda\lambda$ 4026, 4144, 4387, 4471,
and C III $\lambda\lambda$ 4647/4650/4651. The strength of
He II $\lambda$4541 is slightly stronger than He I $\lambda$4388,
and He II $\lambda$4200 is stronger than He I $\lambda$4144.
The strength of C III $\lambda\lambda$ 4647/4650/4651 of
the secondary [see Figure 2 of \citet{bbb16} at phase 0.996]
is slightly stronger than that of He II $\lambda$ 4686, while that
of the primary with longer wavelength, is opposite. These features
indicate that DN Cas is a binary system but with a O8V primary and
O9.5V secondary. The absolute magnitude difference between O8V and
B0V is about 0.85 mag \citep{sos}, and therefore the distance to
DN Cas may be about 2.5 kpc ($V_0 - M_V \approx 12.0$).
And hence the smaller distance obtained by \citet{bbb16} is probably
caused by the misclassification of the spectral type.

Now we should re-consider whether the active SFRs W3-W4-W5
are at the same distance or not. Normally the width of a spiral
arm is considered to be about 500 pc. Currently the number of
SFRs measured with accurate radio VLBI astrometry is
over 100 \citep{rh14}. The distances to many SFRs in the Perseus
arm  were recently published by \citet{chr14}. Using the data
in their Table 5 we calculated the width of the Perseus spiral arm
- $\Delta d$ = 0.9 kpc at $l \approx 95^\circ$, $\Delta d \gtrsim$
1.2 kpc at $l \approx 108^\circ$, and $\Delta d$ = 0.7 kpc at 
$l \approx 111^\circ$. As we did not take into account 
the orientation of the Perseus spiral arm to the line of sight,
the difference $\Delta d$ above is larger than the actual width of the Perseus arm,
however the actual width  may be similar to or larger than 0.5 kpc. The radio VLBI astrometry of three
SFRs in the superbubble around the young open cluster NGC 281
($l \approx 123^\circ$) shows a large difference in 
distance - IRAS 00420+5530: 2.2 $\pm$ 0.05 kpc and 
NGC 281 W: 2.8 $_{-0.22}^{+0.26}$ kpc \citep{ssm13}.
The size of the W4 superbubble is much larger than that of NGC 281,
and so we can expect a much larger size for the W4 region.
\citet{rsh01} could not find any sign of H$\alpha$ line splitting,
and concluded that the large H$\alpha$ emission structure 
has a loop shape, rather than a shell structure. They considered
that the loop could be of
cylindrical shape and its radial extent
similar to its extent on the sky.
In addition, the size of H I holes (H I superbubbles) from
the H I nearby galaxy survey \citep{bbw11} range from
about 100 pc (limited by the resolution of radio observation)
to about 2 kpc. From the information gathered above we can state that
there are no reasonable physical grounds for assuming that 
the giant H II regions W3 and W4 are at the same distance.

\begin{figure}
\epsscale{0.5}
\plotone{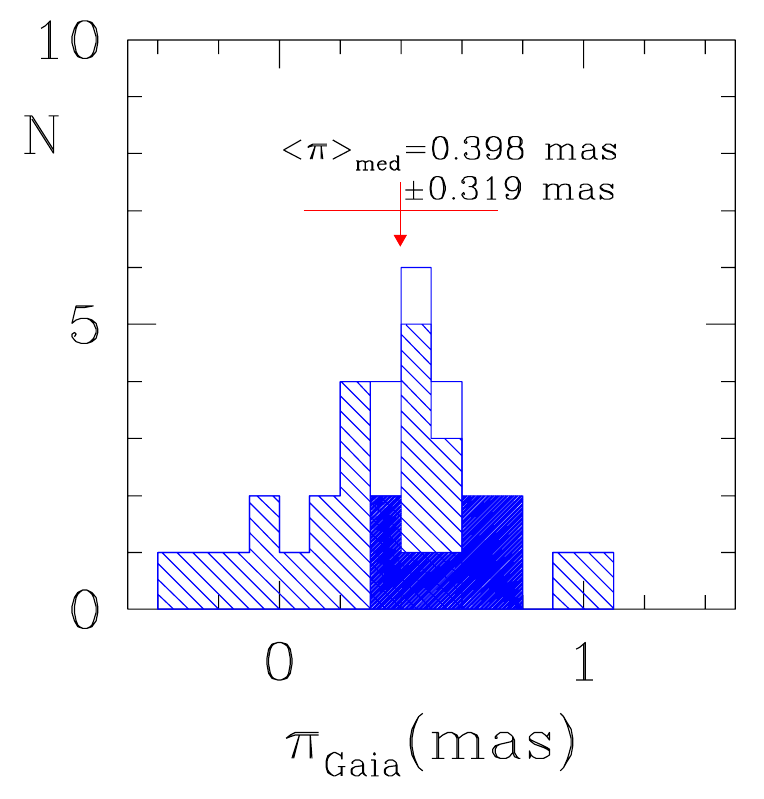}
\caption{The distribution of the TGAS parallaxes of 32 O- and B-type 
stars in IC 1805. The dark, shaded, and open histograms represent
the parallax distribution of O, B, and all 32 stars, respectively.
\label{gaia1} }
\end{figure}

We searched for stars within 30 arcmin of IC 1805 in the Tycho-Gaia
astrometric solution (TGAS) catalog \citep{gaia}, and retrieved data for 103 stars.
Among them, 33 stars are O- or early B-type stars, and one of them 
(ALS 7225 = KM Cas, O9.5V((f))) is close to the western edge of W4.
The latter star is not considered as a member of IC 1805, and is excluded
in the statistics. Figure \ref{gaia1} shows the distribution of the TGAS parallaxes.
The parallaxes for eight bright O-type stars were more concentrated around $\pi$ = 0.56
(mean value) ($\pm$ 0.16) mas, while those for 24 B-type stars scattered
more widely with a median value of 0.32 ($\pm$ 0.33) mas. The median value for
all 32 early-type stars is 0.40 ($\pm$ 0.32) mas (equivalently
$d$ = 2.5 (1.4 -- 13) kpc), which is very similar to the distance
obtained from the ZAMS fitting in section \ref{dist_cmd}. However,
the error of the parallax measurements, even for the TGAS catalog, is still
very high (about 0.27 mas for 8 O type stars), so we will have to wait for a few more years
to get a better distance to IC 1805 from Gaia stellar parallaxes.

As mentioned in section \ref{reddening}
the clouds associated with W3 are in front of IC 1805.
Assuming that the radial distribution of faint MS field stars below the PMS locus
is homogeneous, the surface density difference in Figure \ref{ebvmap} between
IC 1805 and the PAH emission nebula in the west of the observed FOV 
(probably at the same distance as W3) that is
about 15\%, can be converted into 0.36 kpc. This value
is very similar to the difference in distance between W3(OH) and IC 1805. 
From the morphology and distance of the region, we can sketch a
picture of the region. The young open cluster IC 1805,
or more exactly, most of the massive stars, may be at the far
side of the bubble. However, we cannot rule out the possibility
of a large radial scatter of member stars inside the bubble
that can be speculated from the broad MS band of early-B
type stars in the CMDs (see section \ref{hrd}).
The active SFR W3 is at the western edge of the bubble,
and therefore we can expect a large spread in distance
among the stars/SFRs in W3. The reddening-free
CMDs of IC 1795 using the data in \citet{owkw05}
indicate that the distance modulus of the cluster is more
appropriate for 11.9 mag rather than 11.5 mag ($d$ = 2 kpc).

\subsection{Do the O and B stars in IC 1805 have a different origin? \label{propermotion}}

\begin{figure*}
\epsscale{0.9}
\plotone{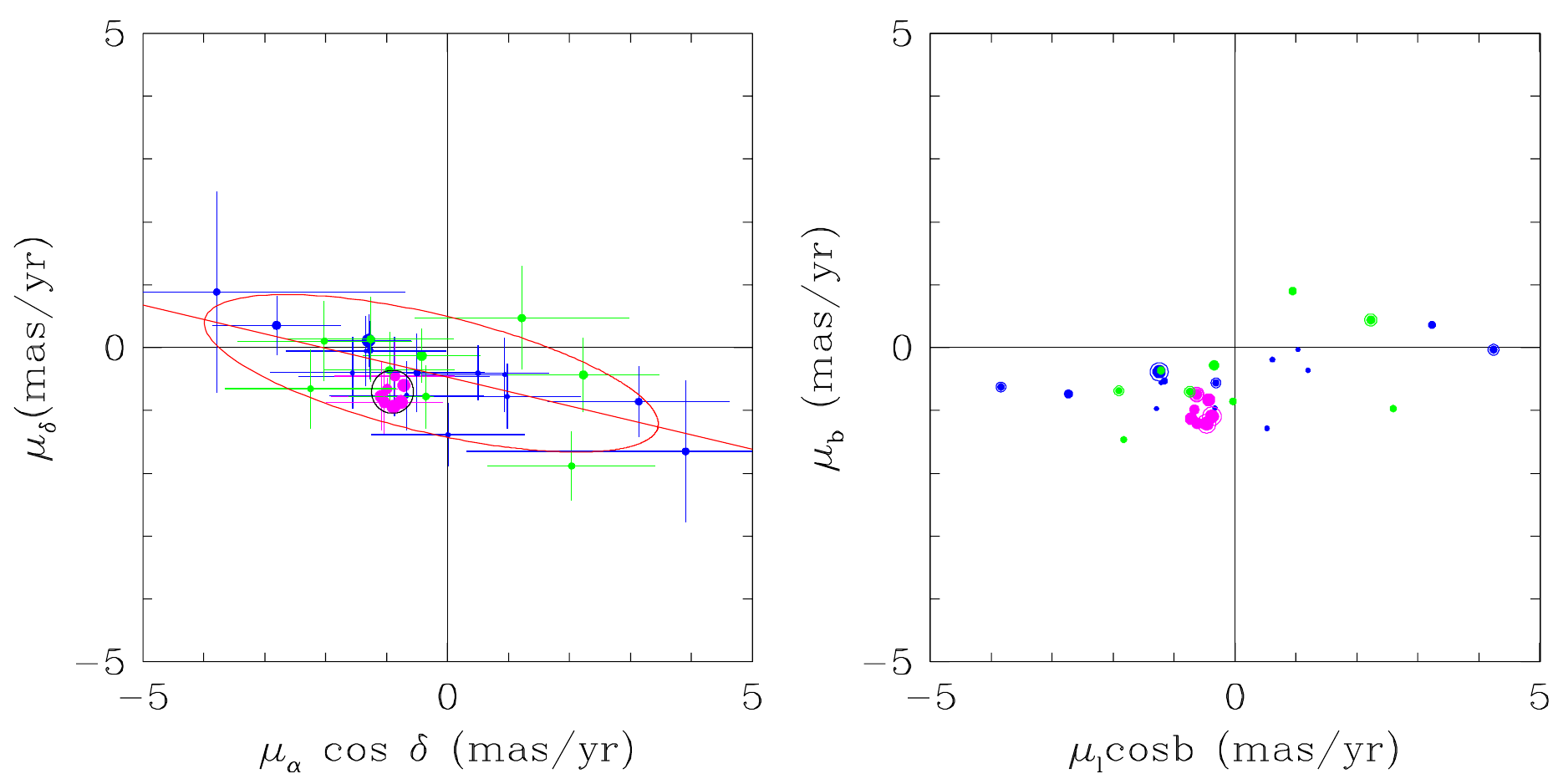}
\caption{TGAS proper motions of O- and B-type stars in IC 1805
in the equatorial coordinate system (left) and in the Galactic coordinate
system (right). Magenta, blue, and
green dots represent O-type stars, normal B-type stars, and
bright B-type stars in IC 1805, respectively.
The ellipses in the left panel show the distribution of the proper 
motion of B-type stars (red) and that of O-type stars (black),
respectively. The red straight line indicates the major axis
of the ellipse.
\label{gaia2} }
\end{figure*}

As mentioned in section \ref{hrd}, there are many early B-type stars that are
apparently brighter than normal stars, and \citet{gv89} considered
them to be an older group with an age of about a few 10 Myr. We looked
for differences in the spatial or kinematic properties among the B-type stars, but
could find no difference in the spatial distribution between the brighter and fainter groups. 
In addition, we
homogenized all available proper motion data (Tycho - \citet{tycho},
PPM - \citet{ppm}, USNOB - \citet{usnob}, USNO ACT - \citet{act}, and 
\citet{vsa65}; all data were linearly transformed to the Tycho system, and then
merged with an appropriate weight), but we could not find any discrete
group among them.

This issue could be disentangled when high quality proper motion and
parallax data from {\it Gaia} become available. However the quality of the currently available
TGAS catalog is not high enough, and we can find no subgroupings among the B-type
stars. Figure \ref{gaia2} shows the TGAS proper motions of O- and early
B-type stars in IC 1805.
Bright B-type stars are colored in green, while
normal (fainter) B-type stars are in blue. Normal B-type stars show a slightly larger
scatter, but the extreme points are due to a larger error. And therefore
we can conclude that there is practically no difference in proper motion
between the bright and normal B-type stars.

The proper motions of the B-type stars show a large scatter as well as an elongated
distribution in the proper motion plane. The principal axis of the distribution is rotated by
approximately 13 degree about an east-west direction, which is nearly parallel to the
Galactic plane as shown in Figure \ref{gaia2} (Right) and the PAH emitting
nebula just behind IC 1805.
And interestingly, the standard deviation along the principal axis is
four times larger than that along the minor axis. The proper
motion of 8 O-type stars, however, is well localized in a small area in the proper
motion plane and the distribution of the O-type stars in the proper motion plane
is nearly circular with an axial ratio of about 1.1. Although the proper motion
of B-type stars has a large error, the difference in the distribution
is remarkable.

If this is a real feature, does it reflect the difference in the formation processes
between O- and B-type stars? A possible explanation for this feature,
if real, may be the differences in the formation environments.
The O-type stars may have formed in the dense central part of a molecular cloud,
and therefore they show a small dispersion in proper motion.
Whereas, the B-type stars may have two different formation scenarios -
one group of B-type stars could have formed from the same molecular cloud as the
O-type stars, while the other group of B-type stars could have formed in small clouds 
dispersed from the center along the Galactic plane. This systematic motion implies
that the formation of these B-type stars were affected by an internal trigger.
The latter case is similar to the star formation activity in the small clouds
scattered in W4 found by \citet{chs00}, but the expanding direction is along
the Galactic plane. 

\subsection{LS I +61 303 and Star Formation in IC 1805}

IC 1805 (or the Cas OB6 association) is considered as the birth
place of the high-mass X-ray binary LS I +61 303
\citep{mrl04}. LS I +61 303 is classified as a Be/X-ray binary
system, but the nature of the compact object is still controversial,
whether it is a neutron star (pulsar) or a black hole (micro-quasar).
The variable radio counterpart of LS I +61 303 has been
resolved as a rapidly precessing relativistic jet \citep{mrp01}.
More recently, \citet{dmr06} could not find any relativistic
motion, and supports a pulsar wind nebula model.
The mass of the system is estimated as 14 $\pm$ 2 $M_\odot$
for the B0Ve primary star and 2 $\pm$ 1 $M_\odot$ for
the compact object, and the mass loss due to the supernova (SN) 
explosion is estimated to be less than 2 $\pm$ 1 $M
_\odot$ \citep{mrl04}. 

The distance estimate to the object was
attempted using radio as well as optical wavelengths.
\citet{fh91} detected an HI 21cm absorption feature at 
$v_r \approx$ -45 km s$^{-1}$, and interpreted the 
velocity component as the spiral arm shock at the nearside
edge of the Perseus arm. As they could not find any
velocity component associated with the cold interstellar
gas in the Perseus arm, they suggested that LS I +61 303
is in the Perseus arm just behind the spiral arm shock,
but in front of the main Perseus arm. They estimated the
distance to LS I +61 303 as 2.0 $\pm$ 0.2 kpc.
Later, \citet{snc98} obtained the same distance
to the object based on spectral classification, reddening
estimation, and the Sp-M$_V$ relation. But as the distance
estimated by \citet{fh91} is based on the spiral arm model
of the Galaxy and that by \citet{snc98} is hampered by
the peculiar nature of the object, the distance to LS I 
+61 303 remains uncertain.

To assess the theories on the origin of LS I +61 303, it is
necessary to check (1) the star formation history in IC 1805
and the surrounding Cas OB6 association, and (2) the astrometric properties of LS I +61 303. 
The probable members of the Cas OB6 association are listed in
\citet{gs92,h78}. We selected only Cas OB6 members distributed
around IC 1805, and placed them in the HRD. The ages of the evolved
association members were between 6.0 and 13 Myr, which is consistent
with the age of the shell structure of the
superbubble \citep{dts97}, while that of the
unevolved members is very similar to that of IC 1805 and IC 1795
\citep{owkw05}. If we focus on the stars in the W3 - W4 region,
the age of the evolved stars is less than about 10 Myr (their age
relies strongly on the membership and spectral type). However,
the most reliable age can be obtained from the highly evolved stars in 
the region [BD +60 493 (B0.5Ia)  and OI 109 (O9.7Ia) in IC 1795 -
\citet{owkw05}], and that is about 7.5 Myr. The age of IC 1805
and IC 1795 is about 3.5 Myr, so the age difference between  the two young clusters 
and these evolved stars is about 4 Myr, which is the lifetime
of a very massive star. The most massive stars among the former generation
of stars (the members of the Cas OB6 association) may have
exploded as SNs about 4 Myr ago and the formation of IC 1805 and IC 1795 in W3
may have been triggered by the SN explosions. The micro-quasar candidate
LS I +61 303 could be a remnant of the previous SN explosions.

The above cluster formation scenario may be feasible if we could find
other independent information supporting the scenario. The angular distance
between LS I +61 303 and the center of IC 1805 is 58$\farcm$22 ($\Delta \alpha$ = 
56$\farcm$24, $\Delta \delta$ = -14$\farcm$23), and therefore the expected proper
motion of LS I +61 303 should be ($\mu_\alpha \approx$ +0.84 mas yr$^{-1}$, 
$\mu_\delta \approx$ -0.21 mas yr$^{-1}$). If the SN explosion have been
exploded somewhere between IC 1805 and IC 1795, then the expected
$\mu_\alpha$ should be larger than the value.
\citet{mrl04} tried to find the birth place of LS I +61 303 using
the proper motion data by \citet{lpj99}. Later,
\citet{dmr06} provided a more accurate proper motion  of LS I 
+61 303 [($\mu_\alpha$, $\mu_\delta$) = (-0.302 $\pm$ 0.07,
-0.257 $\pm$ 0.05) mas yr$^{-1}$], which is very similar to
the proper motion from the TGAS catalog [($\mu_\alpha$,
$\mu_\delta$) =  (-0.354 $\pm$ 0.267, -0.077 $\pm$ 0.211) mas yr$^{-1}$].
Furthermore the proper motion vector of LS I +61 303 is very similar
to that of IC 1805 traced by the O- and early B-type stars.
From the current astrometric data,
there seems to be no causal relationship between the star formation in
IC 1805 and the formation of LS I +61 303.

\section{Summary and Conclusions}

In this paper we presented deep optical and MIR photometry for about 
100,000 stars in a $41' \times 44'$ area of the young open
cluster IC 1805 in the Perseus spiral arm. We selected cluster members
from optical TCDs and CMDs, H$\alpha$ photometry, X-ray emission,
and MIR excess stars from MIR TCDs and SED slope. The low-mass
PMS stars selected from H$\alpha$ emission and/or MIR excess emission
spread over the whole observed region with no strong concentration. 

The total to selective extinction ratio of IC 1805 was determined
from the color-excess ratios of optical to MIR colors, and found 
to be fairly normal ($R_V$ = 3.05 $\pm$ 0.06). The distance modulus of IC 1805
was determined from the reddening-free CMDs, and is 11.9 $\pm$
0.2 mag ($d$ = 2.4 $\pm$ 0.2 kpc) which is about 0.4 kpc farther
than the nearby SFR W3(OH). The massive stars in IC 1805 are
well matched to the isochrone of age 3.5 Myr, while the low-mass
PMS stars have a median age of 2.4 Myr or 1.6 Myr depending
on the adopted PMS evolution models. Although there are
many massive stars in IC 1805, the shape of the IMF is
still bumpy. The slope of the IMF of IC 1805 is nearly Salpeter
value ($\Gamma = -1.3 \pm 0.2$). The shape of the IMF extrapolated
down to the brown dwarf regime and a Monte Carlo simulation of
a model cluster accounting for the binary frequency and mass ratio
distribution of binary system were used to estimate a total mass for IC 1805
of about 2700 $\pm$ 200 M$_\odot$, which is
far smaller than the mass of IC 1805 proposed by \citet{wkb10},
but in agreement with the number of massive O 
stars in the cluster.

Using the recently released astrometric data from the {\it Gaia} mission
we found the median value of the parallaxes of 32 O- and
early B-type members of IC 1805 to be 0.40 ($\pm$ 0.32) mas [$d$
= 2.5 (1.4 -- 13) kpc], which is similar to the parallax from the photometric
distance. In addition, the proper motions of early B-type stars show
an elongated distribution along the Galactic plane, while those of the O-type
stars are well localized. This feature implies that some B-type stars
were likely formed from small clouds dispersed by previous episodes of
massive star formation \citep{chs00} or previous supernova explosions.

\acknowledgments 
The authors thank the anonymous referee who gave many helpful comments
and suggestions. H.S. would like
to express his thanks to Dr. Inwoo Han, the President of Korea Astronomy and
Space Science Institute for hosting him as a Visiting Researcher. 
H.S. acknowledges the support of the National Research Foundation of Korea 
(Grant No. NRF-2015R1D1A1A01058444).

\vspace{5mm}
\facilities{Spitzer (IRAC and MIPS), CFHT, Maidanak:1.5m, XMM-Newton, Gaia}

\software{IRAF, IDL, SM}

\end{document}